\documentclass[epj]{svjour}
\usepackage{graphicx}
\usepackage{color}

\def\mb#1{\setbox0=\hbox{$#1$}\kern-.025em\copy0\kern-\wd0
\kern-0.05em\copy0\kern-\wd0\kern-.025em\raise.0233em\box0}

\begin{document}
   \title{Inhomogeneous Tsallis distributions in the HMF model}

 \author{P.H Chavanis and A. Campa}

\institute{$^1$ Laboratoire de Physique Th\'eorique (IRSAMC), CNRS and UPS, Universit\'e de Toulouse, F-31062 Toulouse, France\\
$^2$ Complex Systems and Theoretical Physics Unit, Health and Technology Department, Istituto Superiore di Sanit\`a, \\
and INFN Roma 1, Gruppo Collegato Sanita, 00161 Roma, Italy}

\titlerunning{Inhomogeneous Tsallis distributions for the HMF model}

   \date{To be included later }

   \abstract{We study the maximization of the Tsallis functional at fixed
   mass and energy in the HMF model. We give a thermodynamical and a
   dynamical interpretation of this variational principle. This leads
   to $q$-distributions known as stellar polytropes in
   astrophysics. We study phase transitions between spatially
   homogeneous and spatially inhomogeneous equilibrium states. We show
   that there exists a particular index $q_c=3$ playing the role of a
   canonical tricritical point separating first and second order phase
   transitions in the canonical ensemble and marking the occurence of
   a negative specific heat region in the microcanonical ensemble. We
   apply our results to the situation considered by Antoni \& Ruffo
   [Phys. Rev. E {\bf 52}, 2361 (1995)] and show that the anomaly
   displayed on their caloric curve can be explained naturally by
   assuming that, in this region, the QSSs are polytropes with
   critical index $q_c=3$. We qualitatively justify the occurrence of
   polytropic (Tsallis) distributions with compact support in terms of
   incomplete relaxation and inefficient mixing (non-ergodicity). Our
   paper provides an exhaustive study of polytropic distributions in
   the HMF model and the first plausible explanation of the surprising
   result observed numerically by Antoni \& Ruffo (1995). In the
   course of our analysis, we also report an interesting situation
   where the caloric curve presents both microcanonical first and
   second order phase transitions.  }

   \maketitle
%

\section{Introduction}
\label{sec_model}

Systems with long-range interactions are numerous in nature. Some
examples include self-gravitating systems (galaxies), two-dimensional
turbulence (vortices), chemotaxis of bacterial populations (clusters)
and some models in plasma physics \cite{houches}. These systems are
fascinating because they present striking features that are absent in
systems with short-range interactions such as negative specific heats
in the microcanonical ensemble, numerous types of phase transitions,
ensembles inequivalence, unusual thermodynamic limit, violent
collisionless relaxation, long-lived quasi stationary states (QSS),
non-Boltzmannian distributions, out-of-equilibrium phase transitions,
re-entrant phases, non-ergodic behavior, slow collisional relaxation,
dynamical phase transitions, algebraic decay of the correlation
functions... We refer to \cite{cdr,tsallisbook} for some recent
reviews on the subject.

In order to understand these strange properties in a simple setting, a
toy model of systems with long-range interactions has been actively
studied. It consists of $N$ particles moving on a ring and interacting
via a cosine potential. This model has been introduced by many authors
[4-8] at about the same period with different
motivations (see a short history in \cite{cvb}). Let us mention for
example that Pichon \cite{pichon} introduced this model to explain the
formation of bars in disk galaxies, giving it a physical
application. This model is now known as the Hamiltonian Mean Field
(HMF) model. This abbreviation was coined by Antoni \& Ruffo \cite{ar}
and it stood at that time either for Hamiltonian or Heisenberg Mean
Field model since the HMF model can be viewed as a mean field XY model
with long-range interactions (i.e. not restricted to the nearest
neighbors). In fact, this model was first introduced much earlier
(this is not well-known) by Messer \& Spohn \cite{ms} and called the
cosine model\footnote{These authors mention that this ``cosine model''
was suggested by G. Battle
\cite{battle}.}. They rigorously studied phase transitions in the canonical ensemble by evaluating the free energy and exhibited a second order phase transition between a homogeneous phase and a clustered phase below a critical temperature $T_c=1/2$.

In their seminal paper, Antoni \& Ruffo \cite{ar} studied the
statistical mechanics of this model in the canonical ensemble (CE)
directly from the partition function and performed $N$-body
simulations in the microcanonical ensemble (MCE). They started from a
waterbag\footnote{A waterbag distribution corresponds to a uniform
distribution function $f(\theta,v)=f_0$ in the domain $[-\theta_m,
\theta_m]\times [-v_m, v_m]$ surrounded by ``vaccum''
$f(\theta,v)=0$. It can have different magnetization $0\le M_0\le 1$.}
initial condition with magnetization $M_0\equiv M(0)=1$ and plotted in
their Fig. 4 the caloric curve giving the averaged kinetic temperature
$T=2E_{kin}/N$ as a function of the energy $U$. They compared their
numerical curve with the Boltzmann prediction of statistical
equilibrium in the canonical ensemble. They found a good agreement at
high and low energies. However, close to the critical energy $U_c=3/4$
(corresponding to $T_c=1/2$), the results differ from the canonical
prediction. In particular, the system bifurcates from the homogeneous
branch at a smaller energy $U\sim 0.6-0.7$ and the caloric curve
$T(U)$ displays a region of negative specific heats. Antoni \& Ruffo
\cite{ar} interpreted this result either as (i) a nonequilibrium
effect, or (ii) as a manifestation of ensembles inequivalence due to
the non-additivity of the energy. However, it is clear from the
previous work of Inagaki
\cite{inagaki} and the application of the Poincar\'e theorem that the
ensembles are equivalent for the HMF model (as confirmed later by
various methods \cite{deviation,cvb,cdr}). Therefore, this negative specific heats region is not
a manifestation of ensembles inequivalence but
rather a nonequilibrium effect.

Latora {\it et al.} \cite{latora,lrt} again performed microcanonical
simulations of the HMF model. They confirmed the ``anomaly'' (negative
specific heats region) reported by Antoni \& Ruffo \cite{ar} and
observed many other anomalies in that region such as non-Gaussian
distributions, anomalous diffusion, L\'evy walks and dynamical
correlations in phase-space. Furthermore, they showed that the
thermodynamic limit $N\rightarrow +\infty$ and the infinite time limit
$t\rightarrow +\infty$ do not commute. They evidenced two regimes in
the dynamics. On a short timescale, of the order of the dynamical time
$t_{D}\sim 1$, the system reaches a quasistationary state (QSS). Then,
on a much longer timescale $t_{relax}(N)$, the system relaxes towards
the Boltzmann distribution of statistical equilibrium. They showed
that the relaxation time increases rapidly (algebraically) with the
number of particles $N$ so that, at the thermodynamic limit
$N\rightarrow +\infty$, the system remains permanently in the QSS. It
is thus clear from this study that the different anomalies mentioned
above characterize the QSS, not the statistical equilibrium state that
is reached much later. In particular, the results of Antoni \& Ruffo
\cite{ar} and Latora {\it et al.} \cite{lrt} show that the QSS is
non-Boltzmannian in the region close to the critical energy $U_c$.
Latora {\it et al.} \cite{lrt} thus proposed to describe the QSS in
terms of Tsallis \cite{tsallis} generalized thermodynamics leading to
$q$-distributions. Note that there is no reason why the QSS should be
Boltzmannian (in the usual sense) since it is an out-of-equilibrium
structure. Therefore, the comparison of the numerical caloric curve
with the Boltzmann equilibrium caloric curve $T(U)$ is not justified
{\it a priori}.

After the conference in Les Houches in 2002, and inspired by the
results in astrophysics and 2D turbulence presented by one of the
authors \cite{chavhouches}, several groups of researchers
\cite{yamaguchi,cvb,epjb,precommun} started to interpret the QSSs
observed in the HMF model in terms of Lynden-Bell's \cite{lb}
statistical theory of violent relaxation based on the Vlasov
equation.
This is a fully predictive theory based on {\it standard
thermodynamics} but taking into account the specificities of the
Vlasov equation (Casimir constraints). Applying the Lynden-Bell theory
to the HMF model (for waterbag initial conditions), an
out-of-equilibrium phase transition was discovered in
\cite{epjb,precommun}. For a given value of the energy $U$, there
exists a critical value of the initial magnetization $(M_0)_{crit}(U)$
such that: for $M_0<(M_0)_{crit}(U)$ the stable Lynden-Bell distribution
(i.e. most probable state) is homogeneous (non-magnetized) and for
$M_0>(M_0)_{crit}(U)$ the stable Lynden-Bell distribution is
inhomogeneous (magnetized). For $U=0.69$, the critical magnetization
is $(M_0)_{crit}=0.897$ \cite{epjb,precommun}. More generally, there
exists a critical line $(M_0)_{crit}(U)$ in the phase diagram
separating homogeneous and inhomogeneous Lynden-Bell distributions. It
was found later \cite{prl2,marseille} that the system displays
first and second order phase transitions separated by a tricritical
point.  There is also an interesting phenomenon of phase reentrance in
the $(f_0,U)$ plane predicted in \cite{epjb} and numerically confirmed
in \cite{reentrance}.  Coming back to the specific value $U=0.69$,
direct numerical simulations of the HMF model for $M_0<(M_0)_{crit}$
have shown that the Lynden-Bell prediction works fairly well
\cite{precommun}. This agreement is remarkable since there is no
fitting parameter in the theory. This led the authors of
\cite{precommun} to argue that ``Lynden-Bell's theory explains
quasistationary states in the HMF model''. A controversy started when some
of these authors \cite{yamaguchi,epnews,cdr} concluded that ``the
approach of Tsallis is unsuccessful to explain QSSs''.

However, caution was made by one author \cite{incomplete,epjb,bbgky} who
argued that the Lynden-Bell theory does not explain
everything. Indeed, for initial magnetization $M_0=1$, the system is
in the non-degenerate limit so that the Lynden-Bell entropy reduces to
the Boltzmann entropy (with a different interpretation). Therefore, in
this limit case, the Lynden-Bell theory leads exactly to the same
prediction as the usual Boltzmann statistical theory but, of course,
for a completely different reason. This observation led to a
re-interpretation \cite{bbgky} of the caloric curve $T(U)$ obtained
by Antoni \& Ruffo \cite{ar} and Latora {\it al.} \cite{lrt}. In this
curve, the theoretical line should not be interpreted as the Boltzmann
statistical equilibrium state but as the Lynden-Bell statistical
equilibrium state\footnote{As mentioned above, there is no reason to
compare the QSS with the Boltzmann prediction that corresponds to the
collisional regime reached for $t\rightarrow +\infty$. By contrast, it
is fully relevant to compare the observed QSS with the Lynden-Bell
prediction that applies to the collisionless regime.}. They turn out
to coincide in the case $M_0=1$ but this is essentially
coincidental. With this new interpretation \cite{bbgky}, the comparison reveals
that the Lynden-Bell theory works well for large and low energies but
that it fails for energies close to the critical energy. Therefore,
{\it the Lynden-Bell theory does not explain the observations in the
range $[0.5,U_c]$}. Chavanis \cite{bbgky} interpreted this
disagreement as a result of {\it incomplete relaxation}. Indeed, it
was emphasized by Lynden-Bell \cite{lb} himself that his approach
implicitly assumes that the system ``mixes efficiently'' so that the
ergodicity hypothesis which sustains his statistical theory
applies. However, it has been observed in many cases of
violent relaxation that the system does not mix efficiently so that
the Lynden-Bell prediction fails (see various examples quoted in
\cite{incomplete,bbgky}).  The qualitative reason is easy to understand
\cite{bbgky}. Since violent relaxation is a purely inertial process
(no collision), mixing is due to the fluctuations of the mean field
potential caused by the fluctuations of the distribution function
itself \cite{lb}. However, as the system approaches metaequilibrium (QSS),
the fluctuations of the distribution function are less and less
efficient (by definition!) and the system can be trapped in a QSS (a
steady solution of the Vlasov equation on the coarse-grained scale)
that is not the most mixed state. This is what happens in astrophysics
(elliptical galaxies are {\it not} described by Lynden-Bell's
distribution that has infinite mass \cite{bt}) and in certain situations
of 2D turbulence [28-30]. The results of Antoni
\& Ruffo \cite{ar} and Latora {\it et al.} \cite{lrt} reveal that the same phenomenon  occurs for the HMF model close to the critical energy\footnote{ Latora {\it et al.} \cite{lrt} find that the Largest Lyapunov exponent for the QSS tends to zero. In this sense, mixing is negligible and one expects anomalies in the relaxation process. This is fully consistent with the idea of incomplete relaxation and inefficient mixing introduced by Lynden-Bell \cite{lb} and further discussed by Chavanis \cite{incomplete,epjb,bbgky}.}.
This is also visible on Fig. 8 of Bachelard {\it et al.}
\cite{bachelard}. There is a huge region in the top right of the phase
diagram where the Lynden-Bell theory does not work. This concerns in
particular the point $U=0.69$ and $M_0=1$, as anticipated in
\cite{epjb,bbgky}.  It is precisely this ``no-man's land'' region
that Tsallis and coworkers have investigated \cite{tsallisbook}. In
this region, standard statistical mechanics (i.e. Lynden-Bell's
theory) does not seem to directly apply\footnote{Quoting Einstein
\cite{einstein} and Cohen
\cite{cohen}, Latora {\it et al.} \cite{lrt} argued that standard
statistical mechanics fails when {\it the dynamics plays a nontrivial
role} (e.g. long-range correlations or fractal structures in phase
space). In our point of view, this is a correct interpretation
although standard thermodynamics should refer here to Lynden-Bell's
theory, which is the proper Boltzmann approach applied to the
Vlasov equation. This subtlety is not addressed in the paper of Latora
{\it et al.} \cite{lrt} since they did not know the theory of
Lynden-Bell at that time. Yet, their general comment can be applied to
Lynden-Bell's theory as well: if the dynamics is nontrivial and the 
system does not mix well, standard (Lynden-Bell) statistical mechanics fails.}.

In their early work, Latora {\it et al.} \cite{lrt} tried to fit the
QSS by a $q$-distribution. They considered a distribution with
$q=-5<1$ (in our notations) leading to power-law tails and introduced
by hand an additional cut-off to make the distribution
normalizable. This procedure is very {\it ad hoc}. Furthermore, even
if we accept it, we can argue that it does not provide an impressive
fit of the QSS. Recently, Campa {\it et al.} \cite{campa1} have
performed new simulations for initial magnetizations $M_0=0$ and
$M_0=1$ and found that the QSS is very well-fitted by a
semi-ellipse\footnote{For $M_0=1$, this differs from the numerical
results of Latora {\it et al.} \cite{lrt}. However, Campa {\it et al.}
\cite{campa1} show that the ordinary waterbag initial condition leads
to the presence of large sample to sample fluctuations so that many
averages are necessary. They proposed to use isotropic waterbag
distributions to reduce the fluctuations.}. Similar results were
obtained earlier by Yamaguchi {\it et al.} \cite{yamaguchi} for the
$M_0=0$ case. They claimed that the QSS is not a $q$-distribution
since it does not have power-law tails. However, Chavanis
\cite{bbgky} remarked that a semi-ellipse is a $q$-distribution
with $q=3$! Since $q>1$, this distribution has a compact
support. Therefore, the numerical results of Campa {\it et al.}
\cite{campa1} and Yamaguchi {\it et al.} \cite{yamaguchi} show that
the system tends to select a particular Tsallis distribution as a
QSS\footnote{Let us be more precise: (i) For $M_0=1$, an isotropic
waterbag distribution violently relaxes towards a $q$-distribution
with $q=3$
\cite{campa1}; (ii) For $M_0=0$, a waterbag
distribution is Vlasov stable and does not undergo violent relaxation
(it is already the Lynden-Bell state). However, in the collisional
regime (due to finite $N$ effects), it becomes a $q$-distribution with
$q=3$ \cite{yamaguchi,campa1}; (iii) for intermediate values of $M_0$,
the system violently relaxes towards the Lynden-Bell distribution
\cite{precommun,campa1}.}. Furthermore, the index $q=3$
seems to play a particular role since Campa {\it et al.} \cite{campa1}
obtained the same index $q=3$ in different situations (see their
Fig. 4). However, until now, the reason for this particular value
remains unknown.  The fact that the QSS has a compact support is
relatively natural in the phenomenology of incomplete violent
relaxation. Indeed, when mixing is not very efficient, we expect that
the high energy states are not sampled by the system. This leads to a
confinement of the distribution which is a virtue of the Tsallis
distributions with $q>1$. A similar confinement was observed in a
plasma experiment \cite{hd} and a good fit was obtained with a
$q$-distribution with index $q=2$ (in our notations)
\cite{boghosian}. This confinement was justified by a lack of
ergodicity in the system \cite{brands}. One can therefore interpret
the Tsallis distributions as an attempt to take into account
incomplete relaxation and non-ergodicity in systems with long-range
interactions.  In this interpretation, the index $q$ could be a
measure of mixing\footnote{This interpretation was proposed by one of
the authors in several papers \cite{grand,epjb,bbgky} and it may be more
accurate than the usual interpretation: ``$q$ is a measure of
nonextensivity'' \cite{tsallisbook}.}. If we assume that the system
mixes efficiently, then $q=1$ and we get the Lynden-Bell theory. If
the system does not mix well, the Lynden-Bell theory fails and $q\neq
1$. Since the value of the $q$ parameter depends on the efficiency of
mixing (which is not known a priori), it appears difficult to
determine its value from first principles. Furthermore, its value can
change from case to case since the degree of mixing can vary depending
on the initial condition (some systems can mix well and others
less). Finally, we can argue that the Tsallis entropy is just one
generalized entropy among many others and that it may not be
universal \cite{nfp}. It may just describe a special type of
non-ergodic behavior but not all of them. In fact, non-ergodic effects
can be so complicated that it is hard to believe that they can be
encapsulated in a simple functional such as the Tsallis functional or
any other \cite{brands,incomplete}. Nevertheless, we must recognize
that some QSSs can be well-fitted by $q$-distributions. Even more
strikingly, following the suggestion of \cite{bbgky}, Campa {\it et
al.}
\cite{campa2} demonstrated numerically that, during the collisional
regime, the time dependent distribution $f(v,t)$ is still very
well-fitted by $q(t)$-distributions with a time-dependent index. When
the index reaches a critical value $q_{crit}(U)$ (predicted by the
theory \cite{bbgky,campa2}), the distribution function becomes Vlasov
unstable and a dynamical phase transition from the homogeneous phase
(non-magnetized) to the inhomogeneous phase (magnetized) is
triggered. This explains previous observations on the evolution of the
magnetization $M(t)$ \cite{lrt,yamaguchi,campa1}. Interestingly, a
very similar behavior has been found by Taruya \& Sakagami
\cite{tsprl} for self-gravitating systems\footnote{There is, however,
a crucial difference in the interpretation. Taruya \& Sakagami
\cite{tsprl} interpret the instability
in terms of Tsallis generalized thermodynamics while Campa {\it et al.}
\cite{campa2} interpret it in terms of Vlasov dynamical instability \cite{aaantonov}.}.

In view of these results, it is interesting to study in more detail
the structure and the stability of $q$-distributions. Note that
$q$-distributions correspond to what have been called stellar
polytropes in astrophysics \cite{bt}. They were introduced long ago by
Eddington \cite{eddington} as particular stationary solutions of the
Vlasov equation. They were used to construct simple self-consistent
mathematical models of galaxies.  At some time, they were found to
provide a reasonable fit of some observed star clusters, the so-called
Plummer \cite{plummer} model. Improved observation of globular
clusters and galaxies showed that the fit is not perfect and more
realistic models have been introduced since then \cite{bt}. However,
stellar polytropes are still important for historical reasons and for
their mathematical simplicity. The stability of polytropic
distributions is an old problem in stellar dynamics \cite{bt}. It has
been reconsidered recently, for box-confined systems, by Taruya \&
Sakagami [42-44] in the framework of Tsallis
generalized thermodynamics and by Chavanis {\it et al.}
\cite{aapoly,grand,lang,cstsallis,aaantonov} in relation to their
nonlinear dynamical stability with respect to the Euler and Vlasov
equations.  It is therefore interesting to extend these studies to the
case of the HMF model. For the moment, only spatially homogeneous
polytropic distributions have been considered
\cite{cvb,cd}. In the present paper, we extend the analysis to
spatially {\it inhomogeneous} polytropes. Specifically, we study the
maximization of the Tsallis functional at fixed mass and energy and
plot the corresponding caloric curves. We give a thermodynamical and a
dynamical interpretation of this variational principle. We find the
existence of a critical polytropic index $q_c=3$ (where results are
analytical) which plays the role of a canonical tricritical point
separating first and second order phase transitions in the canonical
ensemble and marking the onset of negative specific heats in the
microcanonical ensemble (there also exists a microcanonical
tricritical point at $q\simeq 16.9$ and a microcanonical critical
point at $q\simeq 6.55$). Interestingly, this critical value $q_c=3$
turns out to be the one observed by Campa {\it al.} \cite{campa1} in
their numerical simulations. Then, we apply our theory to the
situation considered by Antoni \& Ruffo \cite{ar}. We find that the
structure of their numerical caloric curve $T(U)$ close to the
critical energy can be explained naturally if we assume that the QSSs
in this region are polytropes with the critical index $q_c=3$ (this
result is specifically discussed in Sec. \ref{sec_spec}).  This yields
a transition energy $U'_c=5/8$ lower than $U_c=3/4$ and a region of
negative specific heats in the curve $T(U)$ that are in qualitative
agreement with the numerical results (the agreement could be improved
by allowing $q$ to deviate slightly from the critical value
$q_c=3$). These behaviors had never been explained since the original
paper of Antoni \& Ruffo
\cite{ar}. We provide a plausible explanation in terms of Tsallis
(polytropic) distributions. We qualitatively justify the occurence of
polytropic distributions with a compact support in terms of incomplete
relaxation. Furthermore, the index selected by the system appears to
be the critical one (or close to it). This is the first time that a
sort of prediction of the $q$ index is made in that context. However,
our approach does not explain everything and just opens a direction of
research. Indeed, one has first to confirm the results by more
detailed comparisons with numerical simulations and, in case of
agreement, try to understand why critical polytropes are selected by
the system and if this is a general feature.

\section{Interpretations of the Tsallis functionals}
\label{sec_interpretations}

In order to motivate our study of polytropes, we shall first  recall different interpretations of the Tsallis functionals that have been given previously by one of the authors \cite{cstsallis,assise}.

\subsection{Thermodynamical interpretation}
\label{sec_ti}

There is no reason to describe the QSSs that emerge in Hamiltonian systems with long-range interactions in terms of Boltzmann statistical mechanics since they correspond to out-of-equilibrium structures (in the usual sense). The QSSs are formed during the collisionless regime while Boltzmann statistical equilibrium is reached on a much longer timescale at the end of the collisional regime. Fundamentally, the QSSs are stable steady states of the Vlasov equation (on a coarse-grained scale) and they should be described by Lynden-Bell's statistical mechanics. However, Lynden-Bell's theory, as any statistical theory, assumes efficient mixing and ergodicity. If the system does not mix well, the QSS will differ from the Lynden-Bell prediction (by definition). If we want to apply Tsallis generalized thermodynamics to that context, in order to take into account non ergodic effects and incomplete mixing, we must modify the Lynden-Bell entropy which is  the Boltzmann entropy associated to the Vlasov equation\footnote{Latora {\it et al.} \cite{lrt} did not know the Lynden-Bell theory and proposed to replace the usual Boltzmann entropy $S[f]=-\int f\ln f\, d{\bf r}d{\bf v}$ by the $q$-entropy $S_q[f]=-\frac{1}{q-1}\int (f^q-f)\, d{\bf r}d{\bf v}$. However, as argued long ago in \cite{brands}, this approach is in general incorrect since it does not take into account the constraints of the Vlasov equation. Furthermore, since $S_q[f]$ is rigorously  conserved by the Vlasov equation (it is a particular Casimir), there is no thermodynamical  reason to maximize it.}. Therefore, the proper $q$-entropy to consider is
\begin{equation}
S_q[\rho]=-\frac{1}{q-1}\int (\rho^q-\rho)\, d\eta  d{\bf r} d{\bf v}.
\label{ti1}
\end{equation}
This entropy applies to the distribution $\rho({\bf r},{\bf v},\eta)$ of phase levels $\eta$ (see \cite{assise} for details) so as to take into account the constraints of the Vlasov equation (Casimirs). For $q=1$ (efficient mixing), we recover the Lynden-Bell entropy. For $q\neq 1$, this functional could describe incomplete violent relaxation and non-ergodic effects in the framework of Tsallis thermodynamics. Now, for two-levels initial conditions and in the non degenerate limit (this corresponds to the waterbag model with initial magnetization $M_0=1$ in the HMF model), the Lynden-Bell entropy takes a form similar to the Boltzmann entropy and the constraints reduce to the mass and the energy. In that case, the generalized Lynden-Bell entropy (\ref{ti1}) reduces to
\begin{equation}
S_q[\overline{f}]=-\frac{1}{q-1}\int \left \lbrack\left (\frac{\overline{f}}{\eta_0}\right )^q-\frac{\overline{f}}{\eta_0}\right \rbrack\, d{\bf r} d{\bf v}.
\label{ti2}
\end{equation}
This is similar to the Tsallis entropy considered by Latora {\it et
al.} \cite{lrt} but it arises here for a different reason: note in
particular the bar on $\overline{f}$ (coarse-grained distribution) and
the presence of the initial phase level $\eta_0$. Note that
$S_q[\overline{f}]$ is expected to increase while $S_q[f]$ is
conserved. In this thermodynamical approach, the QSS is obtained by
maximizing the Tsallis entropy at fixed mass and energy
(microcanonical ensemble). This is expected to select the most
probable state {\it under some dynamical constraints} (responsible for
incomplete mixing) that are implicitly taken into account in the form
of the entropy. The canonical ensemble has no justification in the
present context but, as usual, it can be useful to provide a {\it
sufficient} condition of microcanonical stability \cite{ellis}.

Since the Tsallis entropy (\ref{ti1}) includes the Lynden-Bell entropy
as a special case (for $q=1$), it can give at least as good, or
better, results. However, we would like to emphasize several
limitations of Tsallis generalized thermodynamics: (i) at present,
there is no theory predicting the value of $q$. This prediction is of
course very difficult since $q$ is related to non-ergodic effects. For
the moment, $q$ appears as a fitting parameter measuring the degree of
mixing of a system; (ii) it is not clear that all non-ergodic
processes can be described by the Tsallis entropy. Therefore, this
functional is probably not universal. It may, however, describe a
certain class of non-ergodic behaviors\footnote{It has been shown that
the Tsallis entropy satisfies a lot of axioms satisfied by the
Boltzmann entropy. This makes this entropy very ``natural'' to
generalize the Boltzmann entropy. However, the axiomatic approach may
not be the best justification of an entropy. The original
combinatorial approach of Boltzmann seems to be more relevant. The
Boltzmann and the Lynden-Bell entropies can be derived from a
combinatorial analysis by assuming that all the microstates are
equiprobable. It would be interesting to derive the Tsallis entropy
from a combinatorial analysis by putting some constraints on the
availability of the microstates.}; (iii) it is not clear that
non-ergodic effects can be described by a simple entropic functional
such as the Tsallis functional or any other. Maybe, a better approach
is to consider the {\it dynamics of mixing} and develop kinetic
theories and relaxation equations of the process of violent relaxation as
suggested in \cite{csr,incomplete,bbgky,assise}. This kinetic approach
may be closer to the original ideas of Einstein \cite{einstein} and
Cohen \cite{cohen}.

\subsection{Dynamical interpretation}
\label{sec_dyn}

We would like to give another interpretation of Tsallis functional that is not related to thermodynamics and that does not suffer the limitations exposed above. It will  show that the Tsallis formalism can be useful in a dynamical context, using a thermodynamical analogy \cite{cstsallis,aaantonov,assise}.

It has been known for a long time that any distribution function of the form $f=f(\epsilon)$, where $\epsilon=v^2/2+\Phi$ is the individual energy,  is a steady state of the Vlasov equation \cite{jeans}. In particular, $q$-distributions have been introduced long ago by Eddington \cite{eddington} in astrophysics where they are called stellar polytropes\footnote{The connection between $q$-distributions and stellar polytropes was first mentioned by Plastino \& Plastino \cite{pp}. A more detailed discussion has been given recently by Chavanis \& Sire \cite{cstsallis}.}. It is also known that these distributions are critical points of a
certain functional of the form
\begin{equation}
S[f]=-\int C(f)\, d{\bf r}d{\bf v},
\label{dyn1}
\end{equation}
where $C$ is a convex function, at fixed mass and energy. Furthermore, if they are {\it maxima} of this functional, they are nonlinearly dynamically stable with respect to the Vlasov equation. As far as we know, this dynamical stability criterion was first stated by Ipser \& Horwitz \cite{ih} in astrophysics. Furthermore, Ipser \cite{ipser}  introduced the functional $S=-\int f^{1+1/(n-3/2)}\, d{\bf r}d{\bf v}$ (in his notations) associated to stellar polytropes. This is nothing but the Tsallis functional\footnote{Indeed, the second term in the Tsallis functional  (\ref{ps1}) is a constant (proportional to mass) and the first term coincides with the Ipser functional with the notation (\ref{ps8}). As noted in \cite{cstsallis}, the Tsallis functional (\ref{ps1}) is more convenient to take the limit $q\rightarrow 1$ (i.e. $n\rightarrow +\infty$) since it reduces to the Boltzmann functional while the  Ipser functional takes a trivial form. Therefore, the Tsallis functional  allows to make a continuous link between isothermal ($n=\infty$) and polytropic  ($n$ finite) distributions using L'H\^opital's rule.}.

Let us be more precise and more general.  The maximization problem
\begin{eqnarray}
\label{dyn2}
\max_f\left\lbrace S\lbrack f\rbrack\, |\, E\lbrack f\rbrack=E, \, M\lbrack f\rbrack=M\right\rbrace,
\end{eqnarray}
determines a steady state of the Vlasov equation of the form
$f=f(\epsilon)$ with $f'(\epsilon)<0$ that is nonlinearly dynamically
stable. This has been stated by Ellis {\it et al.} \cite{eht} in 2D
turbulence (for the 2D Euler equation) and by Ipser \& Horwitz
\cite{ih}, Tremaine {\it et al.} \cite{thlb} and Chavanis
\cite{aaantonov} in stellar dynamics. The maximization problem
(\ref{dyn2}) is similar to a condition of microcanonical stability in
thermodynamics. Therefore, we can develop a {\it thermodynamical
analogy} \cite{assise} to investigate the nonlinear
dynamical stability problem. In this analogy, $S$ is called a pseudo
entropy. Thus, the Tsallis functional is a particular pseudo entropy
whose maximization at fixed mass and energy determines distributions
(polytropes) that are nonlinearly dynamically stable with respect to
the Vlasov equation \cite{cstsallis,aaantonov}.

The minimization problem
\begin{eqnarray}
\label{dyn3}
\min_f\left\lbrace F\lbrack f\rbrack=E[f]-T S[f]\, |\, M\lbrack f\rbrack=M\right\rbrace.
\end{eqnarray}
also determines a steady state of the Vlasov equation that is
nonlinearly dynamically stable. In fact, (\ref{dyn2}) and (\ref{dyn3})
have the same critical points. However, the criterion (\ref{dyn3}) is
less refined than (\ref{dyn2}). Indeed, the minimization problem
(\ref{dyn3}) is similar to a condition of canonical stability in
thermodynamics. In this analogy, $F$ is called a pseudo free
energy. Now, it is a general result \cite{ellis} that canonical
stability implies microcanonical stability, but the reciprocal is
wrong in case of ensembles inequivalence. Schematically: (\ref{dyn3})
$\Rightarrow$ (\ref{dyn2}). In the present dynamical context, this
means that steady states of the Vlasov equation that are stable
according to the criterion (\ref{dyn3}) are necessarily stable
according to the more constrained criterion (\ref{dyn2}). However,
there may exist steady states of the Vlasov equation that are stable
according to (\ref{dyn2}) while they fail to satisfy (\ref{dyn3}). As
shown in \cite{aaantonov}, this is the case for stellar polytropes with
indices $3<n<5$ in astrophysics.

It can also be shown \cite{assise} that (\ref{dyn3}) is equivalent to
\begin{eqnarray}
\label{dyn4}
\min_\rho\left\lbrace F\lbrack \rho\rbrack\, |\, M\lbrack \rho\rbrack=M\right\rbrace,
\end{eqnarray}
where
\begin{eqnarray}
\label{dyn5}
F=\frac{1}{2}\int \rho\Phi\, d{\bf r}+\int\rho\int^\rho \frac{p(\rho')}{\rho^{'2}}\, d\rho'd{\bf r}.
\end{eqnarray}
More precisely, a distribution function $f({\bf r},{\bf v})$ is solution of (\ref{dyn3}) iff the corresponding density profile $\rho({\bf r})$ is solution of (\ref{dyn4}), where $p(\rho)$ is the equation of state determined by $C(f)$ (see \cite{assise} for more details). Finally, (\ref{dyn4}) is clearly equivalent to
\begin{eqnarray}
\label{dyn6}
\min_{\rho,{\bf u}}\left\lbrace {\cal W}\lbrack \rho,{\bf u}\rbrack\, |\, M\lbrack \rho\rbrack=M\right\rbrace,
\end{eqnarray}
where
\begin{eqnarray}
\label{dyn7}
{\cal W}=\int\rho\int^\rho \frac{p(\rho')}{\rho^{'2}}\, d\rho'd{\bf r}+\frac{1}{2}\int \rho\Phi\, d{\bf r}+\int \rho \frac{{\bf u}^2}{2}\, d{\bf r}.
\end{eqnarray}
Now, it can be shown that this minimization problem determines a
steady solution of the barotropic Euler equation that is formally
nonlinearly dynamically stable \cite{bt}. From the implication
$(\ref{dyn6}) \Leftrightarrow (\ref{dyn4}) \Leftrightarrow
(\ref{dyn3}) \Rightarrow (\ref{dyn2})$, we conclude that a
distribution $f({\bf r},{\bf v})$ is stable with respect to the Vlasov
equation [according to (\ref{dyn2})] if the corresponding density
profile $\rho({\bf r})$ is stable with respect to the Euler equation
[according to (\ref{dyn6})]. As shown in \cite{aaantonov}, this
provides a nonlinear generalization of the Antonov first law in
astrophysics: ``a spherical galaxy $f=f(\epsilon)$ with
$f'(\epsilon)<0$ is nonlinearly dynamically stable with respect to
the Vlasov-Poisson system if the corresponding barotropic star is
nonlinearly dynamically stable with respect to the Euler-Poisson
system''. Interestingly, this also provides a new interpretation
\cite{aaantonov} of this law in terms of ensembles
inequivalence. Indeed, the Antonov first law has the same status as
the fact that ``canonical stability implies microcanonical stability''
in thermodynamics.

{\it Remark:} the ``microcanonical" criterion (\ref{dyn2}) provides itself just a
sufficient condition of nonlinear dynamical stability. There exists an even
more refined criterion of nonlinear dynamical stability taking into account
the conservation of all the Casimirs  (see discussion in \cite{cd}).

\subsection{Generalized H-functions and selective decay}
\label{sec_hf}

A generalized $H$-function is a functional of the coarse-grained distribution function $\overline{f}({\bf r},{\bf v},t)$ of the form
\begin{equation}
H[\overline{f}]=-\int C(\overline{f})\, d{\bf r}d{\bf v},
\label{hf1}
\end{equation}
where $C$ is any convex function.  Tremaine {\it et al.} \cite{thlb}
have shown that the generalized $H$-functions increase during violent
relaxation in the sense that $H(t)\ge H(0)$ for any time $t\ge 0$
where it is assumed that the initial distribution function is not
mixed so that $\overline{f}({\bf r},{\bf v},t=0)=f({\bf r},{\bf
v},t=0)$\footnote{Note that the time evolution of the generalized
$H$-functions is not necessarily monotonic (nothing is implied
concerning the relative values of $H(t)$ and $H(t')$ for
$t,t'>0$).}. By contrast, the energy $E[\overline{f}]$ and
$M[\overline{f}]$ and the mass calculated with the coarse-grained
distribution function are approximately conserved. This suggests a
phenomenological {\it generalized selective decay principle} (for
$-H$): ``due to mixing, the system may tend to a QSS that maximizes a
certain $H$-function (non universal) at fixed mass and energy''
\cite{assise}\footnote{The close relationship between this
phenomenological principle and the criterion of nonlinear dynamical
stability (\ref{dyn2}), as well as the limitations of this
phenomenological principle, are discussed in detail in
\cite{incomplete,assise}.}. In that context, the Tsallis functionals
$H[\overline{f}]=-\frac{1}{q-1}\int (\overline{f}^q-\overline{f})\,
d{\bf r}d{\bf v}$ can be interpreted as particular generalized
H-functions \cite{assise}.

\section{Polytropic distributions: general theory}
\label{sec_pdgt}

For any of the reasons exposed previously, we think that it is useful to study polytropic (Tsallis) distribution functions and investigate their stability through the variational problems (\ref{dyn2}) and (\ref{dyn3}). We first develop a general theory of polytropes following the lines of \cite{lang,cstsallis}. It will be applied specifically to the HMF model in Sec. \ref{sec_application}.

\subsection{Polytropic distributions in phase space}
\label{sec_ps}

Let us consider the Tsallis functional
\begin{equation}
S=-\frac{1}{q-1}\int (f^q-f_0^{q-1} f)\, d{\bf r} d{\bf v}.   \label{ps1}
\end{equation}
We have introduced a constant $f_0$ in order to make the
expression homogeneous. This constant will play no role in the following
since the last term is proportional to the mass that is conserved. Therefore, we could
equally work with the Ipser functional
\begin{equation}
S=-\frac{1}{q-1}\int f^q\, d{\bf r} d{\bf v},  \label{ps2}
\end{equation}
but the first expression allows us to make the connection with isothermal distributions when $q\rightarrow 1$. Indeed, for $q\rightarrow 1$, we recover the Boltzmann functional
\begin{equation}
S=-\int f\ln\left (\frac{f}{f_0}\right )\, d{\bf r} d{\bf v}.   \label{ps3}
\end{equation}
We shall consider the maximization of the Tsallis functional at fixed energy and mass
\begin{equation}
E=\int f\frac{v^2}{2}\, d{\bf r} d{\bf v}+\frac{1}{2}\int \rho\Phi\, d{\bf r},   \label{ps4}
\end{equation}
\begin{equation}
M=\int \rho\, d{\bf r}.   \label{ps5}
\end{equation}
Some interpretations of this variational problem have been given in Sec. \ref{sec_interpretations}. Here, $S$ is  either a generalized entropy (thermodynamical interpretation) or a pseudo entropy (dynamical interpretation). By an abuse of language, and to simplify the terminology, we shall call it simply an entropy. We will work in a space of dimension $d$ since our formalism can have application for different systems.

The critical points of entropy at fixed energy and mass are determined by the condition
\begin{equation}
\delta S-\beta\delta E-\alpha\delta M=0,  \label{ps6}
\end{equation}
where $\beta=1/T$ and $\alpha$  are Lagrange multipliers ($T$ is the inverse temperature and $\alpha$ the chemical potential). This yields the $q$-distributions (or polytropic distributions):
\begin{equation}
f({\bf r},{\bf v})=\left\lbrace \mu-\frac{(q-1)\beta}{q}\left\lbrack \frac{v^2}{2}+\Phi({\bf r})\right\rbrack\right\rbrace_+^{1/(q-1)},  \label{ps7}
\end{equation}
where $\mu=\lbrack f_0^{q-1}-(q-1)\alpha\rbrack/q$. The notation $\lbrack
x\rbrack_+=x$ if $x\ge 0$ and $\lbrack x\rbrack_+=0$ if $x\le 0$.  As is customary in
astrophysics, we define the polytropic index $n$ by the relation\footnote{This relation is sometimes presented as a ``fundamental" relation relating the $q$-parameter to the polytropic index $n$ \cite{tsallisbook}. In our sense, this is just a definition of  notations, no more.}
\begin{equation}
n=\frac{d}{2}+\frac{1}{q-1}.  \label{ps8}
\end{equation}
The distribution function $f({\bf r},{\bf v})$ depends only on the
individual energy $\epsilon={v^2}/{2}+\Phi({\bf r})$,
i.e. $f=f(\epsilon)$. Therefore, it is a steady state of the Vlasov
equation. We shall consider $q>0$ so that $C(f)$ is convex and
$\beta>0$ so that $f'(\epsilon)<0$, which corresponds to the physical
situation. For $n=d/2$ ($q\rightarrow +\infty$), we recover the Fermi
distribution \cite{epjb}. For $n\rightarrow +\infty$ ($q\rightarrow
1$), we recover the isothermal distribution.

We need to distinguish two cases depending on the sign of $q-1$. (i) For $q>1$ ($n\ge d/2$), the
distribution function can be written
\begin{eqnarray}
f=A(\epsilon_{m}-\epsilon)_+^{1\over q-1},
\label{ps9}
\end{eqnarray}
where we have set $A=\lbrack\beta(q-1)/q\rbrack^{1\over q-1}$ and
$\epsilon_{m}=q\mu/\lbrack \beta(q-1)\rbrack$. Such distributions have a compact support in phase space since they
  vanish at $\epsilon=\epsilon_m$.  At a given position, the distribution function vanishes for $v\ge v_{m}({{\bf r}})=\sqrt{2(\epsilon_{m}-\Phi({{\bf r}}))}$.  (ii) For $0<q<1$, the distribution function can be written
\begin{eqnarray}
f=A(\epsilon_{0}+\epsilon)^{1\over q-1},
\label{ps10}
\end{eqnarray}
where we have set $A=\lbrack\beta(1-q)/q\rbrack^{1\over q-1}$ and
$\epsilon_{0}=q\mu/\lbrack\beta(1-q)\rbrack$. Such distributions are defined for all velocities. At a given position, the distribution function behaves, for large velocities, as
$f\sim v^{-2/(1-q)}\sim v^{-(d-2n)}$. We shall only consider distribution functions for which the
density and the pressure
\begin{eqnarray}
\rho=\int f\, d{\bf v},\qquad p=\frac{1}{d}\int f v^2\, d{\bf v},
\label{ps11}
\end{eqnarray}
are finite. This implies $d/(d+2)<q<1$ ($n<-1$).

\subsection{Polytropic equation of state}
\label{sec_es}

For any distribution function of the form $f=f(\epsilon)$, the density and the pressure are functions of $\Phi({\bf r})$
such that $\rho=\rho[\Phi({{\bf r}})]$ and $p=p[\Phi({{\bf r}})]$. Eliminating
$\Phi({{\bf r}})$ between these expressions, we obtain a barotropic
equation of state $p({\bf r})=p[\rho({{\bf r}})]$. Furthermore, it is easy to see that $f=f(\epsilon)$ implies the condition of hydrostatic equilibrium (see Appendix \ref{sec_hydroeq}):
\begin{eqnarray}
\nabla p+\rho\nabla\Phi={\bf 0}.
\label{es0}
\end{eqnarray}
Let us determine the equation of state corresponding to the polytropic distribution (\ref{ps7}).
For $n\ge d/2$, the density and
the pressure can be expressed as (see Appendix \ref{sec_poly}):
\begin{eqnarray}
\rho=AS_d (\epsilon_{m}-\Phi)^{n}2^{d/2-1}{\Gamma(d/2)\Gamma(1-d/2+n)\over \Gamma(n+1)},
\label{es1}
\end{eqnarray}
\begin{eqnarray}
p={AS_d\over n+1}(\epsilon_{m}-\Phi)^{n+1}2^{d/2-1}{\Gamma(d/2)\Gamma(1-d/2+n)\over \Gamma(n+1)},\nonumber\\
\label{es2}
\end{eqnarray}
where $\Gamma(x)$ denotes the Gamma function and $S_d$ the surface of a unit sphere in $d$ dimensions.
For $n<-1$, the density and the pressure can be expressed as (see Appendix \ref{sec_poly}):
\begin{eqnarray}
\rho=AS_d(\epsilon_{0}+\Phi)^{n}2^{d/2-1}{\Gamma(-n)\Gamma(d/2)\over \Gamma(d/2-n)},
\label{es3}
\end{eqnarray}
\begin{eqnarray}
p=-{AS_d\over n+1}(\epsilon_{0}+\Phi)^{n+1}2^{d/2-1}{\Gamma(-n)\Gamma(d/2)\over \Gamma(d/2-n)}.
\label{es4}
\end{eqnarray}
Eliminating the field $\Phi({\bf r})$ between the expressions (\ref{es1})-(\ref{es2}) and (\ref{es3})-(\ref{es4}), one finds that
\begin{eqnarray}
p=K\rho^{\gamma}, \qquad \gamma=1+{1\over n}.
\label{es5}
\end{eqnarray}
This is the  classical polytropic equation of state. This is the reason why the distributions (\ref{ps7}) are called polytropic distributions. Furthermore, $\gamma$ is the ordinary polytropic index and $n$ is a polytropic index commonly used in astrophysics \cite{chandra}. The polytropic constant $K$ is given for $n\ge d/2$ by
\begin{eqnarray}
K={1\over n+1}\biggl \lbrace A S_d 2^{d/2-1}{\Gamma(d/2)\Gamma(1-d/2+n)\over \Gamma(n+1)}
\biggr \rbrace^{-{1\over n}}
\label{es6}
\end{eqnarray}
and for $n<-1$ by
\begin{eqnarray}
K=-{1\over n+1}\biggl \lbrace A S_d 2^{d/2-1}{\Gamma(-n)\Gamma(d/2)\over\Gamma(d/2-n)}\biggr \rbrace^{-{1\over n}}.
\label{es7}
\end{eqnarray}
The polytropic constant $K$ plays the role of the temperature $T_{iso}$ in isothermal systems ($q=1$,
$n=\infty$, $\gamma=1$) and it is sometimes called a ``polytropic
temperature''. For polytropic distributions, the relation between the polytropic temperature $K$ and the thermodynamical temperature  $T=1/\beta$ is
\begin{eqnarray}
T=C_n K^{\frac{2n}{2n-d}},
\label{es8}
\end{eqnarray}
where $C_n$ is given for $n\ge d/2$ by
\begin{eqnarray}
C_n=\frac{2(n+1)^{\frac{2n}{2n-d}}}{2n-d+2}\left\lbrack S_d 2^{d/2-1}{\Gamma(d/2)\Gamma(1-d/2+n)\over \Gamma(n+1)}
\right\rbrack^{{2\over 2n-d}} \nonumber\\
\label{es9}
\end{eqnarray}
and for $n<-1$ by
\begin{eqnarray}
C_n=\frac{2\lbrack -(n+1)\rbrack^{\frac{2n}{2n-d}}}{-2n+d-2}\left\lbrack S_d 2^{d/2-1}{\Gamma(d/2)\Gamma(-n)\over \Gamma(d/2-n)}
\right\rbrack^{{2\over 2n-d}}.\nonumber\\
\label{es10}
\end{eqnarray}
We note that $K$ is a monotonically increasing function of $T$. This remark will have some importance in the following.

\subsection{Polytropic distributions in physical space}
\label{sec_pp}

The density is obtained by integrating Eq. (\ref{ps7}) on the velocity leading to Eqs. (\ref{es1}) and (\ref{es3}). Using Eqs. (\ref{es6}) and (\ref{es7}), we find that the density is related to the  potential $\Phi({\bf r})$  by
 \begin{equation}
\rho({\bf r})=\left\lbrack \lambda-\frac{\gamma-1}{K\gamma}\Phi({\bf r})\right
\rbrack_+^{\frac{1}{\gamma-1}},
\label{pp1}
\end{equation}
where $\lambda=\epsilon_m/(K(n+1))$ for $n\ge d/2$ and $\lambda=\epsilon_0/(-K(n+1))$ for $n<-1$.
As noted in \cite{cstsallis}, the polytropic distribution in phase space $f=f(\epsilon)$ given by Eq. (\ref{ps7})  has the same mathematical form as the polytropic distribution in physical space  $\rho=\rho(\Phi)$ given by Eq. (\ref{pp1}) with  $\gamma$ playing the role of $q$ and $K$ playing the role of $T=1/\beta$. In this correspondence, $\gamma$ is related to $q$ by Eqs. (\ref{es5}) and (\ref{ps8}) leading to $\gamma=((d+2)q-d)/(dq-d+2)$ and $K$ is related to $T$ by Eq. (\ref{es8}). Polytropic distributions (including the isothermal one) are apparently the only
distributions for which $f(\epsilon)$ and $\rho(\Phi)$ have the same mathematical form.

\subsection{Other expressions of the distribution function}
\label{sec_other}

We can write the polytropic distribution function (\ref{ps7})  in different forms that all have their own
interest. This will show that different notions of ``temperature'' exist for
polytropic distributions.

(i) {\it Thermodynamical temperature $T=1/\beta$}: the polytropic distribution (\ref{ps7}) directly comes from the variational principle (\ref{ps6}). Therefore, $\beta=1/T$ is the Lagrange multiplier associated with the conservation of energy. This is the proper thermodynamical temperature to consider, i.e. $\beta=(\partial S/\partial E)_M$ is the conjugate of the energy with respect to the entropy. Note, however, that $T=1/\beta$ does not have the dimension of an ordinary temperature (squared velocity).

(ii) {\it Dimensional temperature $\theta=1/b$}: we can define a quantity that has the dimension of a temperature (squared velocity) by setting $b=\beta/q\mu$. If we define furthermore $f_*=\mu^{1/(q-1)}$, the polytropic distribution (\ref{ps7}) can be rewritten
\begin{eqnarray}
f=f_*\biggl \lbrack 1-{b (q-1)}\epsilon\biggr \rbrack_+^{1\over q-1}.
\label{na1}
\end{eqnarray}
Comparing Eq. (\ref{na1}) with Eqs. (\ref{ps9}) and (\ref{ps10}) we find that $\epsilon_m=1/[b(q-1)]$ and $A=f_*/\epsilon_m^{1/(q-1)}$ for $n\ge d/2$ and $\epsilon_0=1/[b(1-q)]$ and $A=f_*/\epsilon_0^{1/(q-1)}$ for $n<-1$.
Substituting these expressions in Eqs. (\ref{es1}) and (\ref{es3}), we find that the relation between the density and the potential can be written
\begin{eqnarray}
\rho=\rho_{*}\biggl \lbrack 1-{b (q-1)}\Phi\biggr \rbrack_+^{n},
\label{na2}
\end{eqnarray}
where $\rho_*$ is given for $n\ge d/2$ by
\begin{eqnarray}
\rho_*=\frac{S_d}{2} f_{*}\left (\frac{2n-d}{b}\right )^{d/2}\frac{\Gamma(d/2)\Gamma(1-d/2+n)}{\Gamma(n+1)},\nonumber\\
\label{na3}
\end{eqnarray}
and for $n<-1$ by
\begin{eqnarray}
\rho_*=\frac{S_d}{2} f_{*}\left (\frac{d-2n}{b}\right )^{d/2}\frac{\Gamma(d/2)\Gamma(-n)}{\Gamma(d/2-n)}.
\label{na4}
\end{eqnarray}

(iii) {\it Polytropic temperature $K$}: eliminating the potential between Eqs. (\ref{ps9}) and  (\ref{es1}), and between  Eqs. (\ref{ps10}) and (\ref{es3}), and using Eqs. (\ref{es6}) and (\ref{es7}),  we can express the distribution function in terms of the density according to
\begin{eqnarray}
f={1\over Z}\biggl \lbrack \rho({\bf r})^{1/n}-{v^{2}/2\over (n+1)K}\biggr\rbrack_+^{n-d/2},
\label{vb19}
\end{eqnarray}
where $Z$ is given for $n\ge d/2$ by
\begin{eqnarray}
Z=S_d 2^{d/2-1}{\Gamma(d/2)\Gamma(1-d/2+n)\over\Gamma(n+1)}\lbrack K(n+1)\rbrack^{d/2},
\label{vb20}
\end{eqnarray}
and for $n<-1$ by
\begin{eqnarray}
Z=S_d 2^{d/2-1}{\Gamma(-n)\Gamma(d/2)\over\Gamma(d/2-n)}\lbrack -K(n+1)\rbrack^{d/2}.
\label{vb21}
\end{eqnarray}
This is the polytropic counterpart of the
isothermal distribution. The constant $K$ plays a role similar to the temperature $T_{iso}$ in an isothermal distribution. In particular, it is uniform in a polytropic distribution as is the temperature in an isothermal system. This is why $K$  is sometimes called a polytropic temperature.

(iv) {\it Kinetic temperature $T_{kin}({\bf r})$}: the kinetic temperature is defined by
\begin{eqnarray}
T_{kin}({\bf r})=\frac{1}{d}\langle v^2\rangle({\bf r})=p({\bf r})/\rho({\bf r}).
\label{vb21bis}
\end{eqnarray}
For a polytropic distribution, using the equation of state (\ref{es5}), we get
\begin{eqnarray}
T_{kin}({\bf r})=K\rho({\bf r})^{\gamma-1}.
\label{vb21bisb}
\end{eqnarray}
For a spatially  inhomogeneous polytrope, the kinetic temperature $T_{kin}({\bf r})$ is position dependent and differs from the thermodynamical temperature $T=1/\beta$. The velocity of sound $c_s^2=p'(\rho)$ is given by
\begin{eqnarray}
c_s^2({\bf r})=K\gamma\rho({\bf r})^{\gamma-1}=\gamma T_{kin}({\bf r}).
\label{vb21tris}
\end{eqnarray}
It is also position dependent and differs from the velocity dispersion (they differ by a factor $\gamma$). Using Eq. (\ref{vb21bis}), the distribution function
(\ref{vb19}) can be written
\begin{eqnarray}
f=B_{n}{\rho({\bf r})\over \lbrack 2\pi T_{kin}({\bf r})\rbrack^{d/2}}\biggl\lbrack 1-{v^{2}/2\over (n+1)T_{kin}({\bf r})}\biggr\rbrack_+^{n-d/2},
\label{vb23}
\end{eqnarray}
\begin{eqnarray}
B_{n}={\Gamma(n+1)\over\Gamma(1-d/2+n)(n+1)^{d/2}}, \quad (n\ge d/2)
\label{vb24}
\end{eqnarray}
\begin{eqnarray}
B_{n}={\Gamma(d/2-n)\over\Gamma(-n)\lbrack -(n+1)\rbrack^{d/2}} \quad (n<-1).
\label{vb25}
\end{eqnarray}
Note that for $n\ge d/2$, the maximum velocity can be expressed in terms of the kinetic temperature by
\begin{eqnarray}
v_{m}({\bf r})=\sqrt{2(n+1)T_{kin}({\bf r})}.
\label{vb26}
\end{eqnarray}
Using $\Gamma(z+a)/\Gamma(z)\sim z^{a}$ for $z\rightarrow +\infty$, we recover the isothermal distribution for $n\rightarrow +\infty$.
On the other hand, from Eqs. (\ref{vb21bisb}) and (\ref{pp1}), we
immediately get $T_{kin}({\bf r})=K(\lambda-(\gamma-1)\Phi({\bf r})/K\gamma)$ so that
\begin{eqnarray}
\nabla T_{kin}=-{\gamma-1\over\gamma}\nabla\Phi.
\label{vb22}
\end{eqnarray}
This shows that, for a
polytropic distribution, the kinetic temperature (velocity dispersion) is a linear
function of the potential\footnote{For any distribution of the form $f=f(\epsilon)$, we have $\rho=\rho(\Phi)$ and $p=p(\Phi)$ so that the kinetic temperature (velocity dispersion) $T_{kin}=p/\rho$ is a function $T_{kin}=T_{kin}(\Phi)$ of the potential. For a polytropic distribution function, this relation turns out to be linear.}. The coefficient of
proportionality is related
to the polytropic index by $(\gamma-1)/\gamma=1/(n+1)=2(q-1)/\lbrack (d+2)q-d\rbrack$.
The relation (\ref{vb22}) can also be obtained directly from Eq. (\ref{ps9}) [or Eq. (\ref{ps10})] noting that
\begin{eqnarray}
f=A(\epsilon_{m}-\Phi)^{n-d/2}\left \lbrack 1-\frac{v^{2}/2}{\epsilon_{m}-\Phi}\right\rbrack_+^{n-d/2},
\label{vb22ndl}
\end{eqnarray}
and comparing with Eq. (\ref{vb23}).

(v) {\it Energy excitation temperature $T(\epsilon)$}: for any distribution function of the form $f=f(\epsilon)$, one may define a local energy dependent excitation temperature by the relation
\begin{eqnarray}
\frac{1}{T(\epsilon)}=-\frac{d\ln f}{d\epsilon}.
\label{vb22ndlb}
\end{eqnarray}
For the isothermal distribution, $T(\epsilon)$ coincides with the usual temperature $T_{iso}$. For the polytropic distribution (\ref{ps7}), $T(\epsilon)=q\mu/\beta-(q-1)\epsilon$. This excitation temperature has a constant gradient $dT/d\epsilon=1-q$ related to the polytropic index. The other parameter $\mu$ is related to the value of the energy where the temperature reaches zero.

\subsection{Entropy and free energy in physical space}
\label{sec_ef}

Substituting the polytropic distribution function (\ref{vb19}) in the Tsallis entropy (\ref{ps1}), we find after some calculations (see Appendix \ref{sec_deriv}) that
\begin{eqnarray}
S=-\left (n-\frac{d}{2}\right )\left (\beta\int p\, d{\bf r}-f_0^{q-1}M\right ).
\label{ef1}
\end{eqnarray}
On the other hand, the energy (\ref{ps4}) can be written
\begin{eqnarray}
E=\frac{d}{2}\int p\, d{\bf r}+\frac{1}{2}\int \rho\Phi\, d{\bf r}.
\label{ef2}
\end{eqnarray}
Therefore, the free energy $F=E-TS$ is given by
\begin{eqnarray}
F=\frac{1}{2}\int \rho\Phi\, d{\bf r}+n\int p\, d{\bf r},
\label{ef3}
\end{eqnarray}
up to unimportant constant terms. It can also be written
\begin{eqnarray}
F=\frac{1}{2}\int \rho\Phi\, d{\bf r}+\frac{K}{\gamma-1}\int \left (\rho^{\gamma}-\rho_0^{\gamma-1}\rho\right )\, d{\bf r},
\label{ef4}
\end{eqnarray}
where
\begin{eqnarray}
\rho_0^{\gamma-1}=\frac{\gamma-1}{K}\frac{T}{q-1}f_0^{q-1}.
\label{ef5}
\end{eqnarray}
This can be viewed as a free energy $F=E-KS$ associated with a Tsallis entropy in physical space $S=\frac{1}{\gamma-1}\int (\rho^{\gamma}-\rho_0^{\gamma-1}\rho )\, d{\bf r}$ where $\gamma$ plays the role of $q$ and $K$ the role of $T$. Again, it is interesting to note that, for polytropes, the free energy in phase space $F[f]$ has a form similar to the free energy in position space $F[\rho]$ \cite{cstsallis}.

\section{Application to the HMF model}
\label{sec_application}

\subsection{Generalities}
\label{sec_gen}

The HMF model is a system of $N$ particles of unit mass $m=1$ moving on a circle and interacting via a cosine potential. The dynamics of these particles is governed by the Hamilton equations \cite{inagaki,ar,cvb}:
\begin{eqnarray}
\frac{d\theta_i}{dt}=\frac{\partial H}{\partial v_i}, \qquad \frac{d v_i}{dt}=-\frac{\partial H}{\partial \theta_i},\nonumber\\
H=\frac{1}{2}\sum_{i=1}^{N} v_i^2-\frac{k}{4\pi}\sum_{i\neq j}\cos(\theta_i-\theta_j),
\label{gen1}
\end{eqnarray}
where $\theta_i\in [-\pi,\pi]$ and $-\infty<v_i<\infty$ denote  the position (angle) and the velocity of particle $i$ and $k$ is the coupling constant.
The thermodynamic limit corresponds to $N\rightarrow +\infty$ in such a way that the rescaled energy $\epsilon=8\pi E/kM^2$ remains of order unity (this amounts to taking $k\sim 1/N$ and $E/N\sim 1$). In that limit, the mean field approximation becomes exact.  The  total mass and total energy are given by
\begin{equation}
M=\int \rho\, d\theta,   \label{gen2}
\end{equation}
\begin{equation}
E=\int f \frac{v^2}{2}\, d\theta dv+\frac{1}{2}\int \rho\Phi\, d\theta=E_{kin}+W,  \label{gen3}
\end{equation}
where $f(\theta,v)$ is the distribution function, $\rho(\theta)=\int f\, dv$ the density, $E_{kin}$ the kinetic energy and $W$ the potential
energy. The potential is related to the density by
\begin{equation}
\Phi(\theta)=-\frac{k}{2\pi}\int_{0}^{2\pi}\cos(\theta-\theta')\rho(\theta')\, d\theta',\label{gen4}
\end{equation}
and the mean force acting on a particle located in $\theta$ is $\langle F\rangle=-\Phi'(\theta)$. For a cosine interaction, the potential can be written
\begin{equation}
\Phi(\theta)=-B_x\cos\theta-B_y\sin\theta,\label{gen5}
\end{equation}
where
\begin{equation}
B_x=\frac{k}{2\pi}\int_{0}^{2\pi}\rho(\theta)\cos\theta\, d\theta,
\label{gen6}
\end{equation}
\begin{equation}
B_y=\frac{k}{2\pi}\int_{0}^{2\pi}\rho(\theta)\sin\theta\, d\theta,
\label{gen7}
\end{equation}
are the two components of the magnetization vector up to a multiplicative factor (note the change of
sign with respect to \cite{cvb}). Substituting Eq. (\ref{gen5}) in the potential energy (\ref{gen3}), we find that the energy can be rewritten
\begin{equation}
E=\int f \frac{v^2}{2}\, d\theta dv-\frac{\pi B^2}{k},  \label{gen8}
\end{equation}
where $B^2=B_x^2+B_y^2$. It is a nice property of the HMF model that the potential energy can be expressed in terms of the order parameter $B$ played here by the magnetization. This is not the case for more complex potentials such as the gravitational potential. Note, finally, that the kinetic temperature can be expressed in terms of the pressure (\ref{ps11}) so that
\begin{equation}
E=\frac{1}{2}\int_0^{2\pi} p\, d\theta-\frac{\pi B^2}{k}=E_{kin}+W.
\label{gen9}
\end{equation}

\subsection{Complete and incomplete polytropes}
\label{sec_cip}

Let us now study polytropic distributions in the context of the HMF model. According to Eq. (\ref{pp1}), their density profile is given by
\begin{equation}
\rho(\theta)=A\left\lbrack 1-\frac{\gamma-1}{K\gamma A^{\gamma-1}}\Phi(\theta)\right\rbrack_+^{\frac{1}{\gamma-1}},
\label{cip1}
\end{equation}
where $A=\lambda^{1/(\gamma-1)}$. We can assume without loss of generality that the distribution is symmetrical with respect to the $x$ axis ($\theta=0$). In that case, the potential can be written
\begin{equation}
\Phi=-B\cos\theta,
\label{cip2}
\end{equation}
where
\begin{equation}
B=\frac{k}{2\pi}\int_{0}^{2\pi}\rho(\theta)\cos\theta\, d\theta,
\label{cip3}
\end{equation}
is the magnetization ($B_y=0$ and $B=B_x$). The density profile  can be rewritten
\begin{equation}
\rho(\theta)=A\left\lbrack 1+\frac{\gamma-1}{\gamma}x\cos\theta\right\rbrack_{+}^{\frac{1}{\gamma-1}},
\label{cip4}
\end{equation}
where
\begin{equation}
x=\frac{B}{KA^{\gamma-1}}.
\label{cip5}
\end{equation}
For $\gamma\rightarrow 1$, we recover the isothermal distribution (20) and the relation $x=\beta B$ of \cite{cvb}. For $x>0$, the density profile is concentrated around $\theta=0$ and for $x<0$, we get a symmetrical density profile concentrated around $\theta=\pi$.

\begin{figure}
\begin{center}
\includegraphics[clip,scale=0.3]{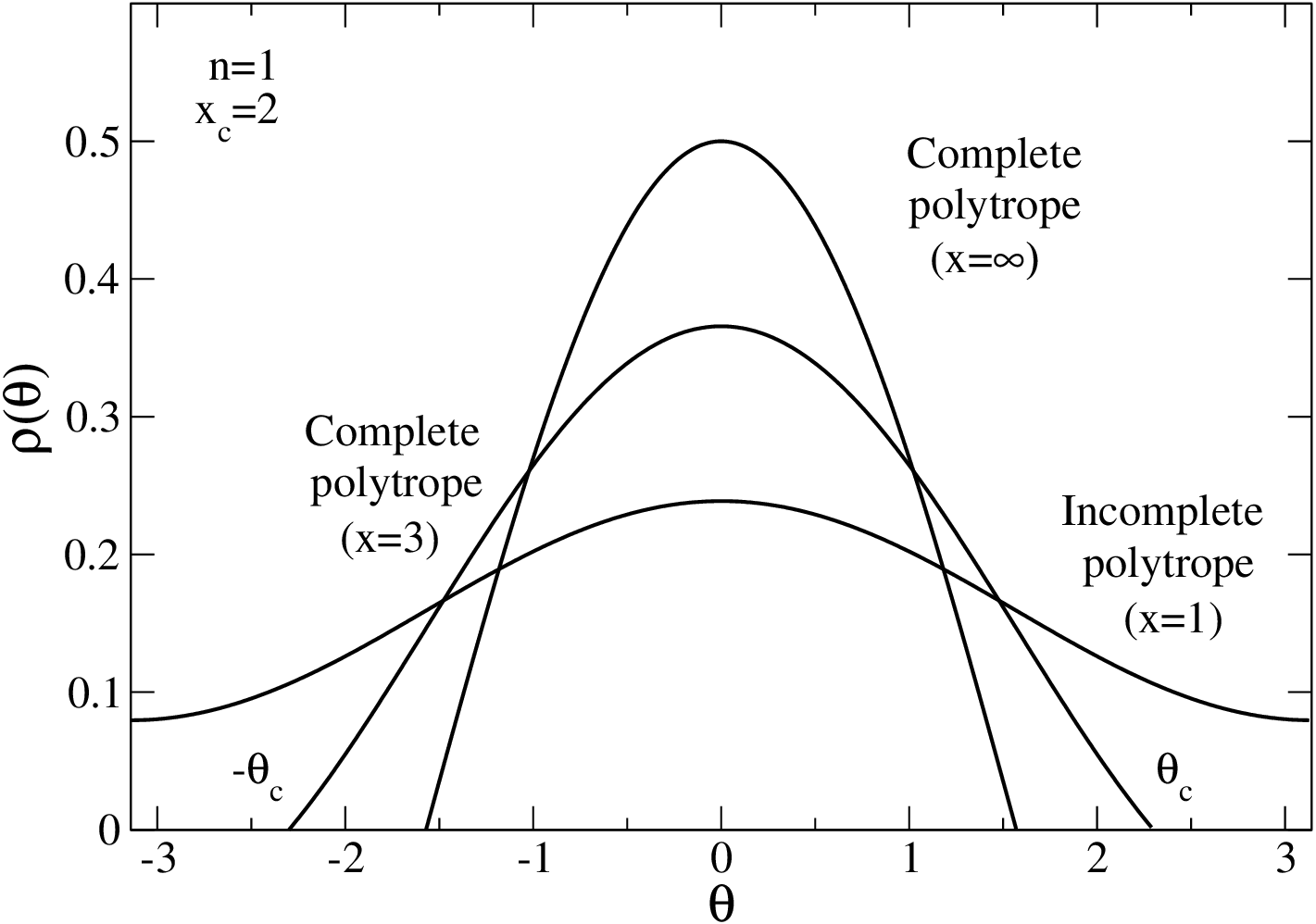}
\caption{Density profile for the polytropes $n=1$. We have represented incomplete ($x>x_c$) and complete ($x<x_c$) polytropes. }
\label{profiles}
\end{center}
\end{figure}

Like for the gravitational interaction, we can have complete or
incomplete polytropes \cite{lang}.  By definition, complete polytropes
have a compact support (the density drops to zero at $\theta_c<\pi$)
while incomplete polytropes extend over the whole domain. Let us first
consider the case $n\ge 1/2$ (i.e. $1\le
\gamma\le 3$). The density profile can be written
\begin{equation}
\rho(\theta)=A\left (1+\frac{x}{n+1}\cos\theta\right )_{+}^{n}.
\label{cip6}
\end{equation}
By symmetry, we can restrict ourselves to the interval $0\le\theta\le\pi$ and to $x\ge 0$. The density profile is monotonically decreasing and it has a compact support iff the term in parenthesis is negative for $\theta=\pi$. This occurs iff
\begin{equation}
x>x_c\equiv n+1=\frac{\gamma}{\gamma-1}.
\label{cip7}
\end{equation}
If $x<x_c$, the density is strictly positive on $0\le\theta\le\pi$ (incomplete polytrope). If $x>x_c$, the density vanishes for $\theta\ge \theta_c$ (complete polytrope) such that
\begin{equation}
1+\frac{1}{n+1}x\cos\theta_c=0.
\label{cip8}
\end{equation}
Therefore, $\theta_c$ is given by
\begin{equation}
\theta_c=\arccos\left (-\frac{x_c}{x}\right ).
\label{cip9}
\end{equation}
Some typical polytropic profiles are represented in
Fig. \ref{profiles}.  We now consider the case $n<-1$ (i.e. $0\le
\gamma\le 1$). The density profile can be written
\begin{equation}
\rho(\theta)=\frac{A}{\left (1-\frac{x}{|n+1|}\cos\theta\right )^{|n|}},
\label{cip10}
\end{equation}
and the solution is physical iff
\begin{equation}
x<x_*=-(n+1)=-\frac{\gamma}{\gamma-1}.
\label{cip11}
\end{equation}
In that case, the  polytropes are always incomplete.

\subsection{The magnetization and the polytropic temperature}
\label{sec_mag}

\begin{figure}
\begin{center}
\includegraphics[clip,scale=0.3]{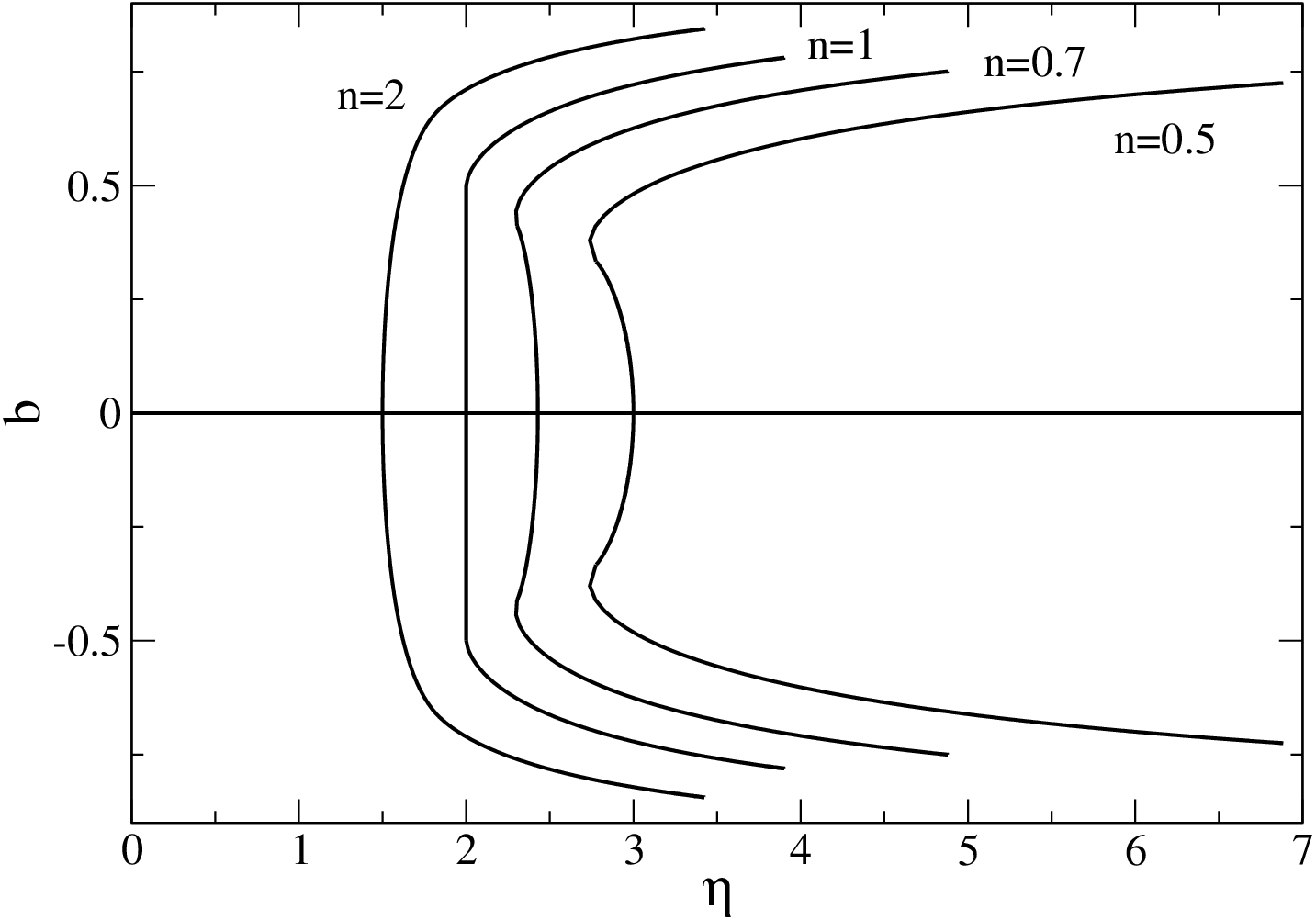}
\caption{Magnetization as a function of the polytropic temperature for $n\ge 1/2$. }
\label{magnetisationNposTemp}
\end{center}
\end{figure}

For the density profile  (\ref{cip4}), the mass and the magnetization are given by
\begin{equation}
M=A\int_{0}^{2\pi}\left (1+\frac{\gamma-1}{\gamma}x\cos\theta\right
)_{+}^{\frac{1}{\gamma-1}}\, d\theta,
\label{mag1}
\end{equation}
\begin{equation}
B=\frac{kA}{2\pi}\int_{0}^{2\pi}\left  (1+\frac{\gamma-1}{\gamma}x\cos\theta\right )_{+}^{\frac{1}{\gamma-1}}\cos\theta\, d\theta.
\label{mag2}
\end{equation}
It is useful to introduce the $\gamma$-deformed Bessel functions
\begin{equation}
I_{\gamma,m}(x)=\frac{1}{2\pi}\int_{0}^{2\pi}\left (1+\frac{\gamma-1}{\gamma}x\cos\theta\right )_{+}^{\frac{1}{\gamma-1}}\cos(m\theta)\, d\theta.
\label{mag3}
\end{equation}
For $\gamma\rightarrow 1$, we recover the Bessel functions $I_m(x)$. For $1\le\gamma\le 3$ ($n\ge 1/2$) and  $x<x_c$, we have
\begin{eqnarray}
I_{\gamma,m}(x)=\frac{1}{2\pi}\int_{-\pi}^{\pi}\left (1+\frac{\gamma-1}{\gamma}x\cos\theta\right )^{\frac{1}{\gamma-1}}\cos(m\theta)\, d\theta,\nonumber\\
\label{mag4}
\end{eqnarray}
and for $x>x_c$, we have
\begin{eqnarray}
I_{\gamma,m}(x)=\frac{1}{2\pi}\int_{-\theta_c}^{\theta_c}\left (1+\frac{\gamma-1}{\gamma}x\cos\theta\right )^{\frac{1}{\gamma-1}}\cos(m\theta)\, d\theta.\nonumber\\
\label{mag5}
\end{eqnarray}
We also have $I_{\gamma,m}(-x)=I_{\gamma,m}(x)$ if $m$ is even and $I_{\gamma,m}(-x)=-I_{\gamma,m}(x)$ if $m$ is odd. In terms of these integrals, we can rewrite the relations (\ref{mag1}) and (\ref{mag2}) in the form
\begin{equation}
A=\frac{M}{2\pi I_{\gamma,0}(x)},
\label{mag6}
\end{equation}
and
\begin{equation}
b\equiv \frac{2\pi B}{kM}=\frac{I_{\gamma,1}(x)}{I_{\gamma,0}(x)}.
\label{mag7}
\end{equation}
Combining Eqs. (\ref{cip5}), (\ref{mag6}) and (\ref{mag7}) and introducing the normalized polytropic temperature \cite{cvb}:
\begin{equation}
\eta\equiv (2\pi)^{\gamma-1}\frac{kM^{2-\gamma}}{4\pi K},
\label{mag8}
\end{equation}
we find that
\begin{equation}
\eta=\frac{x}{2}\frac{I_{\gamma,0}(x)^{2-\gamma}}{I_{\gamma,1}(x)}.
\label{mag9}
\end{equation}
Equations (\ref{mag7}) and (\ref{mag9}) determine the magnetization $B$ as a function of the polytropic temperature $K$ in a parametric form (with parameter $x$). For $\gamma\rightarrow 1$, we recover the  self-consistency relation (26) of \cite{cvb}. Some curves are plotted in Fig. \ref{magnetisationNposTemp} for $n\ge 1/2$. The case of negative indices $n<-1$ is very similar to the isothermal case \cite{cvb} and will not be illustrated in detail. For $x\rightarrow x_*$, $b\to 1$, $\eta\to +\infty$ and $\epsilon\to -2$.

\subsection{The energy}
\label{sec_energy}

\subsubsection{The homogeneous phase}
\label{sec_hom}

In the homogeneous phase, the density is uniform with value $\rho=M/(2\pi)$ and the magnetization vanishes ($B=0$). Therefore, the  energy (\ref{gen9})  reduces to the kinetic energy
\begin{equation}
E=\frac{1}{2}\int_0^{2\pi} p\, d\theta=\frac{1}{2}K\int_0^{2\pi} \rho^{\gamma}\, d\theta=K\pi\left (\frac{M}{2\pi}\right )^{\gamma}.
\label{hom1}
\end{equation}
Introducing the dimensionless energy \cite{cvb}:
\begin{equation}
\epsilon\equiv \frac{8\pi E}{kM^2},
\label{hom2}
\end{equation}
and recalling the expression (\ref{mag8}) of the dimensionless
polytropic temperature, we find that the caloric
curve\footnote{We shall explain in Sec. \ref{sec_tcc} why this is the relevant caloric curve to consider.} for the homogeneous distribution is simply \cite{cvb}:
\begin{equation}
\eta=\frac{1}{\epsilon}.
\label{hom3}
\end{equation}
It has been shown in \cite{cvb,cd} that the homogeneous phase is
stable iff
\begin{equation}
\eta\le \eta_c=\gamma, \qquad \epsilon\ge \epsilon_c=\frac{1}{\gamma}.
\label{hom4}
\end{equation}

\subsubsection{The inhomogeneous phase}
\label{sec_in}

\begin{figure}
\begin{center}
\includegraphics[clip,scale=0.3]{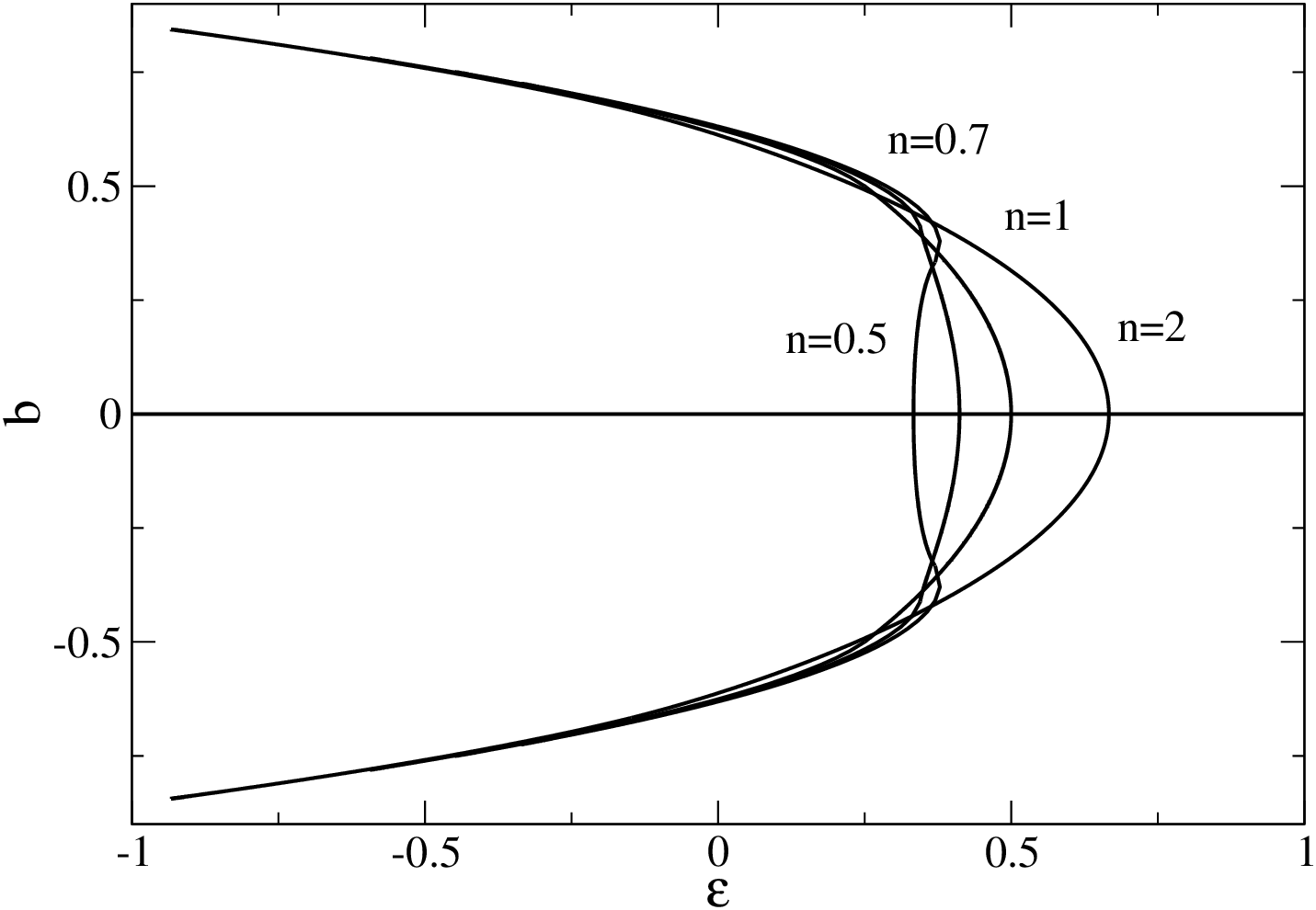}
\caption{Magnetization as a function of the energy for $n\ge 1/2$. }
\label{magnetisationNposEner}
\end{center}
\end{figure}

\begin{figure}
\begin{center}
\includegraphics[clip,scale=0.3]{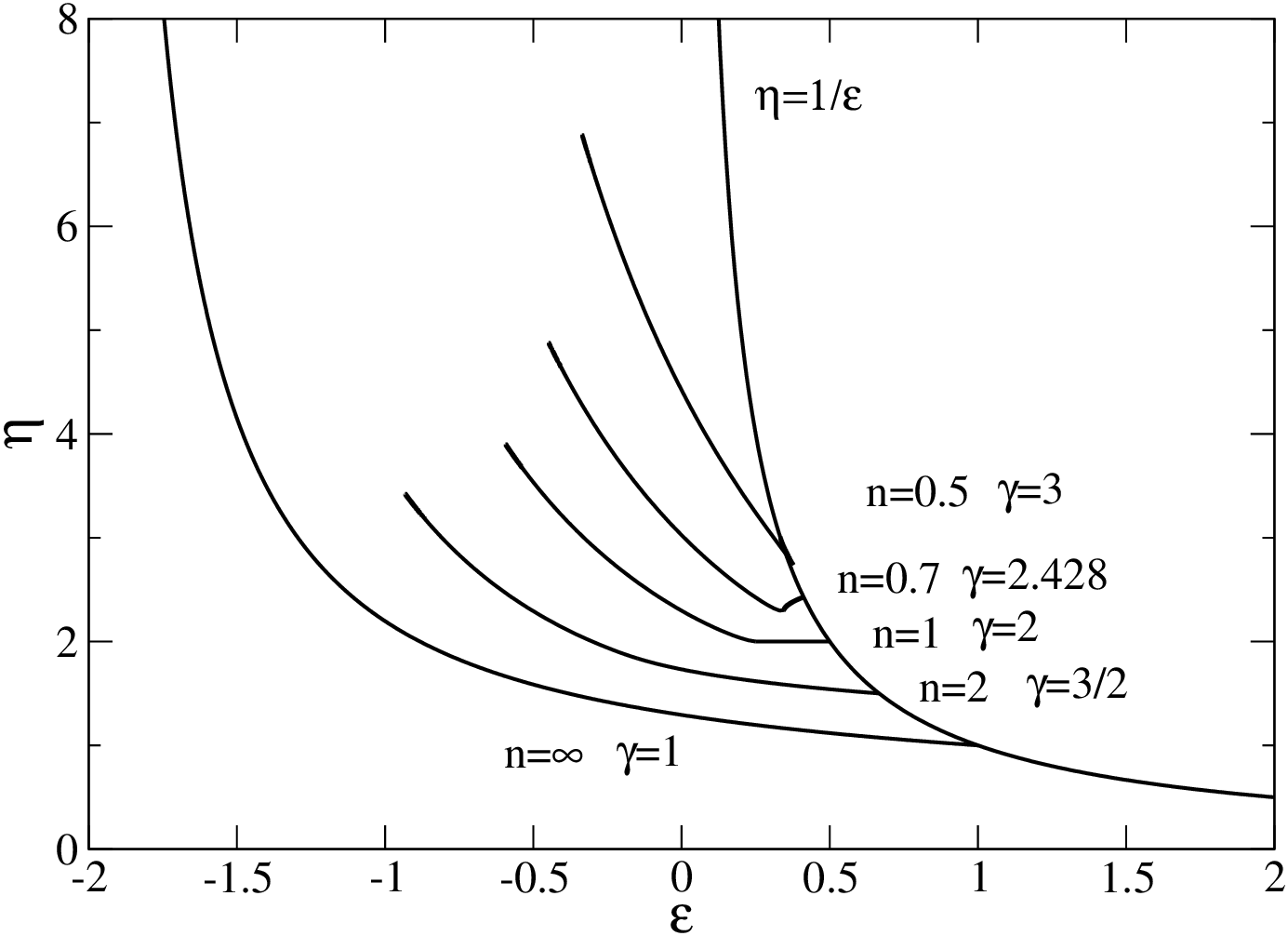}
\caption{Caloric curve for polytropes with positive indices $n\ge 1/2$. }
\label{caloricNpos}
\end{center}
\end{figure}

\begin{figure}
\begin{center}
\includegraphics[clip,scale=0.3]{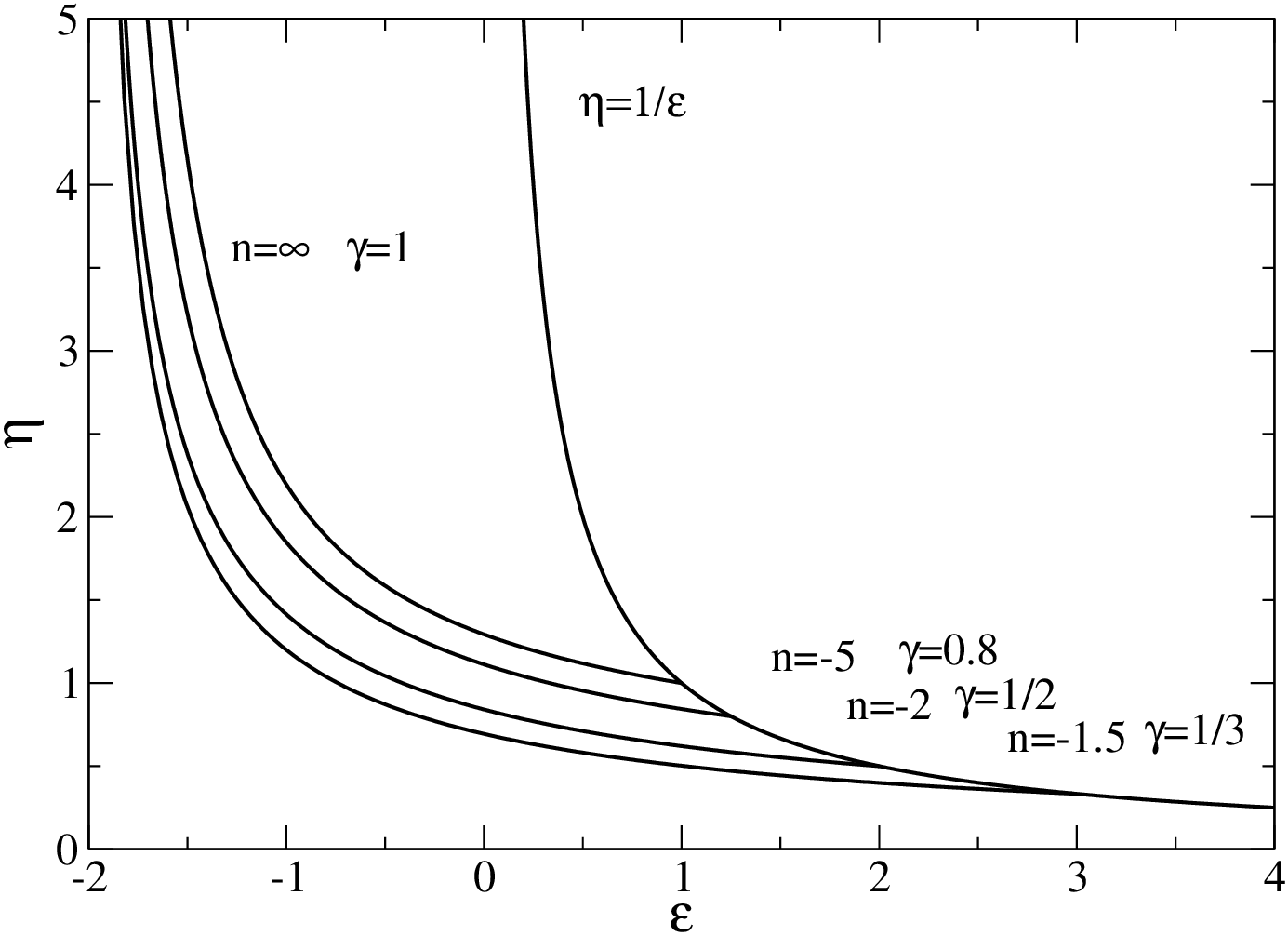}
\caption{Caloric curve for polytropes with negative indices $n<-1$. 
}
\label{caloricNneg}
\end{center}
\end{figure}

To compute the kinetic energy in the inhomogeneous phase, without introducing new integrals, we shall use a trick inspired from the astrophysical problem \cite{lang}. The condition of equilibrium can be written as a condition of hydrostatic balance (see Appendix \ref{sec_hydroeq}) \cite{cvb}:
\begin{equation}
\frac{dp}{d\theta}=-\rho\frac{d\Phi}{d\theta}.
\label{in1}
\end{equation}
For a polytropic equation of state, we have
\begin{equation}
p=K\rho^{\gamma}.
\label{in2}
\end{equation}
Therefore, the foregoing equation can be integrated into
\begin{equation}
\Phi=-\frac{K\gamma}{\gamma-1}\rho^{\gamma-1}+C.
\label{in3}
\end{equation}
We must be careful that this relation is valid only in the region
where $\rho>0$. We must therefore consider two cases. For incomplete polytropes, using $\Phi(\pi)=B$, we get
\begin{equation}
\Phi=-\frac{K\gamma}{\gamma-1}(\rho^{\gamma-1}-\rho(\pi)^{\gamma-1})+B.
\label{in4}
\end{equation}
On the other hand, for complete polytropes, using $\rho(\theta_c)=0$ and $\Phi(\theta_c)=-B\cos\theta_c$, we get
\begin{equation}
\Phi=-\frac{K\gamma}{\gamma-1}\rho^{\gamma-1}-B\cos\theta_c.
\label{in5}
\end{equation}
These two cases can be combined in a single formula
\begin{equation}
\Phi=-\frac{K\gamma}{\gamma-1}(\rho^{\gamma-1}-\rho(\pi)^{\gamma-1})-B\cos\theta_c.
\label{in6}
\end{equation}
For incomplete polytropes $\theta_c=\pi$, and for complete polytropes
$\rho(\pi)=0$.  Substituting Eq. (\ref{in6}) in the potential energy (\ref{gen3}) and using Eq. (\ref{in2}), we obtain
\begin{eqnarray}
W=-\frac{1}{2}\frac{\gamma}{\gamma-1}\int p\, d\theta+\frac{1}{2}\frac{K\gamma}{\gamma-1}M\rho(\pi)^{\gamma-1}\nonumber\\
-\frac{1}{2}MB\cos\theta_c.
\label{in7}
\end{eqnarray}
In the first integral, we recognize the kinetic energy (\ref{gen9}).
Therefore, we obtain
\begin{eqnarray}
E_{kin}=-\frac{\gamma-1}{\gamma}W+\frac{1}{2}KM\rho(\pi)^{\gamma-1}
-\frac{1}{2}\frac{\gamma-1}{\gamma}MB\cos\theta_c.\nonumber\\
\label{in9}
\end{eqnarray}
Using the expression (\ref{gen9})
of the potential energy, we get
\begin{eqnarray}
E_{kin}=\frac{\gamma-1}{\gamma}\frac{\pi B^2}{k}-\frac{1}{2}\frac{\gamma-1}{\gamma}MB\cos\theta_c+\frac{1}{2}KM\rho(\pi)^{\gamma-1}.\nonumber\\
\label{in11}
\end{eqnarray}
Finally, the total energy $E=E_{kin}+W$ is
\begin{equation}
E=-\frac{\pi B^2}{\gamma k}-\frac{1}{2}\frac{\gamma-1}{\gamma}MB\cos\theta_c+\frac{1}{2}KM\rho(\pi)^{\gamma-1}.
\label{in12}
\end{equation}
Using Eqs. (\ref{mag7}) and (\ref{cip4}), the normalized energy (\ref{hom2}) can be written
\begin{equation}
\epsilon=-\frac{2}{\gamma}b^2-2\frac{\gamma-1}{\gamma}b\cos\theta_c+2\frac{b}{x}\left (1-\frac{\gamma-1}{\gamma}x\right )_+.
\label{in13}
\end{equation}
Using Eq. (\ref{mag7}), it can also be written explicitly in terms of $x$ as
\begin{eqnarray}
\epsilon=-\frac{2}{\gamma}\frac{I_{\gamma,1}(x)^2}{I_{\gamma,0}(x)^2}-2\frac{\gamma-1}{\gamma}\frac{I_{\gamma,1}(x)}{I_{\gamma,0}(x)}\cos\theta_c\nonumber\\
+\frac{2}{x}\frac{I_{\gamma,1}(x)}{I_{\gamma,0}(x)}\left (1-\frac{\gamma-1}{\gamma}x\right )_{+}.
\label{in14}
\end{eqnarray}
Equations (\ref{mag7}) and (\ref{in14}) determine the magnetization $b$ as a function of the energy $\epsilon$ in a parametric form. Some curves are plotted  in Fig. \ref{magnetisationNposEner} for $n\ge 1/2$. On the other hand, the caloric curve $\eta(\epsilon)$ is determined by Eqs. (\ref{mag9}) and (\ref{in14}).  Some curves are plotted in Fig. \ref{caloricNpos} for $n\ge 1/2$ and in Fig. \ref{caloricNneg} for $n<-1$. They complete Fig. 12 of \cite{cvb}  by adding the inhomogeneous branches. They will be analyzed in Sec. \ref{sec_tcc}.

\subsection{The averaged kinetic temperature}
\label{sec_kin}

\begin{figure}
\begin{center}
\includegraphics[clip,scale=0.3]{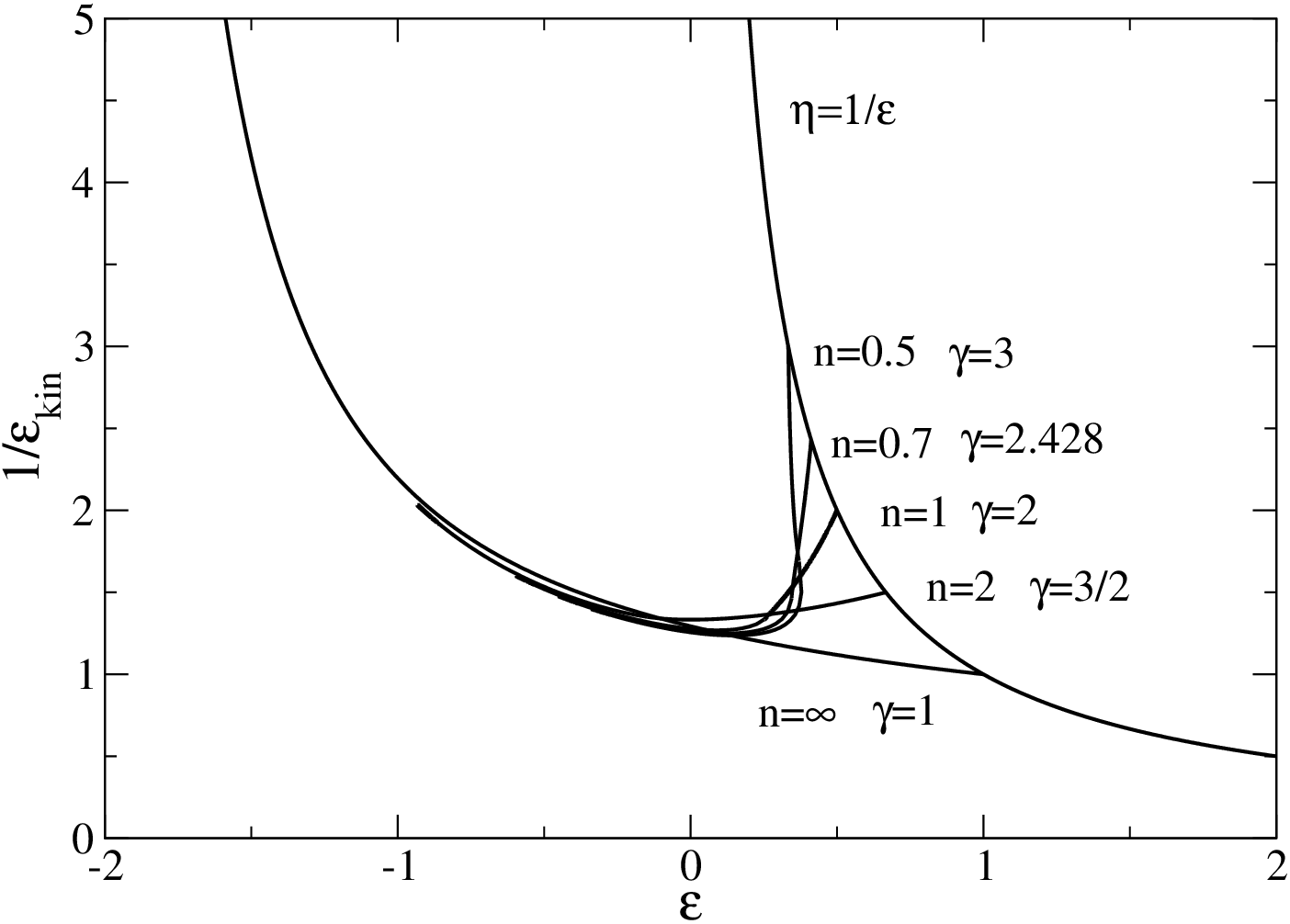}
\caption{Inverse averaged kinetic temperature as a function of the energy for polytropes with positive indices $n\ge 1/2$. }
\label{caloricNposKIN}
\end{center}
\end{figure}

The local kinetic temperature is defined by
\begin{eqnarray}
T_{kin}(\theta)=\langle v^2\rangle(\theta)=\frac{\int f v^2\, dv}{\int f\, dv}.
\label{kin1}
\end{eqnarray}
The averaged kinetic temperature is therefore
\begin{eqnarray}
\langle T_{kin}\rangle=\frac{1}{M}\int_0^{2\pi} \rho(\theta)\langle v^2\rangle(\theta)\, d\theta=\frac{2E_{kin}}{M}.
\label{kin2}
\end{eqnarray}
Defining the normalized kinetic energy and the normalized averaged kinetic temperature by
\begin{eqnarray}
\epsilon_{kin}=\frac{8\pi E_{kin}}{kM^2},\qquad \Theta=\frac{4\pi \langle T_{kin}\rangle}{kM},
\label{kin3}
\end{eqnarray}
we obtain
\begin{eqnarray}
\Theta=\epsilon_{kin}.
\label{kin4}
\end{eqnarray}
If we define the normalized potential energy by
\begin{eqnarray}
w=\frac{8\pi W}{kM^2},
\label{kin5}
\end{eqnarray}
and use Eqs. (\ref{gen9}) and (\ref{mag7}), we obtain
\begin{eqnarray}
w=-2b^2.
\label{kin6}
\end{eqnarray}
Finally, the total energy $E=E_{kin}+W$ can be written in dimensionless form
\begin{eqnarray}
\epsilon=\epsilon_{kin}+w=\epsilon_{kin}-2b^2.
\label{kin7}
\end{eqnarray}
From Eqs. (\ref{in13}) and (\ref{kin7}), we obtain
\begin{eqnarray}
\epsilon_{kin}=\Theta=2\frac{\gamma-1}{\gamma}b^2-2\frac{\gamma-1}{\gamma}b\cos\theta_c\nonumber\\
+2\frac{b}{x}\left (1-\frac{\gamma-1}{\gamma}x\right )_+.
\label{kin8}
\end{eqnarray}
Using Eq. (\ref{mag7}), this can also be written explicitly in terms of $x$ as
\begin{eqnarray}
\epsilon_{kin}=\Theta=2\frac{\gamma-1}{\gamma}\frac{I_{\gamma,1}(x)^2}{I_{\gamma,0}(x)^2}-2\frac{\gamma-1}{\gamma}\frac{I_{\gamma,1}(x)}{I_{\gamma,0}(x)}\cos\theta_c\nonumber\\
+\frac{2}{x}\frac{I_{\gamma,1}(x)}{I_{\gamma,0}(x)}\left (1-\frac{\gamma-1}{\gamma}x\right )_{+}.\qquad
\label{kin9}
\end{eqnarray}
Equations (\ref{kin9})  and (\ref{in14}) determine the physical caloric curve $\epsilon_{kin}(\epsilon)$ or $\Theta(\epsilon)$ in a parametric form. Some curves are represented in Fig. \ref{caloricNposKIN} for $n\ge 1/2$. In the homogeneous phase, we have
\begin{eqnarray}
\epsilon=\epsilon_{kin}=\Theta.
\label{kin10}
\end{eqnarray}
These physical caloric curves will be analyzed in Sec. \ref{sec_phy}.

\subsection{The free energy}
\label{sec_fe}

\begin{figure}
\begin{center}
\includegraphics[clip,scale=0.3]{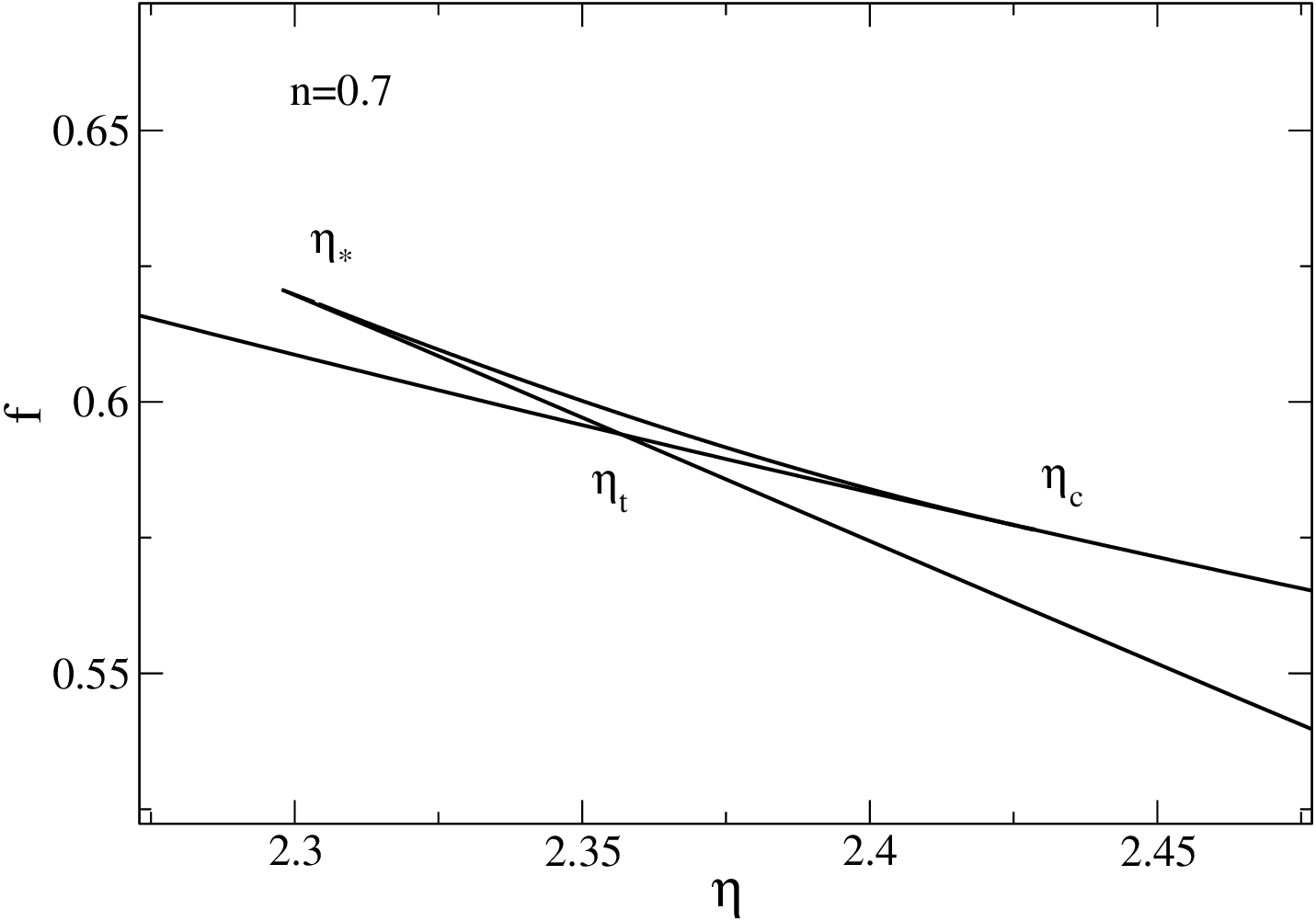}
\caption{Free energy as a function of the inverse polytropic temperature for $n=0.7$. It exhibits a first order canonical phase transition  marked by the discontinuity of $df/d\eta$ at $\eta=\eta_t$.}
\label{fn0.7}
\end{center}
\end{figure}

\begin{figure}
\begin{center}
\includegraphics[clip,scale=0.3]{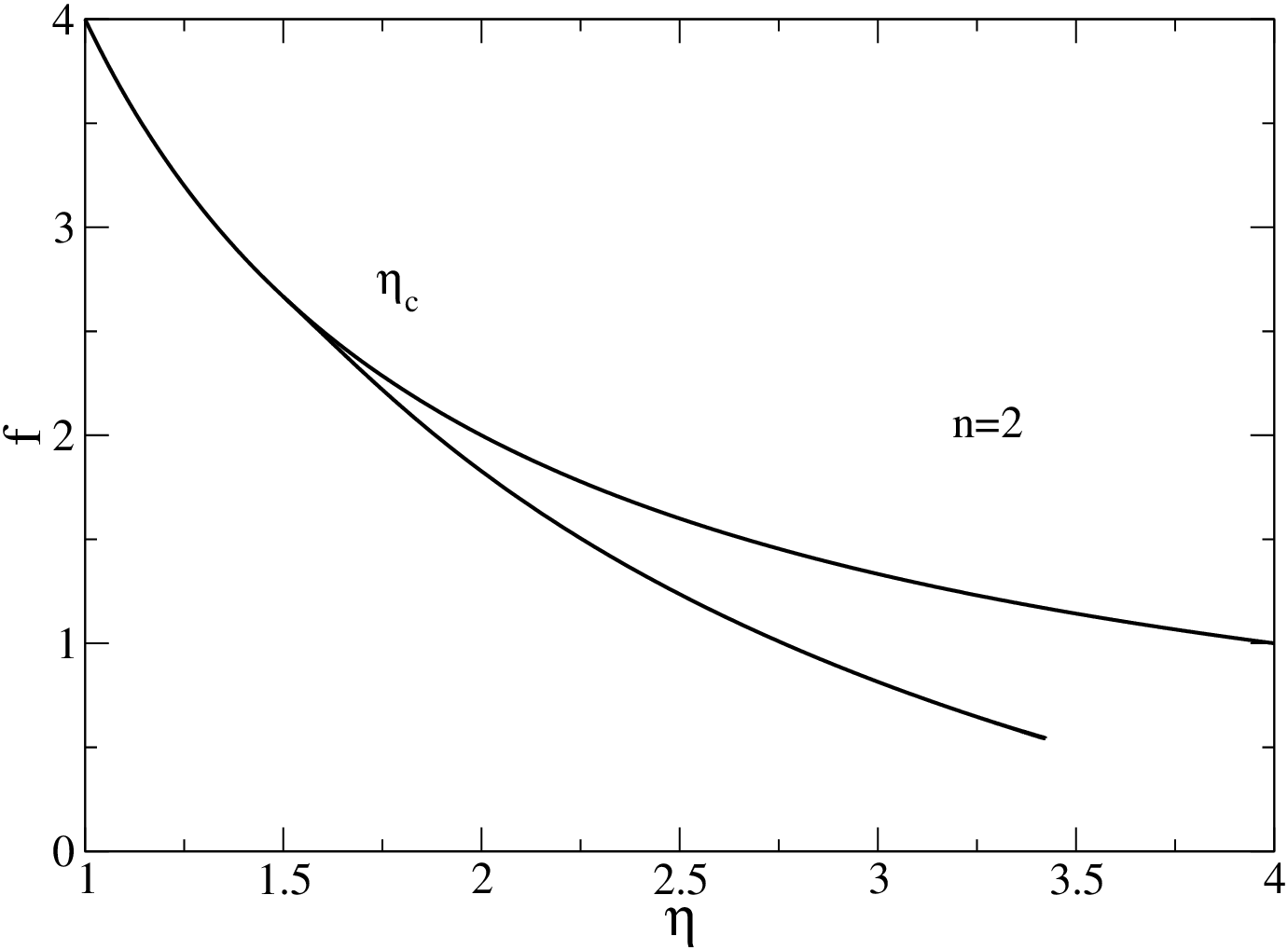}
\caption{Free energy as a function of the inverse polytropic temperature for $n=2$. It exhibits a second order phase transition marked by the discontinuity of $d^2f/d\eta^2$ at $\eta=\eta_c$.}
\label{fn2}
\end{center}
\end{figure}

According to Eq. (\ref{ef3}), the free energy of the polytropes can be written in terms of the density as
\begin{eqnarray}
F=\frac{1}{2}\int \rho\Phi\, d\theta+\frac{K}{\gamma-1}\int \rho^{\gamma}\, d\theta,
\label{fe1}
\end{eqnarray}
up to an unimportant constant. In the second term, we recognize the kinetic energy (\ref{gen9}) while the first term is just the potential energy, so that
\begin{eqnarray}
F=W+\frac{2}{\gamma-1}E_{kin}.
\label{fe2}
\end{eqnarray}
Introducing the normalized free energy
\begin{equation}
f=\frac{8\pi F}{k M^2},
\label{fe3}
\end{equation}
we obtain
\begin{eqnarray}
f=w+\frac{2}{\gamma-1}\epsilon_{kin}.
\label{fe4}
\end{eqnarray}
Substituting Eqs.  (\ref{kin6}) and (\ref{kin8}) in Eq. (\ref{fe4}), we find that
\begin{eqnarray}
f=\frac{2(2-\gamma)}{\gamma}b^2
-\frac{4}{\gamma}b\cos\theta_c
+\frac{4}{\gamma-1}\frac{b}{x}\left (1-\frac{\gamma-1}{\gamma}x\right )_{+}.\nonumber\\
\label{fe5}
\end{eqnarray}
Using Eq. (\ref{mag7}), this can be explicitly written  in terms of $x$ as
\begin{eqnarray}
f=\frac{2(2-\gamma)}{\gamma}\frac{I_{\gamma,1}(x)^2}{I_{\gamma,0}(x)^2}
-\frac{4}{\gamma}\frac{I_{\gamma,1}(x)}{I_{\gamma,0}(x)}\cos\theta_c\nonumber\\
+\frac{1}{\gamma-1}\frac{4}{x}\frac{I_{\gamma,1}(x)}{I_{\gamma,0}(x)}\left (1-\frac{\gamma-1}{\gamma}x\right )_{+}.
\label{fe6}
\end{eqnarray}
Equations (\ref{fe6}) and (\ref{mag9}) determine the free energy   as a function of the  polytropic temperature $f(\eta)$ in a parametric form. Some curves are represented in Figs. \ref{fn0.7} and \ref{fn2} for $n\ge 1/2$. They will be analyzed in Sec. \ref{sec_tcc}. In the homogeneous phase, Eq. (\ref{fe2}) reduces to $F=2E/(\gamma-1)$. Using Eq. (\ref{hom3}), we get
\begin{eqnarray}
f=\frac{2}{\gamma-1}\frac{1}{\eta}.
\label{fe7}
\end{eqnarray}

\subsection{The entropy}
\label{sec_ent}

\begin{figure}
\begin{center}
\includegraphics[clip,scale=0.3]{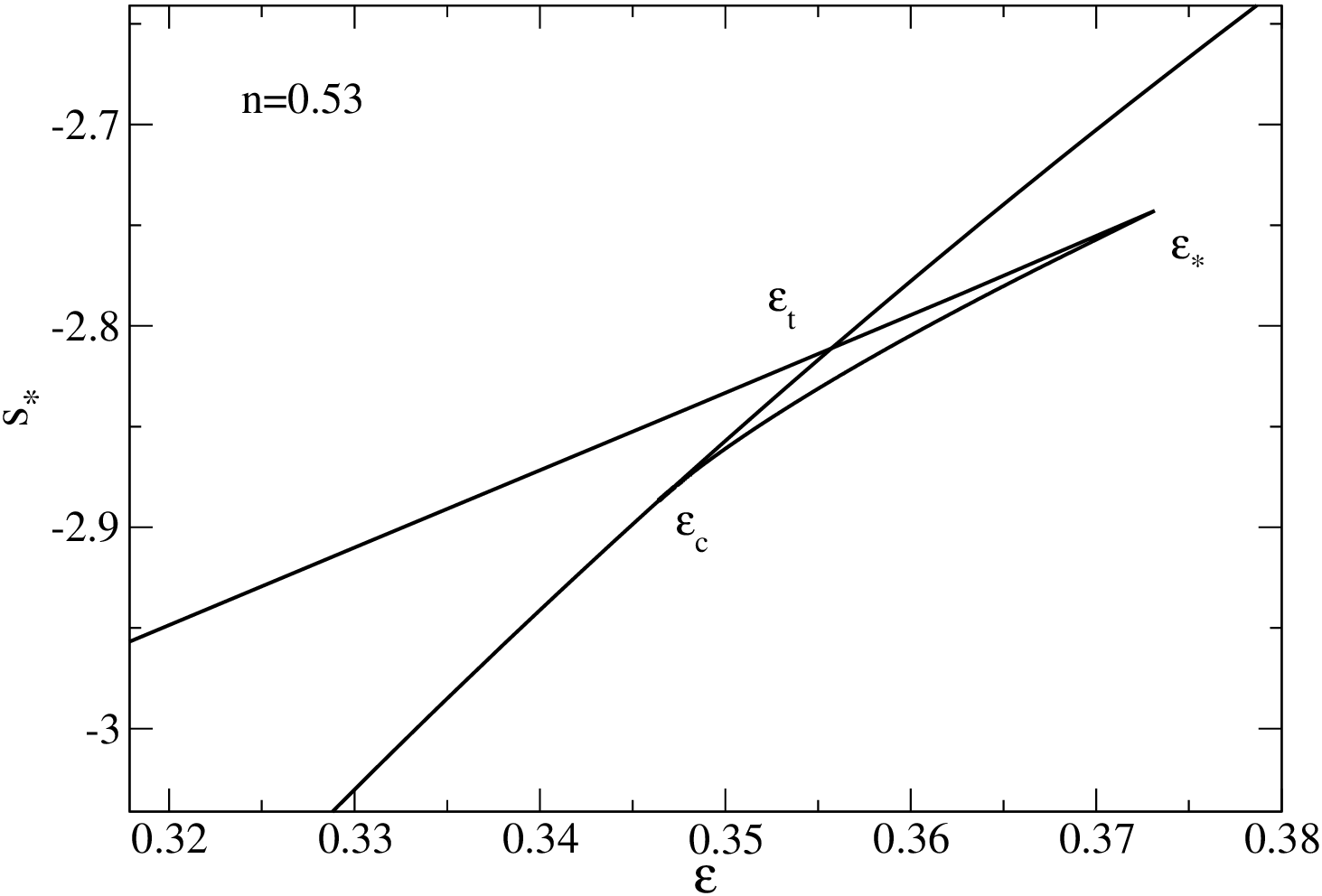}
\caption{Entropy as a function of energy for $n=0.53$. It exhibits a microcanonical first order phase transition marked by the discontinuity of $ds/d\epsilon$ at $\epsilon=\epsilon_t$. To avoid the occurrence of large numbers when $n\simeq 0.5$, we have plotted $s_*=-\epsilon_{kin}^{2n-1}\eta^{2n}$ and $s_*^{homo}=-1/\epsilon$ as a function of $\epsilon$.}
\label{sETOILEn0.53}
\end{center}
\end{figure}

\begin{figure}
\begin{center}
\includegraphics[clip,scale=0.3]{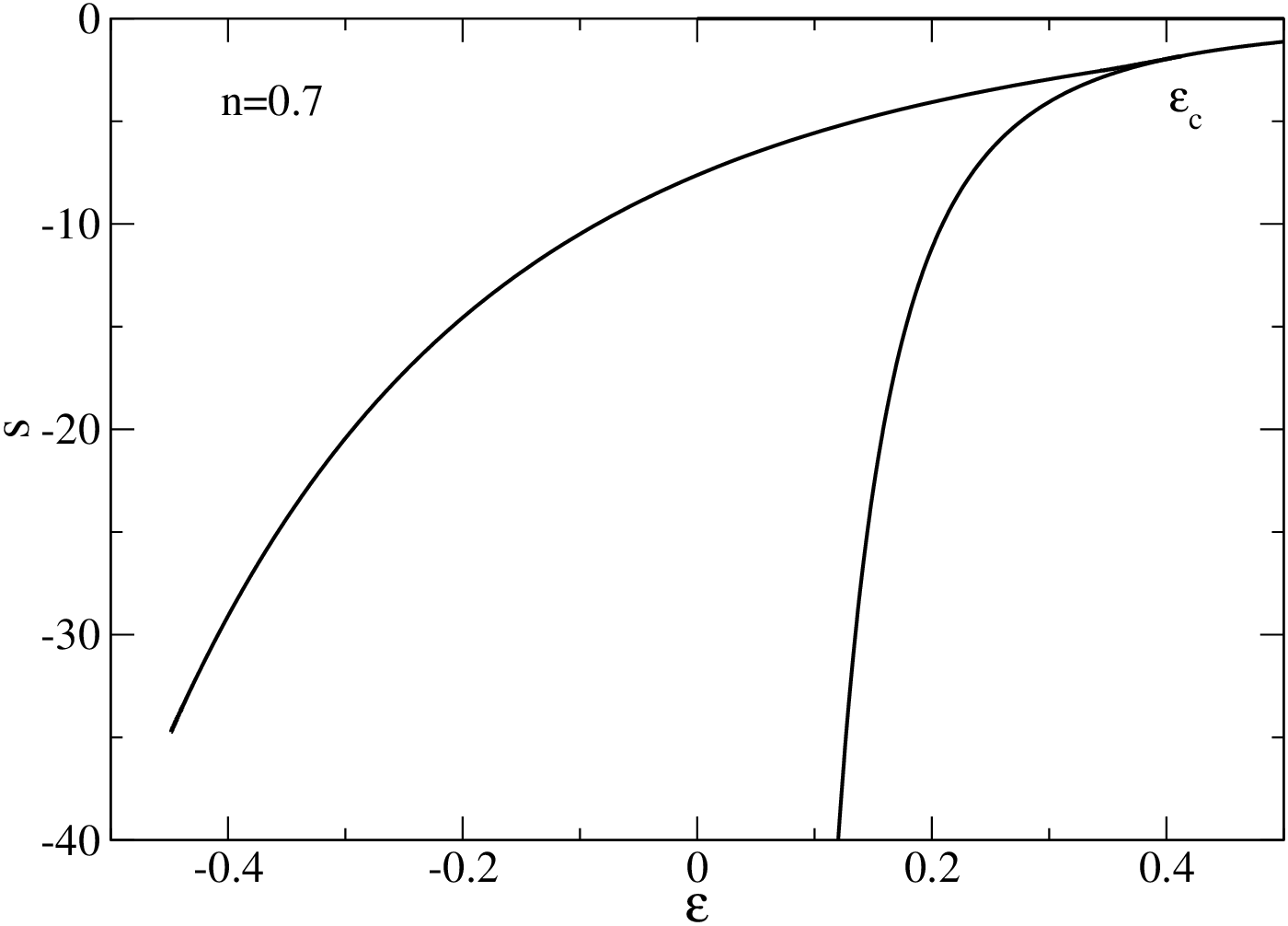}
\caption{Entropy as a function of energy for $n=0.7$. It exhibits a  microcanonical second order phase transition marked by the discontinuity of $d^2s/d\epsilon^2$ at $\epsilon=\epsilon_c$. There is also a small convex dip (hardly visible) evidencing a region of negative specific heats and a canonical first order phase transition. }
\label{sn0.7}
\end{center}
\end{figure}

According to Eq. (\ref{ef1}), the entropy of the polytropes can be written in terms of the density as
\begin{eqnarray}
S=-\left (n-\frac{1}{2}\right )\beta\int p\, d\theta,
\label{ent1}
\end{eqnarray}
up to an unimportant constant. In the second term, we recognize the kinetic energy (\ref{gen9}) so that
\begin{eqnarray}
S=-2\left (n-\frac{1}{2}\right )\beta E_{kin}.
\label{ent2}
\end{eqnarray}
On the other hand, the inverse thermodynamical temperature $\beta=1/T$ is related to the polytropic temperature $K$ by Eq. (\ref{es8}). Therefore, the entropy can be rewritten
\begin{eqnarray}
S=-\left (n-\frac{1}{2}\right )\frac{2E_{kin}}{C_n K^{\frac{2n}{2n-1}}}.
\label{ent3}
\end{eqnarray}
Defining  the normalized entropy by
\begin{eqnarray}
s=\frac{C_n (\pi k)^{\frac{1}{2n-1}}}{M^{\frac{2n}{2n-1}}}S,
\label{ent4}
\end{eqnarray}
we obtain
\begin{eqnarray}
s=-\left (n-\frac{1}{2}\right )\epsilon_{kin}\eta^{\frac{2n}{2n-1}}.
\label{ent5}
\end{eqnarray}
Using Eqs. (\ref{mag9}) and (\ref{kin8}), this gives $s$ as a function of $x$. Finally, Eqs. (\ref{ent5}) and (\ref{in14}) determine the entropy as a function of the  energy  $s(\epsilon)$ in a parametric form. Some curves are represented in Figs. \ref{sETOILEn0.53} and \ref{sn0.7} for $n\ge 1/2$. They will be analyzed in Sec. \ref{sec_tcc}. For the homogeneous phase, using $\epsilon=\epsilon_{kin}$ and $\eta=1/\epsilon$, we get
\begin{eqnarray}
s=-\left (n-\frac{1}{2}\right )\frac{1}{\epsilon^{\frac{1}{2n-1}}}.
\label{ent6}
\end{eqnarray}

\section{The thermodynamical caloric curve}
\label{sec_tcc}

\subsection{General comments}
\label{sec_gc}

Basically, we want to study the optimization problems (\ref{dyn2}) and (\ref{dyn3}) where $S$ is the Tsallis entropy (\ref{ps1}). The maximization problem (\ref{dyn2}) corresponds to a condition of microcanonical stability and the minimization problem (\ref{dyn3}) to a condition of canonical stability\footnote{Recall that these optimization problems can be given a dynamical interpretation (see Sec. \ref{sec_dyn}). For convenience, we use here a thermodynamical language based on the {\it thermodynamical analogy} described in  Sec. \ref{sec_dyn}. We should add the prefix ``pseudo'' (i.e. pseudo entropy, pseudo temperature...) but for simplicity, we avoid it.}. It is well-known that the two optimization problems (\ref{dyn2}) and (\ref{dyn3}) have the same critical points canceling the first order variations of the thermodynamical potential. The critical points of the maximization problem (\ref{dyn2}) are given by Eq. (\ref{ps6}) where $\beta$ and $\alpha$ are Lagrange multipliers associated with the constraints $E$ and $M$. Here, $\beta=1/T$ represents the inverse thermodynamical temperature. It satisfies the thermodynamical relation $\beta=(\partial S/\partial E)_M$. For the Tsallis entropy, the critical points in the microcanonical ensemble are given by Eq. (\ref{ps7}).  The critical points of the minimization problem (\ref{dyn3}) are given by $\delta F+\alpha T\delta M=0$  where  $\alpha$ is a  Lagrange multiplier associated with the constraint $M$. Obviously, these first order variations are equivalent to Eq. (\ref{ps6}) so that the critical points in the canonical ensemble are again given by Eq. (\ref{ps7}). Although the critical points are the same, their stability can differ in the microcanonical and canonical ensembles. From general theorems \cite{ellis}, we know that a critical point is stable in the microcanonical ensemble (maximum of $S$ at fixed $E$ and $M$) if it is stable in the canonical ensemble
(minimum  of $F$ at fixed $M$) but the reciprocal may be wrong. In that case, we have ensembles inequivalence. Related to this property, we can show that the thermodynamical specific heat $C=dE/dT$ is always positive in the canonical ensemble while is can be positive or negative in the microcanonical ensemble. These are general results valid for all entropies of the form (\ref{dyn1}), including the Tsallis entropy (\ref{ps1}), provided that we use the proper thermodynamical variable $E$ and $\beta=(\partial S/\partial E)_M$ that are canonically conjugate with respect to the entropy $S$. Therefore, the  proper thermodynamical caloric curve associated with the optimization problems (\ref{dyn2}) and (\ref{dyn3}) is the one giving the inverse temperature $\beta=1/T$ (Lagrange multiplier) as a function of the energy $E$. It is from this curve that we can study ensembles inequivalence \cite{ellis}. It is also from this curve that we can apply the Poincar\'e theory on the linear series of equilibria \cite{katz,ijmpb}. Now, for polytropic distributions, we have seen in Eq. (\ref{es8}) that the polytropic temperature  $K$ is always a monotonic function of the thermodynamical temperature $T$. Therefore, for convenience, we can equivalently consider the caloric curve giving the polytropic temperature $K$ as a function of the energy $E$ (the preceding properties are unchanged). This is the choice that has been made in the gravitational problem \cite{lang,aaantonov}  and that we shall make here. Therefore, we shall call $K(E)$ the thermodynamical caloric curve. In dimensionless variables, it corresponds to $\eta(\epsilon)$. Several caloric curves $\eta(\epsilon)$ have been plotted in Figs. \ref{caloricNpos} and  \ref{caloricNneg} for different indices $n$. In order to understand their behavior let us first consider the asymptotic limits $x\rightarrow 0$ and $x\rightarrow +\infty$.

\subsection{The limit $x\rightarrow 0$}
\label{sec_ze}

For $x\rightarrow 0$, the deformed Bessel functions (\ref{mag3}) can be approximated by
\begin{eqnarray}
I_{\gamma,0}(x)=1+\frac{2-\gamma}{4\gamma^2}x^2+...,
\label{ze1}
\end{eqnarray}
\begin{eqnarray}
I_{\gamma,1}(x)=\frac{x}{2\gamma}+\frac{(\gamma-2)(2\gamma-3)}{16\gamma^3}x^3+....
\label{ze2}
\end{eqnarray}
Then, we obtain the expansions
\begin{eqnarray}
b=\frac{x}{2\gamma}+\frac{(\gamma-2)(2\gamma-1)}{16\gamma^3}x^3+...
\label{ze3}
\end{eqnarray}
\begin{eqnarray}
\eta=\gamma+\frac{2-\gamma}{8\gamma}x^2+...,
\label{ze4}
\end{eqnarray}
\begin{eqnarray}
\epsilon=\frac{1}{\gamma}+(2\gamma^2-5\gamma-2)\frac{x^2}{8\gamma^3}+...
\label{ze5}
\end{eqnarray}
\begin{eqnarray}
\epsilon_{kin}=\Theta=\frac{1}{\gamma}+(2\gamma^2-\gamma-2)\frac{x^2}{8\gamma^3}+...
\label{ze6}
\end{eqnarray}
From these expressions, we can draw the following conclusions. Let us introduce the critical indices
\begin{eqnarray}
\gamma_*=\frac{5+\sqrt{41}}{4}\simeq 2.8507811...
\label{ze7}
\end{eqnarray}
\begin{eqnarray}
n_*=\frac{4}{1+\sqrt{41}}\simeq 0.54031242...
\label{ze8}
\end{eqnarray}
Let us first consider the case $n\ge 1/2$ so that $1\le \gamma\le
3$. The quantity $2\gamma^2-5\gamma-2$ is positive for $\gamma_*<\gamma\le 3$,
i.e. $1/2\le n<n_*$, showing that the energy increases as $x$
increases.  The quantity $2\gamma^2-5\gamma-2$ is negative for $1\le
\gamma<\gamma_*$, i.e. $n>n_*$. In that case, the energy decreases as
$x$ increases. On the other hand, for $2<
\gamma\le 3$, i.e. $1/2\le n<1$, the inverse temperature $\eta$ decreases as $x$
increases. By contrast, for $1\le  \gamma< 2$, i.e. $n>1$, the inverse
temperature $\eta$ increases as $x$ increases. This explains the
behavior of the caloric curve $\eta(\epsilon)$ close to the
bifurcation point. Close to that point, we have
\begin{eqnarray}
\eta-\eta_c=\frac{(2-\gamma)\gamma^2}{2\gamma^2-5\gamma-2}(\epsilon-\epsilon_c)+...
\label{ze9}
\end{eqnarray}
The specific heat is
\begin{eqnarray}
C_T=\frac{dE}{dT}=-\beta^2\frac{dE}{d\beta}.
\label{ze10}
\end{eqnarray}
Since $T$ is a monotonically increasing  function of $K$, we can equivalently consider the quantity
\begin{eqnarray}
C_K=-\eta^2\frac{d\epsilon}{d\eta}.
\label{ze11}
\end{eqnarray}
At the bifurcation point, we have
\begin{eqnarray}
C_K=-\frac{2\gamma^2-5\gamma-2}{2-\gamma}.
\label{ze12}
\end{eqnarray}
The specific heat close to the bifurcation point is positive for $1/2\le n<n_*$, negative for $n_*<n<1$ and positive again for $n>1$. It vanishes for $n=n_*$ and it is infinite for $n=1$.  Let us now consider the case $n<-1$ so that $0<\gamma\le  1$. In that case,  $\eta$ increases as $x$ increases. On the other hand, the quantity $2\gamma^2-5\gamma-2$ is always negative so that $\epsilon$ decreases as $x$ increases. Therefore, the specific heat is always positive.

\subsection{The limit $x\rightarrow +\infty$}
\label{sec_inf}

Let us first consider $n\ge 1/2$. For $x\rightarrow +\infty$, we find that $\theta_c\rightarrow \frac{\pi}{2}$. The deformed Bessel functions can be approximated by
\begin{equation}
I_{\gamma,0}(x)\sim \frac{1}{2\pi}\frac{1}{(n+1)^n}x^n\sqrt{\pi}\frac{\Gamma\left (\frac{1+n}{2}\right )}{\Gamma\left (\frac{2+n}{2}\right )},
\label{inf1}
\end{equation}
\begin{equation}
I_{\gamma,1}(x)\sim \frac{1}{2\pi}\frac{1}{(n+1)^n}x^n\sqrt{\pi}\frac{\Gamma\left (\frac{2+n}{2}\right )}{\Gamma\left (\frac{3+n}{2}\right )}.
\label{inf2}
\end{equation}
The normalization constant (\ref{mag6}) behaves like
\begin{equation}
A\sim M(n+1)^n \frac{1}{x^n}\frac{1}{\sqrt{\pi}}\frac{\Gamma\left (\frac{2+n}{2}\right )}{\Gamma\left (\frac{1+n}{2}\right )},
\label{inf3}
\end{equation}
and the density profile (\ref{cip4}) tends to the limit distribution
\begin{equation}
\rho(\theta)=\frac{M}{\sqrt{\pi}}\frac{\Gamma\left (\frac{2+n}{2}\right )}{\Gamma\left (\frac{1+n}{2}\right )}\cos^{n}\theta.
\label{inf4}
\end{equation}
Finally, the thermodynamical parameters tend to the limit values:
\begin{eqnarray}
\eta\rightarrow \eta_\infty\equiv\frac{1}{2}(2\pi)^{1/n}\frac{n+1}{(\sqrt{\pi})^{1/n}}\frac{\Gamma\left (\frac{1+n}{2}\right )^{\frac{n-1}{n}}\Gamma\left (\frac{3+n}{2}\right )}{\Gamma\left (\frac{2+n}{2}\right )^{\frac{2n-1}{n}}},\nonumber\\
\label{inf5}
\end{eqnarray}
\begin{equation}
b\rightarrow \frac{\Gamma\left (\frac{2+n}{2}\right )^{2}}{\Gamma\left (\frac{3+n}{2}\right )\Gamma\left (\frac{1+n}{2}\right )},
\label{inf6}
\end{equation}
\begin{equation}
\epsilon\rightarrow \epsilon_\infty\equiv -\frac{2n}{n+1}\frac{\Gamma\left (\frac{2+n}{2}\right )^{4}}{\Gamma\left (\frac{3+n}{2}\right )^2\Gamma\left (\frac{1+n}{2}\right )^2},
\label{inf7}
\end{equation}
\begin{equation}
\epsilon_{kin}=\Theta\rightarrow \frac{2}{n+1}\frac{\Gamma\left (\frac{2+n}{2}\right )^{4}}{\Gamma\left (\frac{3+n}{2}\right )^2\Gamma\left (\frac{1+n}{2}\right )^2}.
\label{inf8}
\end{equation}
This implies in particular that the polytropes $n\ge 1/2$ only exist for $\epsilon\ge \epsilon_{\infty}$ and $\eta\le \eta_{\infty}$. Beyond these points, there is no polytropic distribution anymore. In that case, the system is likely to  converge towards the Lynden-Bell distribution (see Sec. \ref{sec_spec}).

\subsection{Phase transitions}
\label{sec_pt}

We are now in a position to describe in detail the phase transitions that occur in the thermodynamical caloric curve $K(E)$ or, in dimensionless form $\eta(\epsilon)$, plotted in Figs. \ref{caloricNpos} and \ref{caloricNneg}. Note that this curve represents the full series of equilibria containing all critical points of entropy or free energy. We now have to determine which solutions correspond to fully stable (S)  solutions (global maxima of $S$ in MCE or global minima of $F$ in CE), metastable  (M) solutions (local maxima of $S$ in MCE or local minima of $F$ in CE) or unstable (U) solutions (saddle points of $S$ or $F$).
To make this selection, we can compare the equilibrium entropy or free energy of the solutions in competition (see Secs. \ref{sec_fe} and \ref{sec_ent}) or use the Poincar\'e theory of linear series of equilibria \cite{katz,ijmpb}.

\begin{figure}
\begin{center}
\includegraphics[clip,scale=0.3]{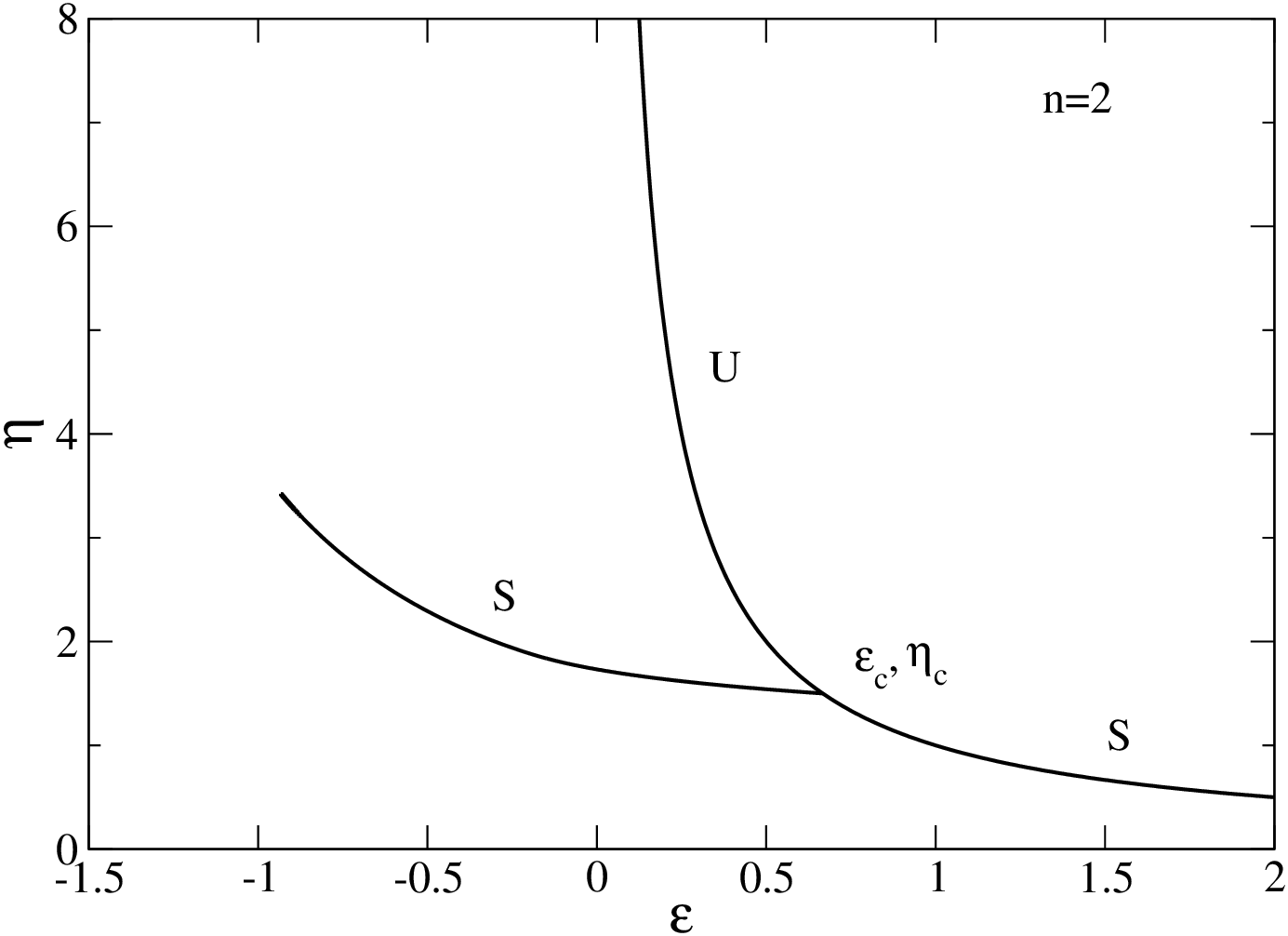}
\caption{For $n>n_{CTP}=1$ (specifically $n=2$), the caloric curve exhibits second order canonical and microcanonical phase transitions. The ensembles are equivalent.}
\label{caloN2}
\end{center}
\end{figure}

\begin{figure}
\begin{center}
\includegraphics[clip,scale=0.3]{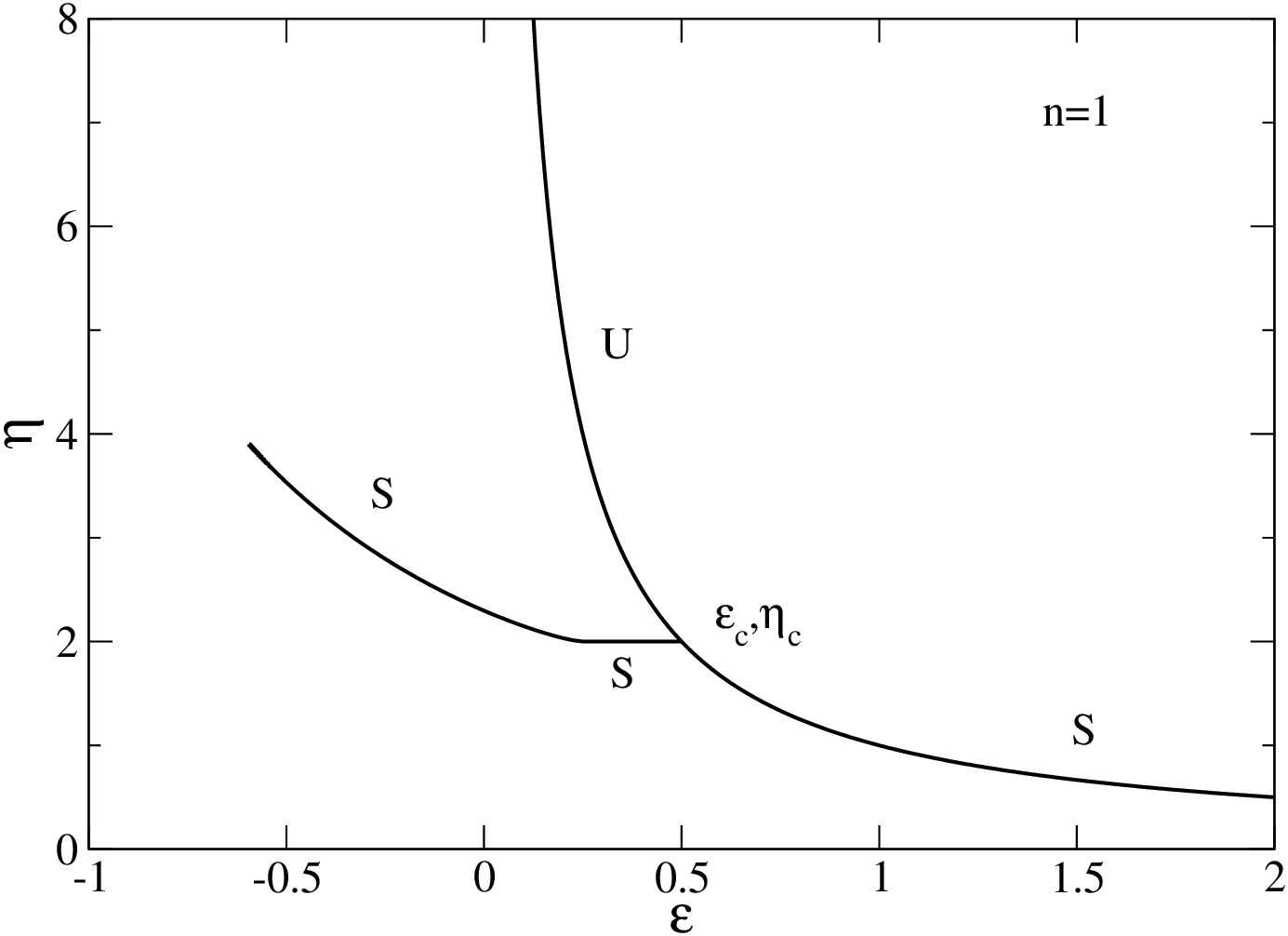}
\caption{For $n=n_{CTP}=1$, there is a degeneracy in the canonical ensemble for $\eta=\eta_c=2$. In the microcanonical ensemble, there is a region of infinite specific heats.}
\label{caloN1}
\end{center}
\end{figure}

\begin{figure}
\begin{center}
\includegraphics[clip,scale=0.3]{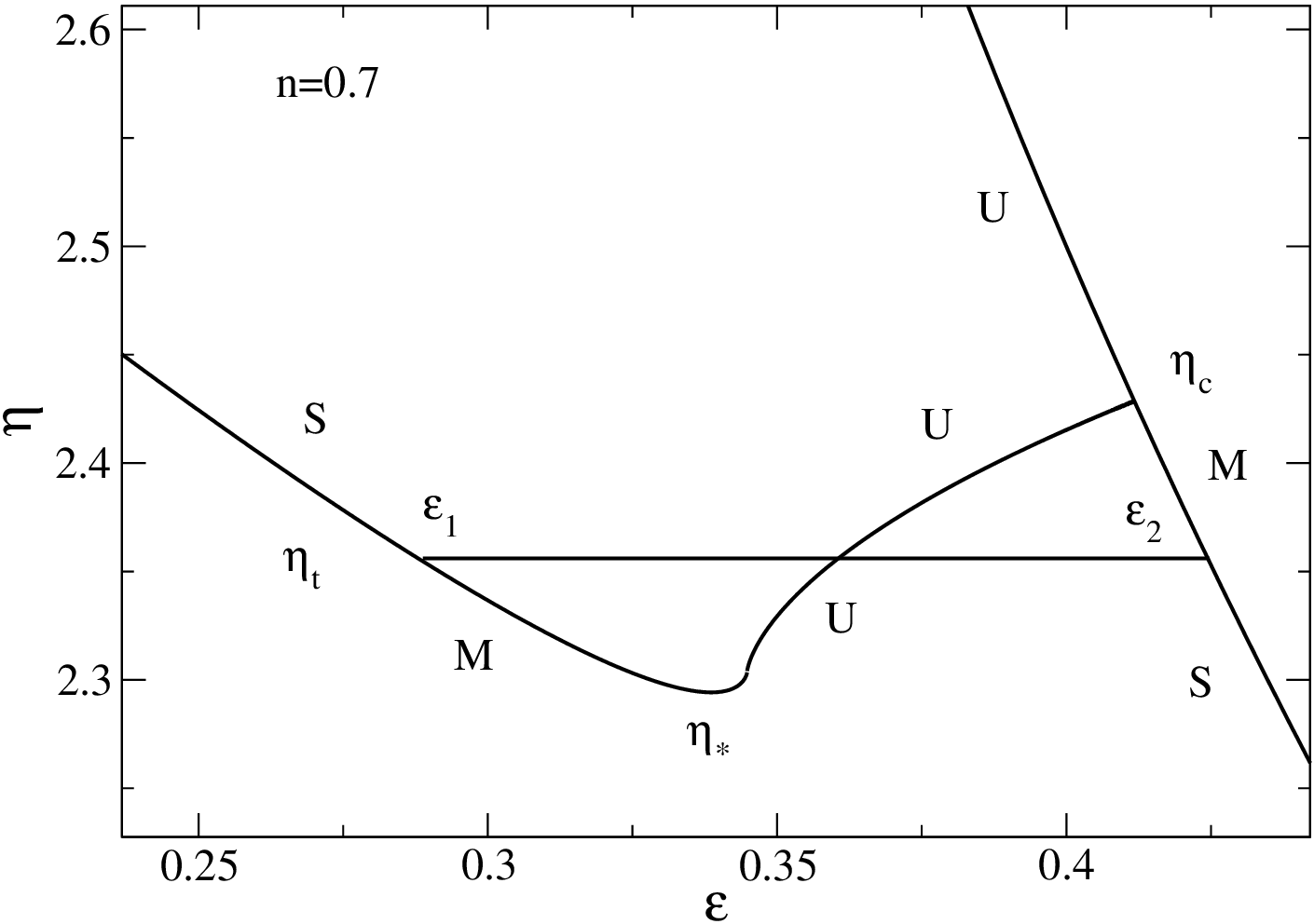}
\caption{Caloric curve in the canonical ensemble for $0.5\le n<n_{CTP}=1$ (specifically $n=0.7$). It exhibits  a canonical first order phase transition marked by the discontinuity of $\epsilon$ at $\eta=\eta_t$. The temperature of transition can be obtained by comparing the free energy of the homogeneous and inhomogeneous phases (see Fig. \ref{fn0.7}). The Maxwell construction does not directly apply in the present situation since we have used $1/K$ instead of $\beta$ in the caloric curve. The stability of the solutions (denoted S, M, and U) can be settled by using the Poincar\'e theorem \cite{katz,ijmpb}. The temperatures $\eta_*$ and $\eta_c$ are spinodal points marking the end of the metastable branches.}
\label{caloN0.7cano}
\end{center}
\end{figure}

\begin{figure}
\begin{center}
\includegraphics[clip,scale=0.3]{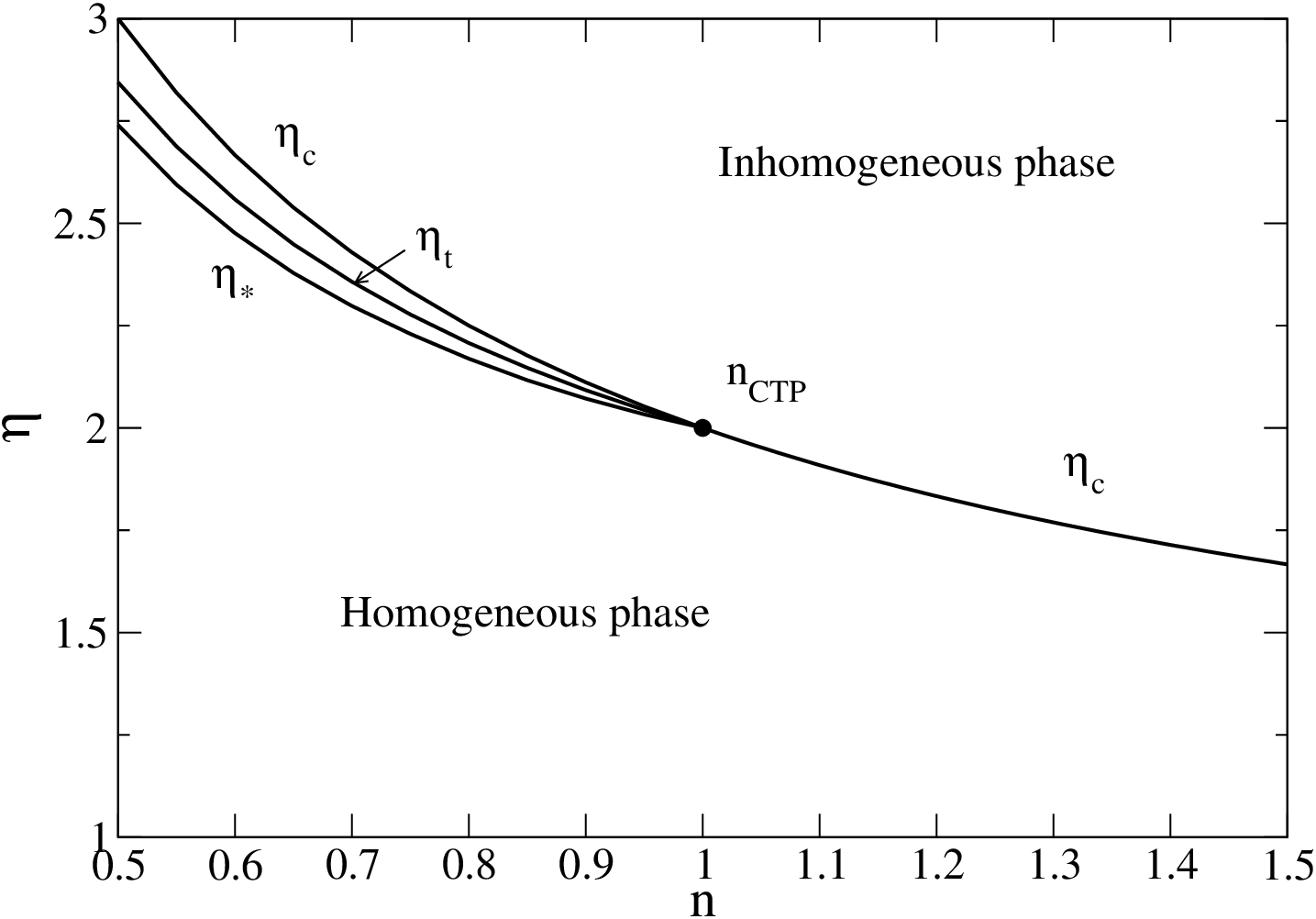}
\caption{Phase diagram in the canonical ensemble. There is a canonical 
tricritical point located at $n_{CTP}=1$ and  $\eta=2$ separating first and second order phase transitions.}
\label{phasecano}
\end{center}
\end{figure}

Let us first consider the canonical ensemble (CE) where the control
parameter is the inverse polytropic temperature $\eta$. For
$n>n_{CTP}=1$, the caloric curve $\epsilon(\eta)$ exhibits a second order
phase transition (see Figs. \ref{caloricNpos}, \ref{fn2} and
\ref{caloN2}) marked by the discontinuity of $d\epsilon/d\eta$ at
$\eta=\eta_c$. The homogeneous solution is fully stable for
$\eta<\eta_c$ and it becomes unstable for $\eta>\eta_c$. At that point
it is replaced by the inhomogeneous branch that is fully stable. The
series of equilibria presents one solution for $\eta<\eta_c$ (the
solution $b=0$ that is fully stable) and three solutions for
$\eta>\eta_c$ (two solutions $\pm b\neq 0$ that are fully stable and
one solution $b=0$ that is unstable); see
Fig. \ref{magnetisationNposTemp}. For $n=n_{CTP}=1$, there is a
degeneracy\footnote{This is a bit similar to critical polytropes
$n_c=3$ in astrophysics \cite{critical}.} at $\eta=\eta_c=2$ since
there exists an infinity of solutions $-1/2\le b\le 1/2$ with the same
free energy (see Figs. \ref{magnetisationNposTemp},
\ref{caloricNpos} and \ref{caloN1}). For $1/2\le n<n_{CTP}=1$, 
the caloric curve $\epsilon(\eta)$ exhibits a first order phase
transition between the homogeneous and inhomogeneous phases (see
Figs. \ref{caloricNpos},  \ref{fn0.7} and
\ref{caloN0.7cano}) marked by the discontinuity of $\epsilon$ at
$\eta=\eta_t$ and the existence of metastable branches. The series of
equilibria presents one solution for $\eta<\eta_*$ (the solution $b=0$
that is fully stable), five solutions for $\eta_*<\eta<\eta_c$ (two
solutions $\pm b\neq 0$ that are fully stable or metastable, two
solutions $\pm b'\neq 0$ that are unstable and one solution $b=0$ that
is fully stable or metastable) and three solutions for $\eta>\eta_c$
(two solutions $\pm b\neq 0$ that are fully stable and one solution
$b=0$ that is unstable); see
Fig. \ref{magnetisationNposTemp}. Therefore, $n_{CTP}=1$ and
$\eta_c=2$ (corresponding to $\epsilon_c=1/2$) is a canonical
tricritical point separating first and second order phase
transitions. The canonical phase diagram in the $(n,\eta)$ plane is
represented in Fig. \ref{phasecano}.

\begin{figure}
\begin{center}
\includegraphics[clip,scale=0.3]{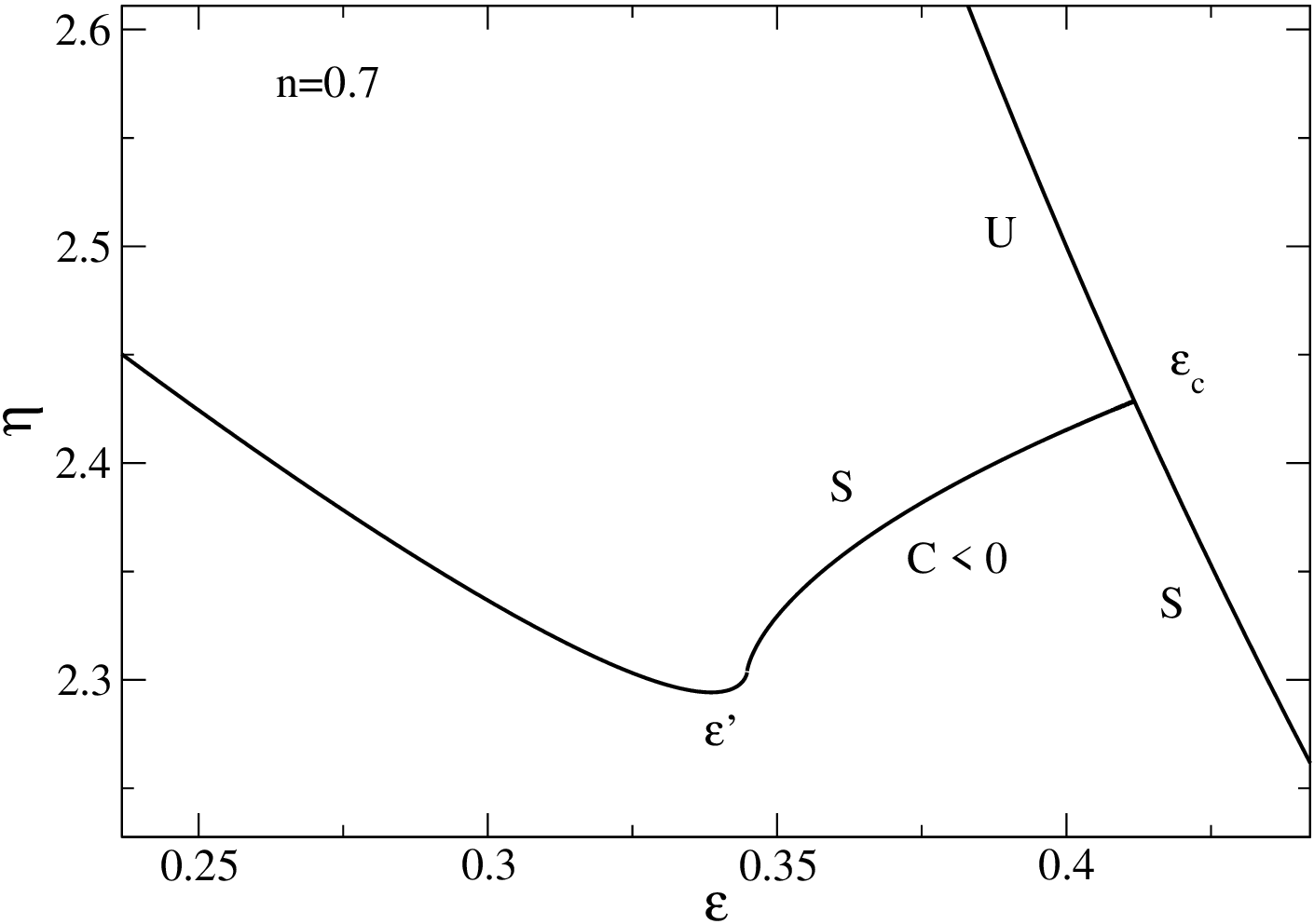}
\caption{Caloric curve in the microcanonical ensemble for $n_{MCP}<n<n_{CTP}=1$ (specifically $n=0.7$). The inhomogeneous states are fully stable. There is a region of negative specific heats between  $\epsilon'$ et $\epsilon_c$. In the canonical ensemble, this region is replaced by a phase transition.  }
\label{caloN0.7micro}
\end{center}
\end{figure}

\begin{figure}
\begin{center}
\includegraphics[clip,scale=0.3]{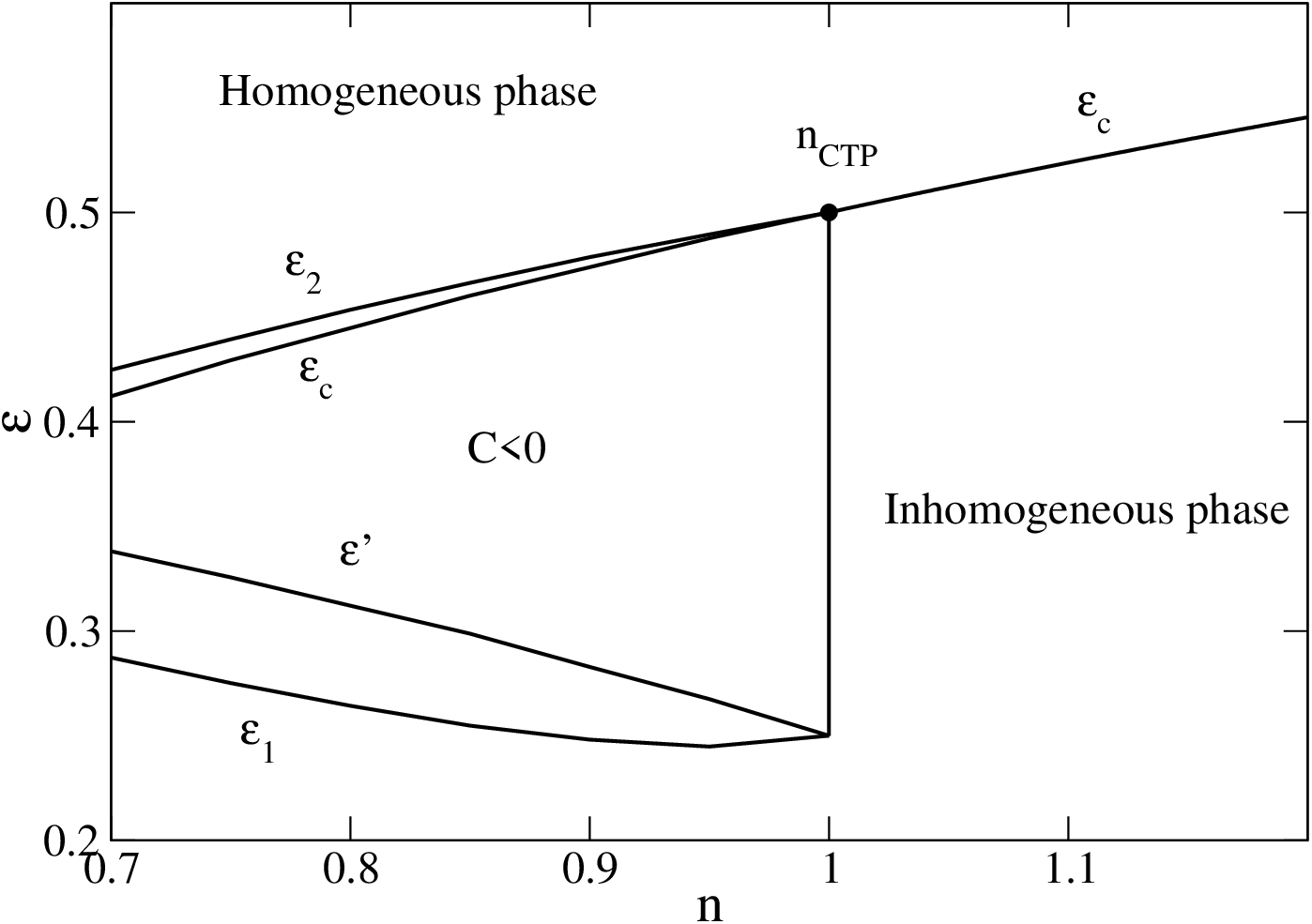}
\caption{Phase diagram in the microcanonical ensemble. In this range 
of indices, there is only a second order phase transition at energy
$\epsilon_c$. The states between $\epsilon'$ et $\epsilon_c$ have
negative specific heats. They are inaccessible in the canonical
ensemble (saddle points of free energy). If we consider only fully
stable states, the ensembles are inequivalent between $\epsilon_1$ and
$\epsilon_2$ (see Fig. \ref{caloN0.7cano}). If we take into account
canonical metastable states, the ensembles are inequivalent between
$\epsilon'$ and $\epsilon_c$.}
\label{phasemicro}
\end{center}
\end{figure}

\begin{figure}
\begin{center}
\includegraphics[clip,scale=0.3]{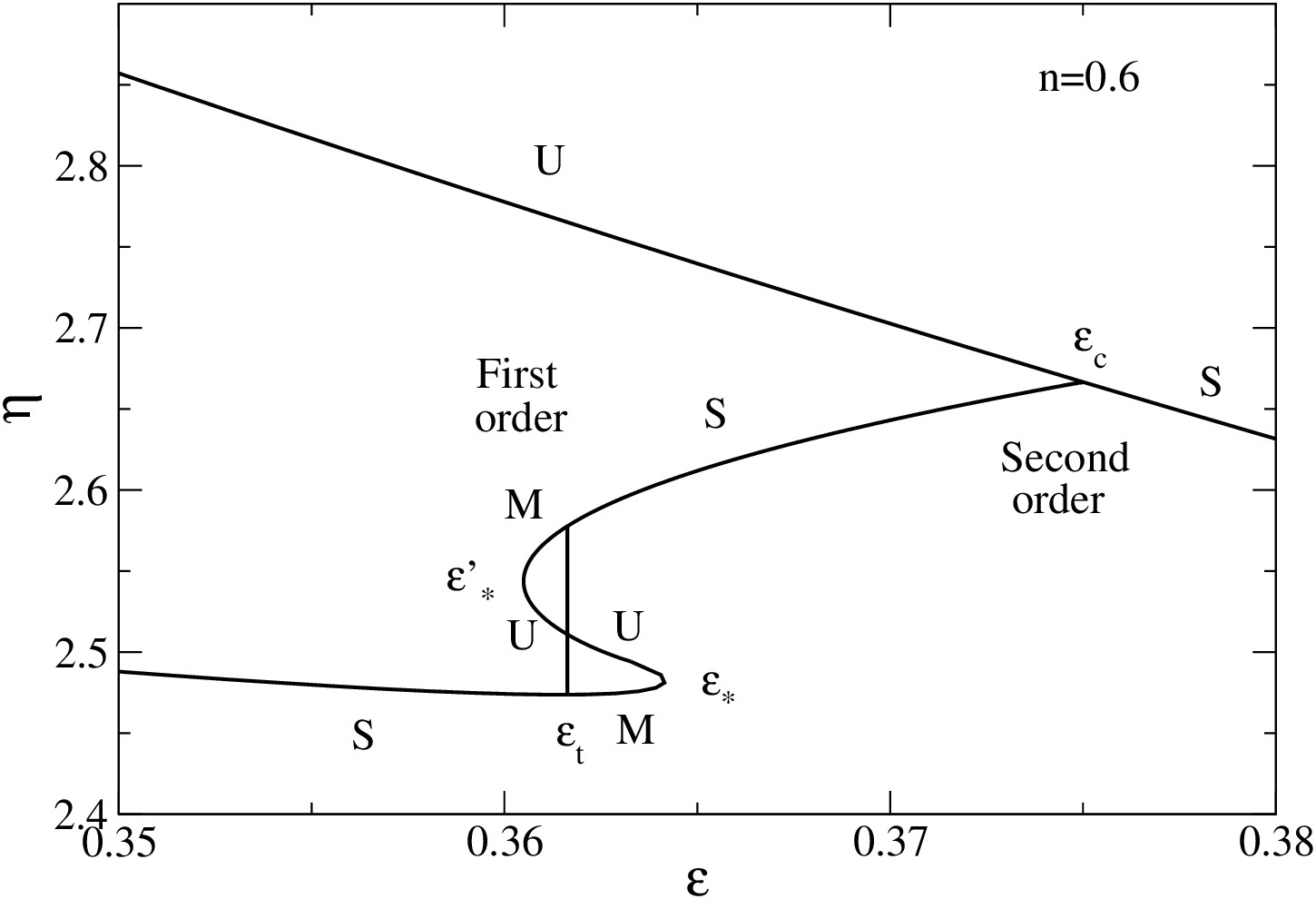}
\caption{Caloric curve in the microcanonical ensemble for $n_{MTP}<n<n_{MCP}$ (specifically $n=0.6$). Interestingly, there exists microcanonical second order {\it and} first order phase transitions on the same curve. The second order phase transition is marked by the discontinuity of $d\eta/d\epsilon$ at $\epsilon=\epsilon_c$. The first order phase transition marked by the discontinuity of $\eta$ at $\epsilon=\epsilon_t$.  The stability of the solutions (denoted S, M, and U) can be settled by using the Poincar\'e theorem \cite{katz,ijmpb}. The energies $\epsilon'_*$ and $\epsilon_*$ are spinodal points marking the end of the metastable branches.}
\label{caloN0.6}
\end{center}
\end{figure}

\begin{figure}
\begin{center}
\includegraphics[clip,scale=0.3]{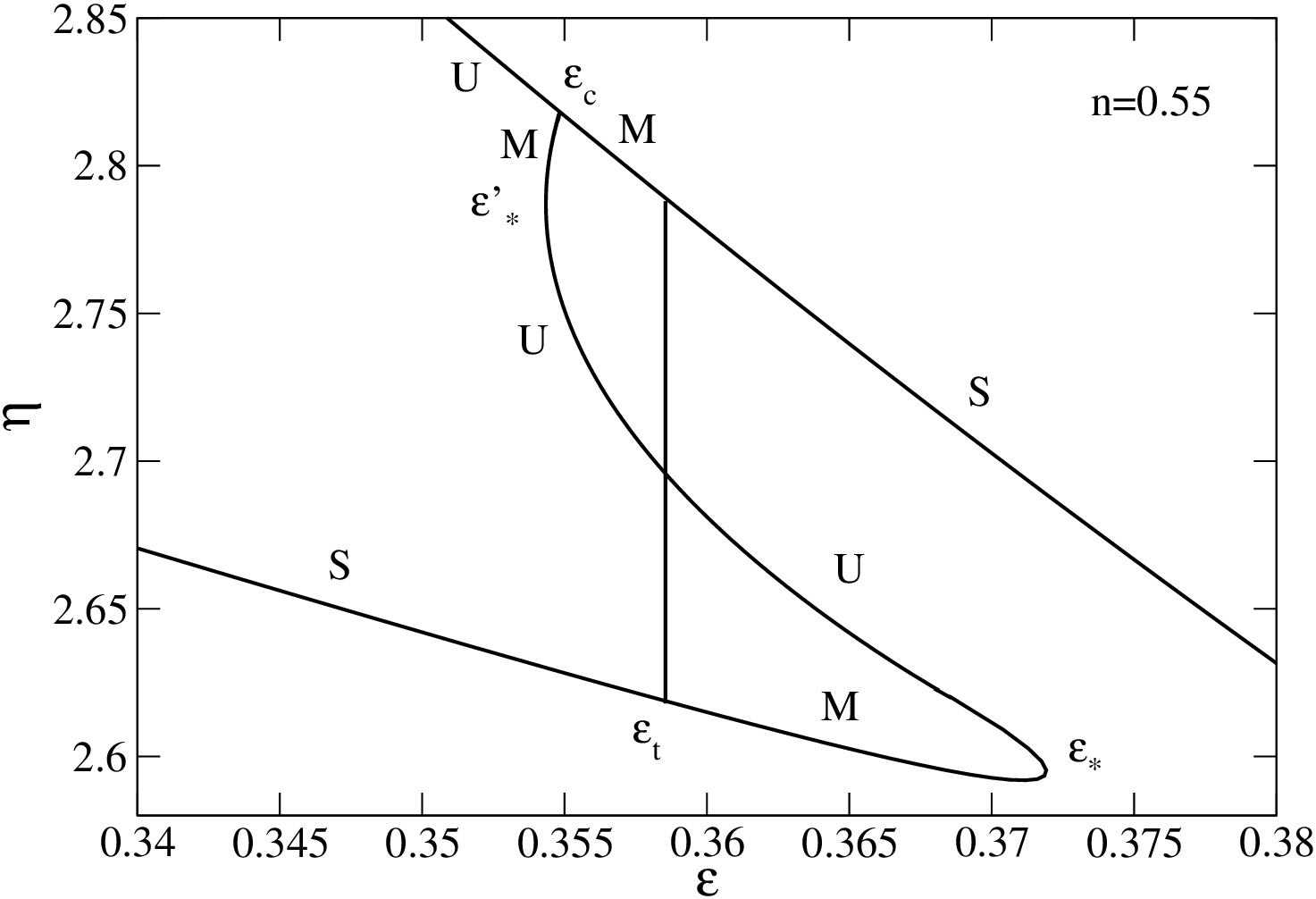}
\caption{Caloric curve in the microcanonical ensemble for $n_{*}<n<n_{MTP}$  (specifically $n=0.55$). If we only consider fully stable states, there is only a first order phase transition at $\epsilon=\epsilon_t$. If we consider the metastable states, there is a second order phase transition at $\epsilon=\epsilon_c$. The spinodal points marking the end of the metastable branches are $\epsilon'_*$ and $\epsilon_*$.}
\label{caloN0.55}
\end{center}
\end{figure}

\begin{figure}
\begin{center}
\includegraphics[clip,scale=0.3]{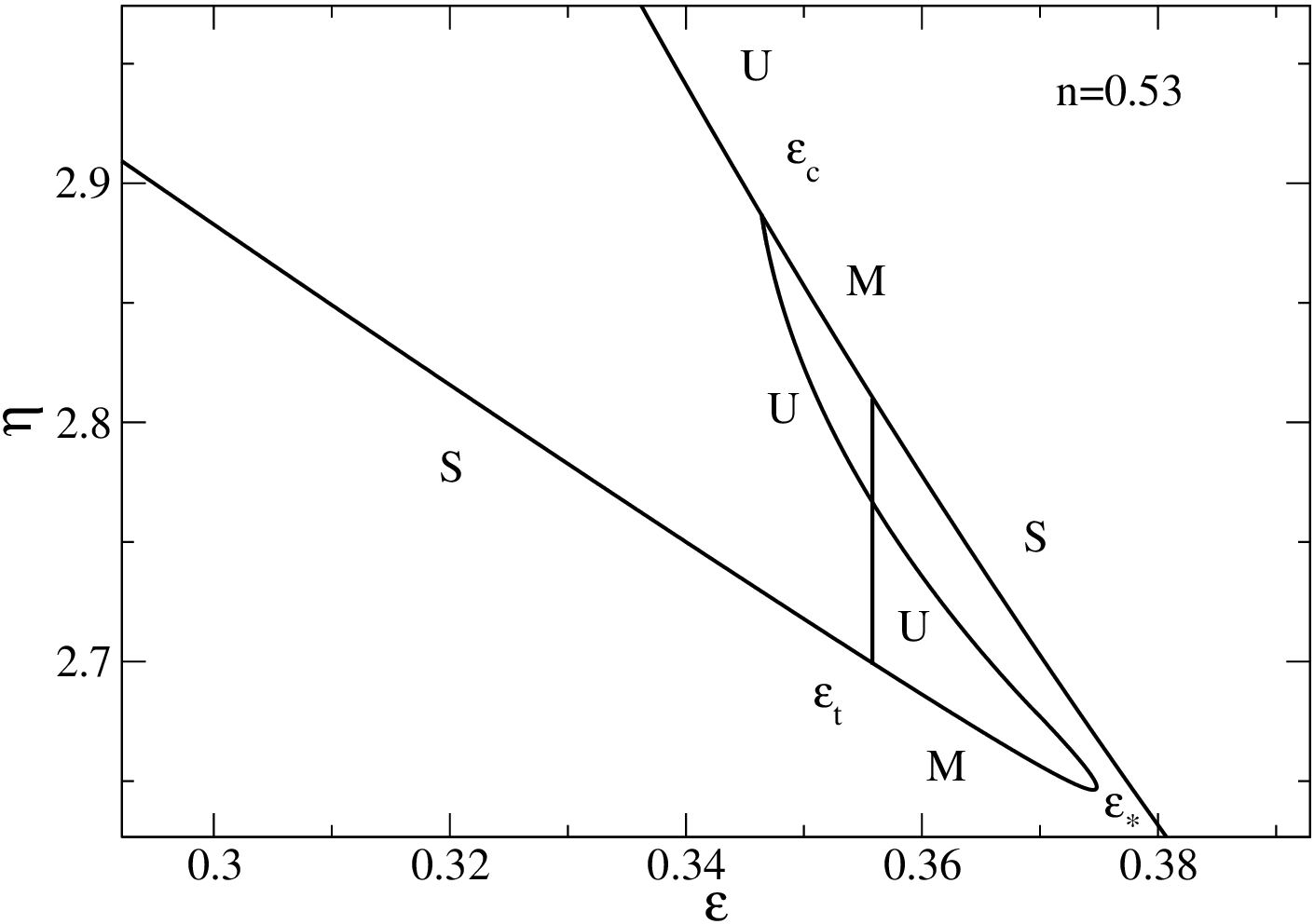}
\caption{Caloric curve in the microcanonical ensemble for $1/2\le n < n_{*}$  (specifically $n=0.53$). There is only a first order phase transition at $\epsilon=\epsilon_t$. The spinodal points marking the end of the metastable branches are $\epsilon_c$ and $\epsilon_*$. Note
that there exists a small region of negative specific heats between the
turning points of energy and temperature that is metastable.}
\label{CALORICn0.53}
\end{center}
\end{figure}

\begin{figure}
\begin{center}
\includegraphics[clip,scale=0.3]{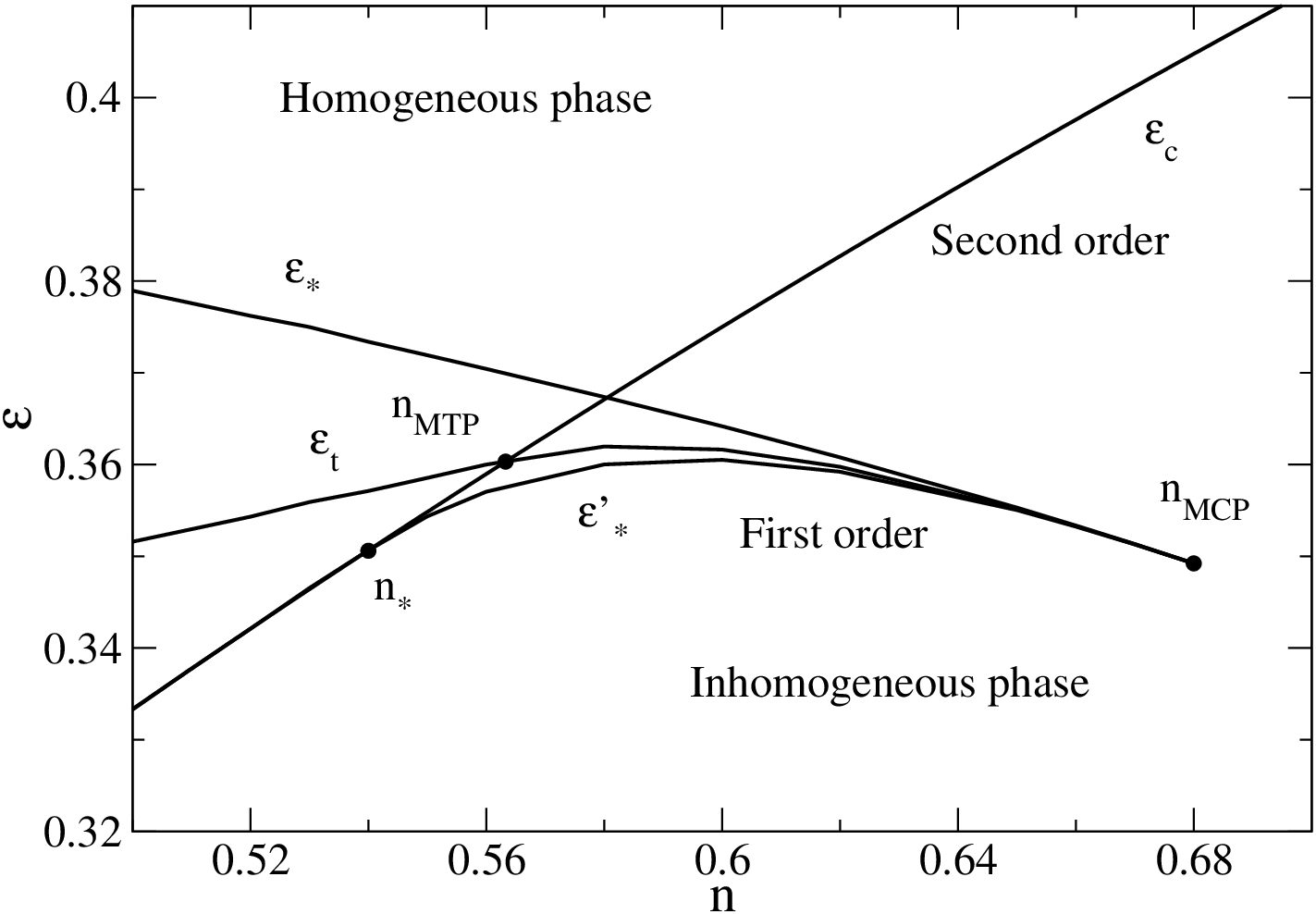}
\caption{Phase diagram in the microcanonical ensemble.  For $n>n_{MCP}$ ($n_{MCP}\simeq 0.68$, $\epsilon_{MCP}=0.3492$), there is only a second order phase transition. For $n_{MTP}<n<n_{MCP}$ ($n_{MTP}\simeq 0.563$, $\epsilon_{MTP}=0.36032$) there is a second order phase transition and a first order phase transition.  For $1/2\le n<n_{MTP}$, there is only a first order phase transition. For  $n_*<n<n_{MTP}$ ($n_*\simeq 0.54$, $\epsilon=0.3508$), the metastable branch presents a second order phase transition.}
\label{phasemicroADD}
\end{center}
\end{figure}

Let us now consider the microcanonical ensemble (MCE) where the
control parameter is the energy $\epsilon$. For $n>n_{MCP}\simeq
0.68$, the caloric curve $\eta(\epsilon)$ exhibits a second order
phase transition marked by the discontinuity of $d\eta/d\epsilon$ at
$\epsilon=\epsilon_c$ (see Figs. \ref{caloricNpos}, \ref{sn0.7}, 
\ref{caloN2}, \ref{caloN1} and \ref{caloN0.7micro}). The homogeneous solution 
is fully stable for $\epsilon>\epsilon_c$ and it becomes unstable for
$\epsilon<\epsilon_c$. At that point it is replaced by the
inhomogeneous branch that is fully stable. For $n>n_{CTP}=1$, the
specific heat is positive and the ensembles are equivalent (see
Fig. \ref{caloN2}). For $n_{MCP}<n<n_{CTP}=1$, there is a region of
negative specific heat in the microcanonical ensemble between
$\epsilon'$ and $\epsilon_c$ (see Fig. \ref{caloN0.7micro}) that is
replaced by a phase transition in the canonical ensemble (see
Fig. \ref{caloN0.7cano}). The series of equilibria presents one
solution for $\epsilon>\epsilon_c$ (the solution $b=0$ that is fully
stable) and three solutions for $\epsilon<\epsilon_c$ (two solutions
$\pm b\neq 0$ that are fully stable and one solution $b=0$ that is
unstable); see Fig. \ref{magnetisationNposEner}. The corresponding
microcanonical phase diagram in the $(n,\epsilon)$ plane is
represented in Fig. \ref{phasemicro}. For $n_{MTP}\simeq 0.563<
n<n_{MCP}\simeq 0.68$, there is a very interesting
situation\footnote{To our knowledge, this is the first example of that
kind.} in which the caloric curve exhibits a second order phase
transition at $\epsilon_c$ between the homogeneous phase and the
inhomogeneous phase (as before) and a first order phase transition at
$\epsilon_t$ between two inhomogeneous phases (see
Fig. \ref{caloN0.6}).  The first order phase transition is marked by
the discontinuity of $\eta$ at $\epsilon=\epsilon_t$ and the existence
of metastable branches.  For $n=0.6$, the series of equilibria
presents one solution for $\epsilon>\epsilon_c$ (the solution $b=0$
that is fully stable), three solutions for
$\epsilon_*<\epsilon<\epsilon_c$ (two solutions $\pm b\neq 0$ that are
fully stable and one solution $b=0$ that is unstable), seven solutions
for $\epsilon'_*<\epsilon<\epsilon_*$ (four solutions $\pm b\neq 0$
and $\pm b''\neq 0$ that are fully stable or metastable, two solutions
$\pm b'\neq 0$ that are unstable and one solution $b=0$ that is
unstable) and three solutions for $\epsilon<\epsilon'_*$ (two
solutions $\pm b\neq 0$ that are fully stable and one solution $b=0$
that is unstable). Therefore, $n_{MCP}\simeq 0.68$ and $\epsilon\simeq
0.3492$ (corresponding to $\eta=2.336$) is a microcanonical critical
point marking the appearance of the first order phase transition. At
$n=n_{MTP}\simeq 0.563$, the energies of the first and second order
phase transitions coincide ($\epsilon_t=\epsilon_c$) and, for $1/2\le
n<n_{MTP}\simeq 0.563$, there is only a first order phase transition
at $\epsilon=\epsilon_t$ between homogeneous and inhomogeneous
states. Therefore, $n_{MTP}\simeq 0.563$ and $\epsilon\simeq 0.3603$
(corresponding to $\eta=2.776$) is a microcanonical tricritical point
separating first and second order phase transitions. For $n_*\simeq
0.54<n<n_{MTP}\simeq 0.563$, there remains a sort of second order
phase transition at $\epsilon=\epsilon_c$ for the {\it metastable}
states (see Fig. \ref{caloN0.55}). Between $\epsilon'_*$ and
$\epsilon_c$, the specific heat is negative. For $n=n_*\simeq 0.54$,
the energies $\epsilon'_*$ and $\epsilon_c$ coincide and for $1/2\le
n<n_*$, the ``metastable'' second order phase transition
disappears. In that case, there is only a first order phase transition
(see Figs. \ref{sETOILEn0.53} and \ref{CALORICn0.53}). The series of
equilibria presents one solution for $\epsilon>\epsilon_*$ (the
solution $b=0$ that is fully stable), five solutions for
$\epsilon_c<\epsilon<\epsilon_*$ (two solutions $\pm b\neq 0$ that are
fully stable or metastable, two solutions $\pm b'\neq 0$ that are
unstable and one solution $b=0$ that is fully stable or metastable)
and three solutions for $\epsilon<\epsilon_c$ (two solutions $\pm
b\neq 0$ that are fully stable and one solution $b=0$ that is
unstable); see Fig. \ref{magnetisationNposEner}.  The microcanonical
phase diagram in the $(n,\epsilon)$ plane, summarizing all these
results, is represented in Fig. \ref{phasemicroADD}.

In conclusion, for $1/2\le n<n_{MTP}$, we have canonical and microcanonical first order phase transitions, for  $n_{MTP}<n<n_{MCP}$, we have canonical first order transitions, microcanonical first order transitions and microcanonical second order transitions, for $n_{MCP}<n<n_{CTP}=1$, we have canonical first order  phase transitions and microcanonical second order  phase transitions and for $n>n_{CTP}=1$, we have canonical and microcanonical second order phase transitions. The ensembles are equivalent for $n>n_{CTP}=1$ and inequivalent for $n<n_{CTP}=1$ in some range of energies. The microcanonical and canonical tricritical points do not coincide. Similar observations have been made for other models [62-65]. For $n<-1$,  we have canonical and microcanonical second order phase transitions and the ensembles are equivalent.

Using the thermodynamical analogy of Sec. \ref{sec_dyn}, we conclude that the polytropes denoted (S) or (M) that are maxima of pseudo entropy $S$ at fixed mass and energy are nonlinearly dynamically stable according to the ``microcanonical'' criterion (\ref{dyn2}). We cannot conclude that the polytropes denoted (U) are dynamically unstable since (\ref{dyn2}) provides just a sufficient condition of nonlinear dynamical stability.

\section{The physical caloric curve}
\label{sec_phy}

We have explained previously that the proper thermodynamical caloric
curve to consider when we study phase transitions and ensembles
inequivalence is the curve $T(E)$ where $T=1/\beta$ is the
thermodynamical temperature with $\beta=(\partial S/\partial
E)_M$. However, in practice, the caloric curve that is experimentally
or numerically accessible is the curve $\langle T_{kin}\rangle(E)$
that gives the averaged kinetic temperature as a function of the
energy $E$. This is the one, for example, that is represented in
Fig. 4 of Antoni \& Ruffo \cite{ar}. It is therefore useful to study
this curve specifically. We will call it the physical caloric
curve. In dimensionless form, it is given by
$\Theta(\epsilon)$. Several caloric curves $1/\Theta(\epsilon)$ have
been plotted in Fig. \ref{caloricNposKIN} for different indices $n\ge
1/2$. In order to understand their behavior, let us consider the
asymptotic limit $x\rightarrow 0$.

From expressions (\ref{ze5}) and (\ref{ze6}), we can draw the following conclusions. Let us introduce the critical indices
\begin{eqnarray}
\gamma_0=\frac{1+\sqrt{17}}{4}\simeq 1.2807764...
\label{phy1}
\end{eqnarray}
\begin{eqnarray}
n_0=\frac{4}{\sqrt{17}-3}\simeq 3.5615528...
\label{phy2}
\end{eqnarray}
Let us first consider the case $n\ge 1/2$ so that $1\le \gamma\le
3$. The quantity $2\gamma^2-5\gamma-2$ is positive for $\gamma_*<\gamma<3$,
i.e. $1/2\le n<n_*$, showing that the energy increases as $x$
increases.  The quantity $2\gamma^2-5\gamma-2$ is negative for $1\le
\gamma<\gamma_*$, i.e. $n>n_*$, showing that the energy decreases as
$x$ increases. On the other hand, the quantity $2\gamma^2-\gamma-2$ is
positive for $\gamma_0<\gamma\le 3$, i.e. $1/2\le n<n_0$, showing that
the inverse kinetic temperature decreases as $x$ increases.  The
quantity $2\gamma^2-\gamma-2$ is negative for $1\le
\gamma<\gamma_0$, i.e. $n>n_0$, showing that the inverse kinetic temperature increases as $x$ increases. This explains the
behavior of the physical caloric curve $1/\Theta(\epsilon)$ close to the
bifurcation point.  Close to that point, we have
\begin{eqnarray}
\Theta-\Theta_c=\frac{2\gamma^2-\gamma-2}{2\gamma^2-5\gamma-2}(\epsilon-\epsilon_c)+...
\label{phy3}
\end{eqnarray}
The physical specific heat is
\begin{eqnarray}
C_{kin}=\frac{dE}{d\langle T_{kin}\rangle}=\frac{M}{2}\frac{d\epsilon}{d\Theta}.
\label{phy4}
\end{eqnarray}
At  the bifurcation point, we have
\begin{eqnarray}
C_{kin}=\frac{M}{2}\frac{2\gamma^2-5\gamma-2}{2\gamma^2-\gamma-2}.
\label{phy5}
\end{eqnarray}
The physical specific heat close to the bifurcation point is positive for $1/2\le n<n_*$, negative for $n_*<n<n_0$ and positive again for $n>n_0$. It vanishes for $n=n_*$ and is infinite for $n=n_0$. Let us now consider the case $n<-1$ so that $0<\gamma\le 1$. In that case, the term $2\gamma^2-\gamma-2$ is always negative so that $1/\Theta$ increases as $x$ increases. On the other hand, the term $2\gamma^2-5\gamma-2$ is always negative so that $\epsilon$ decreases as $x$ increases. Therefore, the physical specific heat is always positive.

{\it Remark:} We must be careful that we cannot deduce any stability result from this curve. For example, the Poincar\'e theorem \cite{katz,ijmpb} and the general results on ensembles inequivalence \cite{ellis} apply to the thermodynamical caloric curve $T(E)$, not to the physical caloric curve $\langle T_{kin}\rangle(E)$. Furthermore, the notion of canonical ensemble is
only defined in terms of the variables $(E,T)$ via the Legendre transform $F=E-TS$, not in terms of the variables $(E,\langle T_{kin}\rangle)$. Note, in particular, that for $1<n<n_0$, the solutions close to the bifurcation point have negative physical  specific heat  $C_{kin}=dE/d\langle T_{kin}\rangle$ although they are stable both in canonical and microcanonical ensembles (they have positive specific heat $C=dE/dT$)\footnote{A similar result has been observed recently for the Lynden-Bell distribution (in preparation).}.

\section{The critical index $n_c=1$}
\label{sec_crit}

\subsection{Analytical expressions}
\label{sec_ae}

We have seen that the index $n_c=1$ (i.e. $\gamma_c=2$, $q_c=3$) is
particular because it corresponds to a canonical tricritical
point. Furthermore, we will see that for this particular index, the
algebra greatly simplifies and analytical results can be
obtained. This is because the relationship (\ref{cip1}) between the
density and the potential is linear\footnote{Note that a similar
linear relationship occurs in 2D turbulence between the vorticity and
the stream function for minimum enstrophy states. Now, the enstrophy
$\Gamma_2=\int
\omega^2\, d{\bf r}$, which is a quadratic functional, can be
interpreted as a particular Tsallis functional with $q=2$ (in our
notations). This precisely corresponds to the case $\gamma=2$
considered here (in 2D turbulence, we directly work in physical space
so that $q$ plays the role of $\gamma$). We may note that an
approximate linear $\omega-\psi$ relationship has been observed in a
plasma experiment \cite{hd,boghosian} and interpreted as a case of
incomplete violent relaxation \cite{brands}. It is interesting to note
that the same index $\gamma=2$ arises in the HMF model (see
Sec. \ref{sec_spec}). Note, however, that in many other situations of
2D turbulence, the $\omega-\psi$ relationship is not linear, so that
this relation is not universal and may lead to physical
inconsistencies \cite{brands}.}. For $n_c=1$, the density profile is
\begin{equation}
\rho(\theta)=A\left (1+\frac{x}{2}\cos\theta\right )_{+}.
\label{ae1}
\end{equation}
For $x<x_c=2$, the deformed Bessel functions take the simple form
\begin{equation}
I_{2,0}(x)=1, \qquad I_{2,1}(x)=\frac{x}{4}.
\label{ae2}
\end{equation}
Then, we get
\begin{equation}
A=\frac{M}{2\pi},\quad \eta=\eta_c=2,\quad b=\frac{x}{4},
\label{ae3}
\end{equation}
\begin{eqnarray}
\epsilon=\frac{1}{2}-\frac{x^2}{16},\quad \Theta=\frac{1}{2}+\frac{x^2}{16}.
\label{ae4}
\end{eqnarray}
We note in particular that the polytropic temperature is constant $\eta=\eta_c=2$ (i.e. $K_c={k}/{4}$ in dimensional form) so that there exists an infinity of solutions (the range of magnetizations $-1/2\le b\le 1/2$) with the same temperature. However, they have different energies (see Fig. \ref{caloricNpos}) ranging from $1/4$ to $1/2$. We will show analytically in Sec. \ref{sec_stab} that these solutions are stable in the microcanonical ensemble while they are degenerate in the canonical ensemble. In the range $1/4\le \epsilon\le 1/2$, the thermodynamical caloric curve $\eta=\eta_c$  has an infinite specific heat $C=dE/dT=\infty$.  The corresponding physical caloric curve is given by
\begin{eqnarray}
\Theta=1-\epsilon.
\label{ae5}
\end{eqnarray}
It has a  constant negative specific heat
\begin{eqnarray}
C_{kin}=\frac{M}{2}\frac{d\epsilon}{d\Theta}=-\frac{M}{2}.
\label{ae6}
\end{eqnarray}
This is an example where the thermodynamical caloric curve and the physical caloric curve give very different results.

For $x> x_c=2$, we have $\cos\theta_c=-2/x$ and the deformed Bessel functions take the form
\begin{equation}
I_{2,0}(x)=\frac{1}{2\pi}\left\lbrack \sqrt{x^2-4}+2\arccos\left (-\frac{2}{x}\right )\right\rbrack,
\label{ae7}
\end{equation}
\begin{equation}
I_{2,1}(x)=\frac{1}{2\pi}\left\lbrack \frac{1}{x}\sqrt{x^2-4}+\frac{x}{2}\arccos\left (-\frac{2}{x}\right )\right\rbrack.
\label{ae8}
\end{equation}
Then, we obtain
\begin{equation}
A=\frac{M}{\sqrt{x^2-4}+2\arccos\left (-\frac{2}{x}\right )},\quad \eta=\frac{x}{2I_{2,1}(x)},
\label{ae9}
\end{equation}
\begin{eqnarray}
b=\frac{I_{2,1}(x)}{I_{2,0}(x)},\quad \epsilon=-\frac{I_{2,1}(x)^2}{I_{2,0}(x)^2}
+\frac{2}{x}\frac{I_{2,1}(x)}{I_{2,0}(x)},
\label{ae10}
\end{eqnarray}
\begin{eqnarray}
\Theta=\frac{I_{2,1}(x)^2}{I_{2,0}(x)^2}
+\frac{2}{x}\frac{I_{2,1}(x)}{I_{2,0}(x)}.
\label{ae11}
\end{eqnarray}
For $x\rightarrow +\infty$, we have the equivalents $I_{2,0}(x)\sim \frac{x}{2\pi}$ and $I_{2,1}(x)\sim \frac{x}{8}$. Therefore, the normalization constant behaves like $A\sim \frac{M}{x}$ and the density tends to the limit distribution
\begin{equation}
\rho(\theta)=\frac{M}{2}\cos\theta,
\label{ae12}
\end{equation}
for $\theta\le \pi/2$. On the other hand, the thermodynamical parameters tend to
\begin{equation}
\eta\rightarrow 4,\quad b\rightarrow \frac{\pi}{4},\quad \epsilon\rightarrow -\frac{\pi^2}{16},\quad \Theta\rightarrow \frac{\pi^2}{16}.
\label{ae13}
\end{equation}

\subsection{Application to the case $M_0=1$}
\label{sec_spec}

We shall now apply our theory of polytropes  to interpret the physical caloric curve shown in Fig. 4 of Antoni \& Ruffo \cite{ar}. It is obtained from an initial waterbag distribution with magnetization $M_0=1$ and it has been confirmed by other groups \cite{latora,lrt}. Up to date, there is no explanation of the ``anomalies'' that take place close to the critical energy $U_c=3/4$. These anomalies are: (i) the bifurcation from the homogeneous branch to the inhomogeneous branch takes place at a lower value of the energy, and (ii) the physical caloric curve presents a region of negative specific heat $U'(T)<0$. It was noted in \cite{epjb,bbgky} that these anomalies cannot be explained in terms of Lynden-Bell's theory so that other approaches could be considered... We here argue that these anomalies can be explained naturally by assuming that the QSSs in that region are polytropic distributions with an index close to the critical one $n_c=1$ (i.e. $\gamma_c=2$ or $q_c=3$).

\begin{figure}
\begin{center}
\includegraphics[clip,scale=0.3]{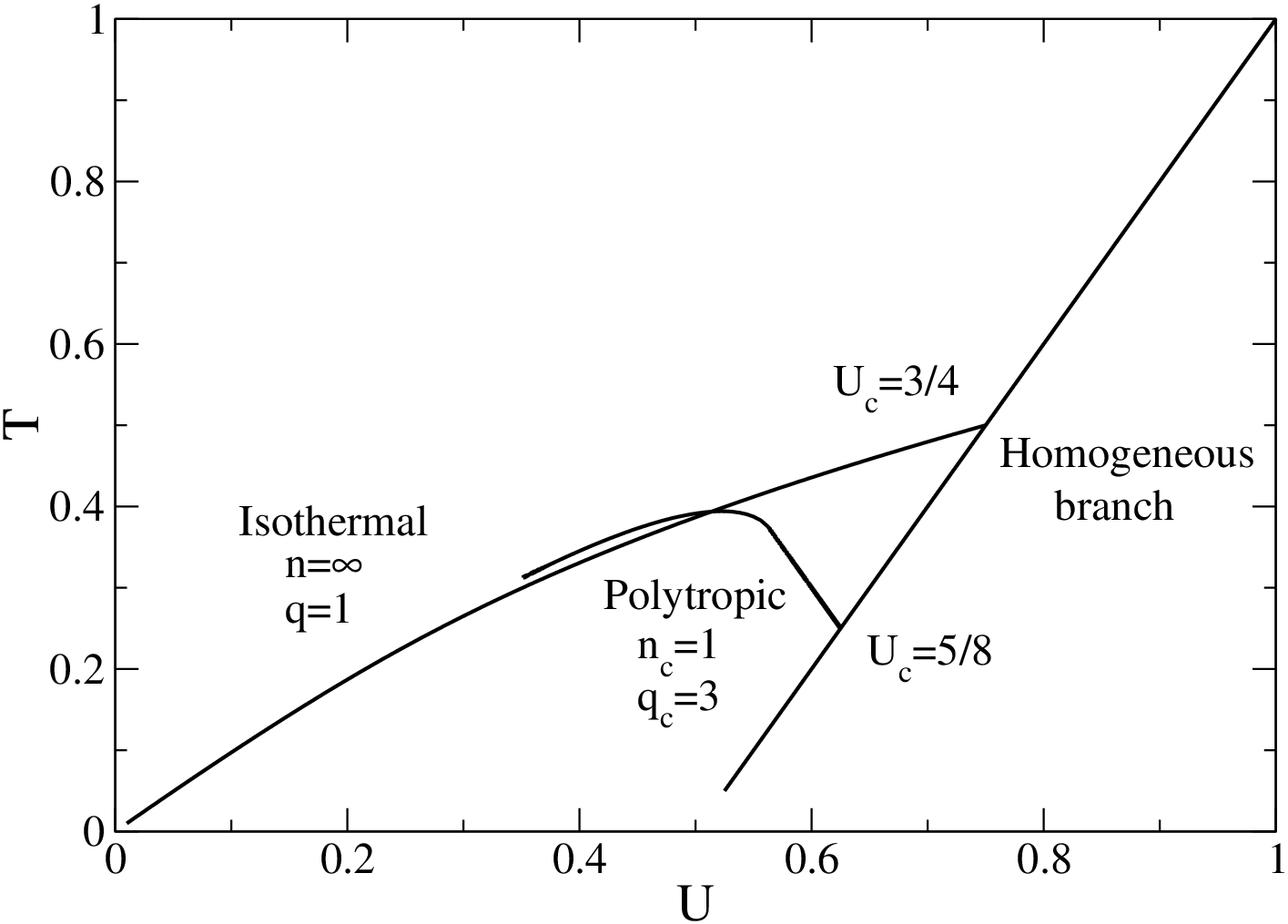}
\caption{Kinetic temperature $T$ as a function of the energy $U$ for $n=1$. Our theory of polytropes seems to explain the anomalies displayed on the numerical caloric curve of Antoni \& Ruffo \cite{ar} close to the critical energy, unlike the Lynden-Bell approach which leads to isothermal distributions.}
\label{chiDpetitt}
\end{center}
\end{figure}

We make the link between  the notations of Antoni \& Ruffo \cite{ar} and our notations \cite{cvb} by setting $U=\epsilon/4+1/2$ and $T=\Theta/2$, where $U$ is the energy and $T$ the kinetic temperature. This amounts to taking $k=2\pi/N$ and $M=N$ in the previous formulae and adding a constant  in the energy (ground state). For $n=1$, we find that the transition from the homogeneous states to the inhomogeneous states takes place for $\epsilon_c=1/2$, i.e. $U_c=5/8\simeq 0.625$. On the other hand, for indices $n_*<n<n_0$, the physical specific heat $U'(T)$ is negative. In particular, for $n=1$, we have $C_{kin}=\frac{dU}{dT}=\frac{1}{2}\frac{d\epsilon}{d\Theta}=-\frac{1}{2}$. This is at variance with Lynden-Bell's prediction. Note, however, that the branch of inhomogeneous polytropes exists only above the minimum energy $\epsilon=-\pi^2/16$, i.e. above $U=-{\pi^2}/{64}+{1}/{2}\simeq 0.345$ (corresponding to $T\rightarrow {\pi^2}/{32}\simeq 0.308$). Below this energy, we expect that the system reaches the Lynden-Bell distribution. The numerical caloric curve shows a good agreement with the Lynden-Bell prediction (equivalent to Boltzmann) in that region. However, this agreement should be ascertained by explicitly computing the distribution functions. Similarly, it should be checked whether the homogeneous states correspond to Lynden-Bell distributions (equivalent to Boltzmann distributions), polytropic distributions or something else.

In conclusion, our theory of polytropes qualitatively explains the ``anomalies'' displayed by the caloric curve of Antoni \& Ruffo \cite{ar}. As in previous works \cite{epjb,bbgky}, we justify these polytropic distributions  as a result of incomplete violent relaxation and lack of ergodicity (inefficient mixing). We stress, however, that polytropes are not  universal attractors in case of non-ergodicity and that other distributions could emerge. However, it turns out that polytropes play a special role. We have not tried to compare in more detail the theory and the observations in order to determine the best value of $q$ (it may not be exactly $q=3$ but it happens to  be close). This is left for a future study where we will also determine numerically the distribution functions to check whether they can be fitted by $q\simeq 3$ polytropes. The simulations of Campa {\it et al.} \cite{campa1}  go in that direction but, since there are differences with the simulations of Latora {\it et al.} \cite{lrt}, a detailed study must be performed. Nevertheless, our approach brings a plausible explanation of the numerical results of Antoni \& Ruffo \cite{ar} in terms of Tsallis distributions (polytropes). Furthermore, the apparently selected index $q_c=3$ is not completely arbitrary. First, it leads to distributions with a compact support, which is consistent  with the phenomenology of incomplete mixing. Secondly, it happens to coincide with (or be close to) the critical index that appears in the theory. It is interesting to note that this value was found numerically \cite{campa1} before the theory showed that it has a special meaning. All these results have to be confirmed.

\section{Stability analysis}
\label{sec_stab}

In the previous sections, we have determined critical points of entropy (or free energy) and we have plotted the corresponding series of equilibria. For some indices, we have found multiple equilibria for the same value of $E$ (or $K$)  and, in case of competition, we have compared the equilibrium entropy (or the equilibrium  free energy) of the different solutions in order to select the optimal one. However, a more precise analysis should determine whether the selected solutions really are entropy maxima (or free energy minima) by investigating the sign of the second order variations of $S$ (or $F$). This is a more complicated problem that will be solved only partially in the following sections.

\subsection{Variational principles: general theory}
\label{sec_vpgt}

\subsubsection{Microcanonical ensemble}
\label{sec_me}

Basically, we have to solve the maximization problem
\begin{eqnarray}
\label{me1}
\max_f\left\lbrace S\lbrack f\rbrack\, |\, E\lbrack f\rbrack=E, \, M\lbrack f\rbrack=M\right\rbrace,
\end{eqnarray}
where $S[f]$ is the Tsallis functional (\ref{ps1}). To solve this problem, we can proceed in two steps. {\it First step:} we first maximize $S$ at fixed $E$, $M$ {\it and} $\rho({\bf r})$. Since the specification of $\rho({\bf r})$ determines $M$ and $W$, this is equivalent to maximizing $S$ at fixed $E_{kin}$ and $\rho({\bf r})$. Writing the first variations as
\begin{eqnarray}
\label{me2}
\delta S-\beta\delta E_{kin}-\int\alpha({\bf r})\delta\left (\int f\, d{\bf v}\right )\, d{\bf r}=0,
\end{eqnarray}
we obtain
\begin{equation}
f({\bf r},{\bf v})=\left\lbrack \mu({\bf r})-\frac{(q-1)\beta}{q} \frac{v^2}{2}\right\rbrack_+^{1/(q-1)}. \label{me3}
\end{equation}
This critical point is the global maximum of entropy with the previous
constraints since $\delta^2S=-\frac{q}{2}\int f^{q-2}(\delta f)^2\,
d{\bf r}d{\bf v}\le 0$ (the constraints are linear in $f$ so their
second variations vanish). The function $\mu({\bf r})$ is determined
by the density $\rho=\int f\, d{\bf v}$. This leads to the expression
(\ref{vb19}). From this expression (\ref{vb19}), we can compute the
pressure $p=\frac{1}{d}\int fv^2\, d{\bf v}$ and we obtain the
polytropic equation of state (\ref{es5}). We can now express the
energy and the entropy as a function of $\rho$ and $p$ and we obtain
expressions (\ref{ef1}) and (\ref{ef2}). These expressions
are valid here out-of-equilibrium. {\it Second step:} we now have to
solve the variational problem
\begin{eqnarray}
\label{me4}
\max_\rho\left\lbrace S\lbrack \rho\rbrack\, |\, E\lbrack \rho\rbrack=E |\, M\lbrack \rho\rbrack=M\right\rbrace,
\end{eqnarray}
where $S$, $E$ and $M$ are given by Eqs. (\ref{ef1}), (\ref{ef2}) and (\ref{ps5}). {\it Conclusion:} the solution of (\ref{me1}) is given by Eq. (\ref{me3}) [or (\ref{vb19})] where $\rho$ is the solution of (\ref{me4}).  Therefore, we have reduced the study of the initial variational problem (\ref{me1}) for the distribution function $f({\bf r},{\bf v})$ to the study of a simpler variational problem (\ref{me4}) for the density $\rho({\bf r})$. Note that the two problems (\ref{me1}) and (\ref{me4}) are equivalent in the sense that the distribution function $f({\bf r},{\bf v})$ is a maximum of $S[f]$ at fixed $E$ and $M$ iff the corresponding density profile  $\rho({\bf r})$ is a maximum of $S[\rho]$ at fixed $E$ and $M$.

The variational problem (\ref{me4}) has been studied by Chavanis \& Sire \cite{lang} in arbitrary dimension of space. The first variations return the relation (\ref{pp1}). Then, this distribution is a local maximum of entropy at fixed mass and energy iff (see Appendix D of \cite{lang}):
\begin{eqnarray}
\label{me5}
\frac{\delta^2S}{\beta}=-\frac{1}{2}\gamma\int p\frac{(\delta\rho)^2}{\rho^2}\, d{\bf r}-\frac{1}{2}\int \delta\rho\delta\Phi\, d{\bf r}\nonumber\\
-\frac{2n}{d(2n-d)}\frac{1}{\int p\, d{\bf r}}\left\lbrack\int\left (\Phi+\frac{d}{2}\gamma\frac{p}{\rho}\right )\delta\rho\, d{\bf r}\right\rbrack^2\le 0,
\end{eqnarray}
for all perturbations $\delta\rho$ that do not change mass at first order (the conservation of the energy has been taken into account in obtaining Eq. (\ref{me5})).

\subsubsection{Canonical ensemble}
\label{sec_ce}

In the canonical ensemble, we have to solve the minimization problem
\begin{eqnarray}
\label{ce1}
\min_f\left\lbrace F\lbrack f\rbrack\, |\, M\lbrack f\rbrack=M\right\rbrace,
\end{eqnarray}
where $F=E-TS$ and $T=1/\beta$.  To solve this problem, we can proceed in two steps. {\it First step:} we first minimize $F$ at fixed $M$ {\it and} $\rho({\bf r})$. Since the specification of $\rho({\bf r})$ determines $M$, this is equivalent to minimizing $F$ at fixed $\rho({\bf r})$. Writing the first variations as
\begin{eqnarray}
\label{ce2}
\delta F+T\int\alpha({\bf r})\delta\left (\int f\, d{\bf v}\right )\, d{\bf r}=0,
\end{eqnarray}
we obtain
\begin{equation}
f({\bf r},{\bf v})=\left\lbrack \mu({\bf r})-\frac{(q-1)\beta}{q} \frac{v^2}{2}\right\rbrack_+^{1/(q-1)}.  \label{ce3}
\end{equation}
This critical point is the global minimum of free energy with the
previous constraint since $\delta^2F=\frac{q}{2}T\int f^{q-2}(\delta
f)^2\, d{\bf r}d{\bf v}\ge 0$ (the constraint is linear in $f$ so its
second variations vanish). The function $\mu({\bf r})$ is determined
by the density $\rho=\int f\, d{\bf v}$. This leads to the expression
(\ref{vb19}). From this expression (\ref{vb19}), we can compute the
pressure $p=\frac{1}{d}\int fv^2\, d{\bf v}$ and we obtain the
polytropic equation of state (\ref{es5}). We can now express the free
energy as a function of $\rho$ and $p$ and we obtain expression
(\ref{ef3}). These expressions are valid here out-of-equilibrium. {\it
Second step:} we now have to solve the variational problem
\begin{eqnarray}
\label{ce4}
\min_\rho\left\lbrace F\lbrack \rho\rbrack\, |\, M\lbrack \rho\rbrack=M\right\rbrace,
\end{eqnarray}
where $F$ and $M$ are given by Eqs. (\ref{ef3}) and (\ref{ps5}). {\it Conclusion:} the solution of (\ref{ce1}) is given by Eq. (\ref{ce3})  [or (\ref{vb19})] where $\rho$ is the solution of (\ref{ce4}).  Therefore, we have reduced the study of the initial variational problem (\ref{ce1}) for the distribution function $f({\bf r},{\bf v})$ to the study of a simpler variational problem (\ref{ce4}) for the density $\rho({\bf r})$. Note that the two problems (\ref{ce1}) and (\ref{ce4}) are equivalent in the sense that the distribution function $f({\bf r},{\bf v})$ is a minimum of $F[f]$ at fixed $M$  iff the corresponding density profile  $\rho({\bf r})$ is a minimum of $F[\rho]$ at fixed $M$.

The variational problem (\ref{ce1}) has been studied by Chavanis \& Sire \cite{lang} in arbitrary dimensions of space. The first variations immediately return the relation (\ref{pp1}). Then, this distribution is a local minimum of free energy at fixed mass iff
\begin{eqnarray}
\label{ce5}
{\delta^2F}=\frac{1}{2}\gamma\int p\frac{(\delta\rho)^2}{\rho^2}\, d{\bf r}+\frac{1}{2}\int \delta\rho\delta\Phi\, d{\bf r}\ge 0
\end{eqnarray}
for all perturbations $\delta\rho$ that do not change mass at first order.

{\it Remark 1:} from the stability criteria (\ref{me5}) and (\ref{ce5}), we clearly see that canonical stability implies microcanonical stability (but not the converse). Indeed, since the last term in Eq. (\ref{me5}) is negative, it is clear that if inequality (\ref{ce5}) is satisfied, then inequality (\ref{me5}) is automatically satisfied. In general, this is not reciprocal and we may have ensembles inequivalence. However, if we consider a spatially homogeneous system for which $\Phi$ is uniform, the last term in Eq. (\ref{me5}) vanishes (since the mass is conserved) and the stability criteria (\ref{me5}) and (\ref{ce5}) coincide. Therefore, for spatially homogeneous systems, we have ensembles equivalence.

{\it Remark 2:} the approach developed previously in the canonical
ensemble can be generalized to any functional of the form (\ref{dyn1})
\cite{assise}. The minimization problem (\ref{ce1}) is equivalent to
(\ref{ce4}) with the free energy $F[\rho]$ given by
\begin{eqnarray}
F=\frac{1}{2}\int \rho\Phi\, d{\bf r}+\int\rho\int^\rho \frac{p(\rho')}{\rho^{'2}}\, d\rho'd{\bf r},
\label{ce6}
\end{eqnarray}
where $p(\rho)$ is the equation of state associated to $C(f)$. A critical point of (\ref{ce6}) satisfies the condition of hydrostatic balance
\begin{eqnarray}
\nabla p=-\rho\nabla\Phi.
\label{ce7}
\end{eqnarray}
Then, this critical point is a local minimum of free energy at fixed mass iff
\begin{eqnarray}
\label{ce8}
{\delta^2F}=\int \frac{p'(\rho)}{2\rho}(\delta\rho)^2\, d{\bf r}+\frac{1}{2}\int \delta\rho\delta\Phi\, d{\bf r}\ge 0
\end{eqnarray}
for all perturbations $\delta\rho$ that do not change mass at first order. For the polytropic equation of state (\ref{es5}), this returns the previous results.

\subsection{Application to the HMF model}
\label{sec_az}

Let us now apply the preceding results to the HMF model. We start by the canonical ensemble which is simpler in a first step.

\subsubsection{Canonical ensemble}
\label{sec_azc}

According to the results of Sec. \ref{sec_ce}, we have to solve
\begin{eqnarray}
\label{azc1}
\min_\rho\left\lbrace F\lbrack \rho\rbrack\, |\, M\lbrack \rho\rbrack=M\right\rbrace,
\end{eqnarray}
with
\begin{eqnarray}
\label{azc2}
F[\rho]=-\frac{\pi B^2}{k}+\frac{K}{\gamma-1}\int \rho^\gamma\, d\theta.
\end{eqnarray}
To solve this problem, we can again proceed in two steps. We first minimize $F[\rho]$ at fixed $M$ {\it and} $B$. This gives
\begin{eqnarray}
\label{azc3}
\rho_1(\theta)=A\left\lbrack 1+\frac{\gamma-1}{\gamma}\lambda\cos\theta\right\rbrack_{+}^{\frac{1}{\gamma-1}},
\end{eqnarray}
where the Lagrange multipliers $A$ and $\lambda$ are determined by the constraints $M$ and $B$ through
\begin{equation}
A=\frac{M}{2\pi I_{\gamma,0}(\lambda)},
\label{azc4}
\end{equation}
and
\begin{equation}
\frac{2\pi B}{kM}=\frac{I_{\gamma,1}(\lambda)}{I_{\gamma,0}(\lambda)}.
\label{azc5}
\end{equation}
This is the global minimum  of $F$ with the previous constraints since $\delta^2F\ge 0$ (this can be deduced from Eq. (\ref{ce5}) by taking $\delta\Phi=0$ since $B$ is fixed). Then, we can express the free energy $F$ as a function of $B$ by writing $F(B)\equiv F[\rho_1]$. This gives
\begin{eqnarray}
F(B)=-\frac{\pi B^2}{k}+\frac{K}{\gamma-1}\frac{M^\gamma}{\lbrack 2\pi I_{\gamma,0}(\lambda)\rbrack^{\gamma-1}}\nonumber\\
\times\left ( 1+\frac{\gamma-1}{\gamma} \lambda \frac{2\pi B}{kM}\right )_+,
\label{azc6}
\end{eqnarray}
where $\lambda$ is related to $B$ by Eq. (\ref{azc5}). Finally,  (\ref{azc1}) is equivalent to
\begin{eqnarray}
\label{azc7}
\min_B\left\lbrace F(B)\right\rbrace,
\end{eqnarray}
in the sense that the solution of (\ref{azc1}) is given by Eqs. (\ref{azc3})-(\ref{azc5}) where $B$ is the solution of (\ref{azc7}). For given $K$ and $M$ (canonical ensemble), we just have to determine the minimum of a {\it function} $F(B)$ of the magnetization $B$. The critical points $F'(B)=0$ should return the condition  (\ref{mag9}) of equilibrium. Furthermore, the  minima correspond to $F''(B)>0$. For isothermal distributions, the equations take a simple form and we can show \cite{prepa} through simple graphical constructions (by studying the slopes of simple curves) that they return the well-known stability results. For polytropic distributions, it seems more difficult to study (\ref{azc7}) at a general level.

Note that in terms of $\lambda$, related to the magnetization by (\ref{azc5}), the normalized free energy (\ref{azc6}) can be written
\begin{eqnarray}
\label{azc8}
f(\lambda)=-\left \lbrack\frac{I_{\gamma,1}(\lambda)}{I_{\gamma,0}(\lambda)}\right \rbrack^2
+\frac{1}{\gamma-1}\frac{1}{\eta}\frac{1}{I_{\gamma,0}(\lambda)^{\gamma-1}}\nonumber\\
\times\left\lbrack 1+\frac{\gamma-1}{\gamma} \lambda \frac{I_{\gamma,1}(\lambda)}{I_{\gamma,0}(\lambda)}\right\rbrack_+.
\end{eqnarray}
Therefore, for each value of the polytropic temperature $\eta$, it suffices
to study this function and determine its minima.
Of course, this can be done easily for
each $\eta$ but it seems difficult to find a general criterion of stability. It
seems even difficult to check (in the non-isothermal case) that
$f'(\lambda)=0$ leads to $\lambda=x$ where $x$ is given by Eq. (\ref{mag9}).

However, some analytical results can be obtained for the critical index $n=1$ ($\gamma=2$). In that case, using Eq. (\ref{ae2}) for $\lambda\le 2$,  the expression (\ref{azc8}) of the free energy becomes
\begin{eqnarray}
\label{azc9}
f(\lambda)=-\frac{\lambda^2}{16}+\frac{1}{\eta}\left (1+\frac{\lambda^2}{8}\right ).
\end{eqnarray}
Its derivative is
\begin{eqnarray}
\label{azc10}
f'(\lambda)=-\frac{\lambda}{8}+\frac{\lambda}{4\eta}.
\end{eqnarray}
For $\eta\neq \eta_c=2$, the only critical point $f'(\lambda)=0$ is $\lambda=0$ (homogeneous phase). This is a minimum for $\eta<\eta_c$ ($f''(0)>0$)  and a maximum for $\eta>\eta_c$ ($f''(0)<0$). If we now consider the critical temperature $\eta=\eta_c=2$, we get
\begin{eqnarray}
\label{azc11}
f(\lambda)=\frac{1}{2}.
\end{eqnarray}
Therefore,  the solutions $0\le \lambda\le 2$ have the same free energy and the inhomogeneous branch is degenerate.

\subsubsection{Microcanonical ensemble}
\label{sec_azm}

Let us now consider the microcanonical ensemble. According to the results of Sec. \ref{sec_me}, we have to solve
\begin{eqnarray}
\label{azm1}
\max_\rho\left\lbrace S\lbrack \rho\rbrack\, | \, E\lbrack \rho\rbrack=E, \, M\lbrack \rho\rbrack=M\right\rbrace,
\end{eqnarray}
with
\begin{eqnarray}
\label{azm2}
S[\rho]=-\left (n-\frac{1}{2}\right )\beta\int p\, d\theta,
\end{eqnarray}
\begin{eqnarray}
\label{azm3}
E[\rho]=\frac{1}{2}\int p\, d\theta-\frac{\pi B^2}{k}.
\end{eqnarray}
To solve this problem, we can again proceed in two steps. We first maximize $S[\rho]$ at fixed $E$, $M$ {\it and} $B$. This gives the optimal  distribution (\ref{azc3}) as in the canonical ensemble. This is the global maximum of $S$ with the previous constraints since $\delta^2S\le 0$ (this can be deduced from Eq. (\ref{me5}) by taking $\delta\Phi=0$ since $B$ is fixed). Then, we can express the entropy $S$ and the energy $E$  as a function of $B$ by writing  $S\equiv S[\rho_1]$ and $E=E[\rho_1]$. This gives
\begin{eqnarray}
E=\frac{KM^\gamma}{2[2\pi I_{\gamma,0}(\lambda)]^{\gamma-1}}\left ( 1+\frac{\gamma-1}{\gamma} \lambda \frac{2\pi B}{kM}\right )_+-\frac{\pi B^2}{k},\nonumber\\
\label{azm4}
\end{eqnarray}
and
\begin{eqnarray}
S=-\frac{2n-1}{C_n}\frac{1}{K^{\frac{2n}{2n-1}}}\left (E+ \frac{\pi B^2}{k}\right ),
\label{azm5}
\end{eqnarray}
where $\lambda$ is related to $B$ by Eq. (\ref{azc5}). The first equation determines $K$ as a function of $B$, $E$ and $M$.  Then, $S$ becomes a function of $B$, $E$ and $M$. Finally, (\ref{azm1}) is equivalent to
\begin{eqnarray}
\label{azm6}
\max_B\left\lbrace S(B)\right\rbrace,
\end{eqnarray}
in the sense that the solution of (\ref{azm1}) is given by Eqs. (\ref{azc3})-(\ref{azc5}) where $B$ is the solution of (\ref{azm6}).
For given $E$ and  $M$, we just have to determine the maximum of a {\it function} $S(B)$ of the magnetization $B$. The critical points $S'(B)=0$ should return the condition  (\ref{in14}) of equilibrium. Furthermore, the  maxima correspond to $S''(B)<0$. Unfortunately, it seems difficult to study the maximization problem (\ref{azm6}) at a general level, except in the isothermal case $\gamma=1$ \cite{prepa}.

Note that in terms of $\lambda$, related to the magnetization by (\ref{azc5}), the normalized entropy can be written
\begin{eqnarray}
\label{azm7}
s(\lambda)=-\left (n-\frac{1}{2}\right )\frac{\left (1+\frac{\gamma-1}{\gamma}\lambda \frac{I_{\gamma,1}(\lambda)}{I_{\gamma,0}(\lambda)}\right )_+^{\frac{2n}{2n-1}}}{(\epsilon+2 \frac{I_{\gamma,1}(\lambda)^2}{I_{\gamma,0}(\lambda)^2})^{\frac{1}{2n-1}}I_{\gamma,0}(\lambda)^{\frac{2}{2n-1}}}.
\nonumber\\
\end{eqnarray}
Therefore, for each value of the energy $\epsilon$ it suffices
to study this function and determine its maxima. Of course, this can be done easily for
each $\epsilon$ but it seems difficult to find a general criterion of stability. It
seems even difficult to check (in the non-isothermal case) that
$s'(\lambda)=0$ leads to $\lambda=x$ where $x$ is given by Eq. (\ref{in14}).

However, some analytical results can be obtained for the critical index $n=1$ ($\gamma=2$). In that case, using Eq. (\ref{ae2}) for $\lambda\le 2$, the expression (\ref{azm7}) of the entropy becomes
\begin{eqnarray}
\label{azm8}
s(\lambda)=-\frac{1}{2}\frac{\left (1+\frac{\lambda^2}{8}\right )^{2}}{\epsilon+\frac{\lambda^2}{8}}.
\end{eqnarray}
We easily obtain
\begin{eqnarray}
\label{azm9}
s'(\lambda)=-\frac{\lambda\left (1+\frac{\lambda^2}{8}\right )\left (2\epsilon-1+\frac{\lambda^2}{8}\right )}{8\left (\epsilon+\frac{\lambda^2}{8}\right )^2}.
\end{eqnarray}
The critical points $s'(\lambda)=0$  are $\lambda=0$, corresponding to the homogeneous state, and $\lambda^2=x^2=8(1-2\epsilon)$, corresponding to the inhomogeneous solutions (\ref{ae4}). Then, we find that
\begin{eqnarray}
\label{azm10}
s''(x)=-\frac{1-2\epsilon}{2(1-\epsilon)}.
\end{eqnarray}
For $x\le x_c=2$, $\epsilon$ goes from $1/4$ to $1/2$. For these values, $s''(x)<0$ implying that  the inhomogeneous solutions $\lambda=x$ are entropy maxima (stable). On the other hand, for $\lambda=0$, we have
\begin{eqnarray}
\label{azm11}
s''(0)=\frac{1-2\epsilon}{8\epsilon^2}.
\end{eqnarray}
For $1/4\le \epsilon\le 1/2$, we find $s''(0)>0$ implying that the homogeneous solution $\lambda=0$ is an entropy minimum   (unstable).

\section{Conclusion}
\label{sec_conclusion}

The nature of quasi stationary states (QSS) in Hamiltonian systems
with long-range interactions has created a lively debate in the
statistical mechanics community \cite{assisebook}. Some researchers
have argued that these QSSs could not be explained in terms of
ordinary statistical mechanics and that it was necessary to use a new
thermodynamics (Tsallis generalized thermodynamics)
\cite{epnewsbis,tsallisbook}. Other researchers have argued, on the
contrary, that Tsallis thermodynamics was unsuccessful to explain QSSs
\cite{epnews,cdr}. This led to a violent polemic. We have since the
start adopted an intermediate position
\cite{chavhouches,incomplete,epjb,bbgky,campa2}. In principle, the
QSSs can be explained in terms of Lynden-Bell's theory, i.e. ordinary
thermodynamics adapted to the Vlasov equation. However, this approach,
as usual, {\it assumes ergodicity and efficient mixing}. There are
cases where the system mixes well so that Lynden-Bell's prediction
works well \cite{precommun}. However, there are known situations were
mixing is not efficient enough so that other distributions emerge
\cite{lrt,campa1}. This is always the case in astrophysics
(Lynden-Bell's theory fails to describe galaxies) and in some
situations of 2D turbulence [28-30]\footnote{In the
plasma experiment, although Lynden-Bell's prediction fails to describe
the details of the distribution, it provides however a {\it fair}
first order prediction \cite{brands}. The success or failure of
Lynden-Bell's prediction can be more or less severe depending on the
initial conditions and it is not possible to draw a general conclusion
(see the analogy with the meandering course of the Mississippi River
in \cite{bt}).}. Since there is no rigorous theory of non-ergodic
(partially mixed) systems, we are open to new suggestions including
the one by Tsallis.

The idea of Tsallis generalized thermodynamics, as we understand it,
is to account for non ergodicity and incomplete mixing. It is based on
the postulate that, in case of incomplete mixing, the system still
maximizes an entropy but this is not the Lynden-Bell entropy.  The
Lynden-Bell entropy is obtained from a combinatorial analysis assuming
that all microstates are equiprobable. This is the main postulate of
statistical mechanics.  Tsallis generalized thermodynamics rejects
this equiprobability postulate and argues that complex systems
``prefer'' some regions of phase space rather than others. Then,
Tsallis form of entropy is introduced as a postulate or justified from
an axiomatic basis. We think that some works remains to be done in
this area to determine to which situations of incomplete mixing
Tsallis thermodynamics applies. Tsallis entropy certainly does not
describe all types of incomplete mixing (e.g. galaxies are not stellar
polytropes) but it may describe some of them. Therefore, Tsallis
distributions (polytropes) are not universal attractors, but they are
not useless neither. The results of this paper, and those reported in
\cite{campa1,campa2}, suggest that Tsallis distributions can be useful
to describe QSSs in some situations where the Lynden-Bell theory
fails. This happens precisely in the anomalous region discovered long
ago by Antoni \& Ruffo \cite{ar}, which stimulated all the activity on
the HMF model. Therefore, our findings tend to show that the
approaches of ordinary thermodynamics (Lynden-Bell) and generalized
thermodynamics (Tsallis) are complementary rather than antagonistic
(this is often the final outcome of a conflicting situation) \cite{epjb}.

We can also avoid entering in the polemic by taking the following
simple position \cite{cstsallis,aaantonov,incomplete,assise}. We can
argue that, in case of incomplete violent relaxation, the system
reaches a QSS that is a steady solution of the Vlasov equation on the
coarse-grained scale. This QSS differs from the Lynden-Bell prediction
but it is not possible to predict it as it depends strongly on the
dynamics.  Of course, we must select only nonlinearly
dynamically stable steady states and this leads to investigate the
variational problems of the form (\ref{dyn2})-(\ref{dyn3}). These
variational problems are {\it similar} to thermodynamical variational
problems\footnote{This is the ``Canada-Dry'' analogy: in these
variational problems, $S$ has the color of an entropy, but it is not
an entropy!} but they lead to dynamical stability criteria, not
thermodynamical stability criteria. This dynamical approach does not
attempt to explain {\it how} the QSSs are selected. It just provides
formal dynamical stability criteria for a large class of distribution
functions. In that dynamical approach, polytropic distributions just
appear as {\it particular steady states of the Vlasov equation}
\cite{cstsallis}. By contrast, the thermodynamical approach of
Lynden-Bell, or its generalization \cite{tsallis}, seeks to explain
{\it how} the QSSs are selected and {\it predict} them from general
(yet possibly arguable) considerations.  We think that dynamics and
thermodynamics are intermingled \cite{cstsallis,assise} and that much
work remains to be done in this area. Also, the methods of chaos can
be very useful to explain incomplete relaxation and lack of ergodicity
\cite{bachelardprl,firpo}. The subject is certainly not closed.

\appendix

\section{The condition of hydrostatic equilibrium}
\label{sec_hydroeq}

For any distribution function of the form $f=f(\epsilon)$ where $\epsilon=v^2/2+\Phi({\bf r})$ is the individual energy, using Eq. (\ref{ps11}), one has
\begin{eqnarray}
\label{hydroeq1}
\nabla p=\frac{1}{d}\nabla\Phi\int f'(\epsilon) v^2\, d{\bf v}=\frac{1}{d}\nabla\Phi\int \left (\frac{\partial f}{\partial {\bf v}}\cdot {\bf v}\right )\, d{\bf v}\nonumber\\
=-\frac{1}{d}\nabla\Phi \int f \nabla_{\bf v}\cdot {\bf v}\, d{\bf v}=-\nabla\Phi \int f\, d{\bf v}=-\rho\nabla\Phi,\nonumber\\
\end{eqnarray}
which is the condition of hydrostatic equilibrium (\ref{es0}).

\section{Derivation of the polytropic density and pressure laws}
\label{sec_poly}

For $n>d/2$, the polytropic DF can we written
\begin{eqnarray}
\label{pol1}
f=A(\epsilon_{m}-\epsilon)_{+}^{n-d/2}.
\end{eqnarray}
The density and the pressure can be expressed as
\begin{eqnarray}
\label{pol2}
\rho=AS_{d}Q_{0}(\Phi), \qquad p=\frac{1}{d}AS_{d}Q_{2}(\Phi),
\end{eqnarray}
with
\begin{eqnarray}
\label{pol3}
Q_{k}=\int_{0}^{\sqrt{2(\epsilon_{m}-\Phi)}}\left (\epsilon_{m}-\Phi-\frac{v^{2}}{2}\right )^{n-d/2}v^{k+d-1}dv.\quad 
\end{eqnarray}
Setting $x=v^{2}/\lbrack 2(\epsilon_{m}-\Phi)\rbrack$, we obtain
\begin{eqnarray}
\label{pol4}
Q_{k}=2^{(k+d-2)/2}(\epsilon_{m}-\Phi)^{n+k/2}\nonumber\\
\times\int_{0}^{1}(1-x)^{n-d/2}x^{(k+d-2)/2}dx. 
\end{eqnarray}
The integral can be expressed in terms of Gamma functions leading to 
\begin{eqnarray}
\label{pol5}
Q_{k}=2^{(k+d-2)/2}(\epsilon_{m}-\Phi)^{n+k/2}\nonumber\\
\times\frac{\Gamma((d+k)/2)\Gamma(1-d/2+n)}{\Gamma(1+k/2+n)}.
\end{eqnarray}
Then, the density and the pressure can be expressed in terms of the potential $\Phi$ as in Eqs. (\ref{es1})-(\ref{es2}).

For $n<-1$, the polytropic DF can we written
\begin{eqnarray}
\label{pol8}
f=A(\epsilon_{0}+\epsilon)^{n-d/2}.
\end{eqnarray}
The density and the pressure can be expressed as
\begin{eqnarray}
\label{pol9}
\rho=AS_{d}R_{0}(\Phi), \qquad p=\frac{1}{d}AS_{d}R_{2}(\Phi),
\end{eqnarray}
with
\begin{eqnarray}
\label{pol10}
R_{k}=\int_{0}^{+\infty}\left (\epsilon_{0}+\Phi+\frac{v^{2}}{2}\right )^{n-d/2}v^{k+d-1}dv.\quad 
\end{eqnarray}
Setting $x=v^{2}/\lbrack 2(\epsilon_{0}+\Phi)\rbrack$, we obtain
\begin{eqnarray}
\label{pol11}
R_{k}=2^{(k+d-2)/2}(\epsilon_{0}+\Phi)^{n+k/2}\nonumber\\
\times\int_{0}^{+\infty}(1+x)^{n-d/2}x^{(k+d-2)/2}dx. 
\end{eqnarray}
The integral can be expressed in terms of Gamma functions leading to 
\begin{eqnarray}
\label{pol12}
R_{k}=2^{(k+d-2)/2}(\epsilon_{0}+\Phi)^{n+k/2}\nonumber\\
\times\frac{\Gamma((d+k)/2)\Gamma(-k/2-n)}{\Gamma(d/2-n)}.
\end{eqnarray}
Then, the density and the pressure can be expressed in terms of the potential $\Phi$ as in Eqs. (\ref{es3})-(\ref{es4}).

\section{Derivation of the expression (\ref{ef1}) of the entropy}
\label{sec_deriv}

In the expression (\ref{ps1}) of the entropy, we have to evaluate $\int f^q\, d{\bf v}$. Let us first consider the case $q>1$. Using Eq. (\ref{vb19}) and defining $x=v^2/\lbrack 2(n+1)K\rho^{1/n}\rbrack$, we get
\begin{eqnarray}
\label{deriv1}
\int f^q\, d{\bf v}=\frac{1}{Z^q}\rho^\gamma S_d 2^{\frac{d-2}{2}}\left\lbrack (n+1)K\right\rbrack^{d/2}\nonumber\\
\times\int_0^1 (1-x)^{n+1-\frac{d}{2}} x^{\frac{d-2}{2}}\, dx.
\end{eqnarray}
The integral can be evaluated  in terms of $\Gamma$ functions and we obtain
\begin{eqnarray}
\label{deriv2}
\int f^q\, d{\bf v}=\frac{1}{Z^q}\rho^\gamma S_d 2^{\frac{d-2}{2}}\left\lbrack (n+1)K\right\rbrack^{d/2}\nonumber\\
\times \frac{\Gamma(d/2)\Gamma(2-d/2+n)}{\Gamma(2+n)}.
\end{eqnarray}
Using the property $\Gamma(x+1)=x\Gamma(x)$ and the expression (\ref{vb20}) of $Z$, the foregoing equation reduces to
\begin{eqnarray}
\label{deriv3}
\int f^q\, d{\bf v}=\frac{1}{Z^{q-1}}\rho^\gamma \frac{1-d/2+n}{1+n}.
\end{eqnarray}
Using the equation of state (\ref{es5}), we obtain
\begin{eqnarray}
\label{deriv4}
\int f^q\, d{\bf v}=\frac{1-d/2+n}{(n+1)KZ^{q-1}}p.
\end{eqnarray}
Combining Eqs. (\ref{vb20}) and (\ref{es6}), it is easy to establish that
\begin{eqnarray}
\label{deriv5}
(n+1)KZ^{q-1}=\frac{1}{A^{q-1}}.
\end{eqnarray}
Substituting this identity in Eq. (\ref{deriv4}), we find that
\begin{eqnarray}
\label{deriv6}
\int f^q\, d{\bf v}=\left (1-\frac{d}{2}+n\right )A^{q-1} p.
\end{eqnarray}
Recalling the definition of $A$ after Eq. (\ref{ps9}), we obtain
\begin{eqnarray}
\label{deriv7}
\int f^q\, d{\bf v}=\left (1-\frac{d}{2}+n\right )\beta \frac{q-1}{q} p.
\end{eqnarray}
After simplification, this yields the simple result
\begin{eqnarray}
\label{deriv8}
\int f^q\, d{\bf v}=\beta p.
\end{eqnarray}
Substituting this identity in Eq. (\ref{ps1}), we obtain Eq. (\ref{ef1}). The case $q<1$ can be treated similarly and yields the same final result.

\end{document}